\documentclass[%
 aip,
 amsmath,amssymb,
reprint,%
]{revtex4-1}

\usepackage{graphicx}%
\usepackage{dcolumn}%
\usepackage{bm}%
\usepackage[utf8]{inputenc}
\usepackage[T1]{fontenc}
\usepackage{mathptmx}
\usepackage{etoolbox}

\usepackage{amsbsy,booktabs}
\usepackage[dvipsnames]{xcolor}
\usepackage{comment}
\usepackage{bm}

\newcommand{\edit}[1]{#1}
\newcommand{\editr}[1]{#1}

\newcommand{\on}{q_{\rm on}}

\newcommand{\off}{q_{\rm off}}

\newcommand{\Off}{q_{\rm off}}
\newcommand{\kT}{ k_BT }
\newcommand{\dd}{\text{d}}

\newcommand{\kl}{ k_{\ell} }
\newcommand{\kb}{ k_{b} }
\newcommand{\lb}{ \ell_b} %
\newcommand{\xl}{ x_{\ell} }
\newcommand{\xlO}{x_{\ell,0}}
\newcommand{\Dl}{ D_{\ell} }

\newcommand{\etal}{ \eta_{\ell} }
\newcommand{\gammal}{ \gamma_{\ell} }

\newcommand{\piu}{ \pi_{\rm u} }
\newcommand{\pib}{ \pi_{\rm b} }
\newcommand{\Pu}{ \Pi_{\rm u} }
\newcommand{\Pb}{ \Pi_{\rm b} }
\newcommand{\dl}{\partial_{\ell}}
\newcommand{\dll}{\partial_{\ell\ell}}
\newcommand{\ks}{ k_{\rm slip} }
\newcommand{\Fb}{\mathcal{F}_{\rm b}}
\newcommand{\Fu}{\mathcal{F}_{\rm u}}
\newcommand{\homogenization}{averaging~} %
\newcommand{\Homogenization}{Averaging~}

\makeatletter
\def\@email#1#2{%
 \endgroup
 \patchcmd{\titleblock@produce}
  {\frontmatter@RRAPformat}
  {\frontmatter@RRAPformat{\produce@RRAP{*#1\href{mailto:#2}{#2}}}\frontmatter@RRAPformat}
  {}{}
}%
\makeatother
\begin{document}

\preprint{AIP/123-QED}

\title[Coarse-graining fast linkers]{Coarse-grained dynamics of transiently-bound fast linkers }

\author{Sophie Marbach}
 \affiliation{CNRS, Sorbonne Universit\'{e}, Physicochimie des Electrolytes et Nanosyst\`{e}mes Interfaciaux, F-75005 Paris, France }
  \affiliation{%
 Courant Institute of Mathematical Sciences, New York University,
NY, 10012, U.S.A.
}%
\author{Christopher E. Miles}%
 \email{sophie@marbach.fr, chris.miles@uci.edu; Both authors contributed equally to this work.}
\affiliation{ Department of Mathematics, University of California, Irvine, Irvine, USA, 92697.
}%

\date{April 25, 2023}%

\begin{abstract}
Transient bonds between fast linkers and slower particles are widespread in physical and biological systems. In spite of their diverse structure and function, a commonality is that the linkers diffuse on timescales much faster compared to the overall motion of the particles they bind to. This limits numerical and theoretical approaches that need to resolve these diverse timescales with high accuracy. Many models, therefore, resort to effective, yet \textit{ad-hoc}, dynamics, where linker motion is only accounted for when bound. This paper provides a mathematical justification for such coarse-grained dynamics that preserves detailed balance at equilibrium. Our derivation is based on multiscale \homogenization techniques and is broadly applicable. We verify our results with simulations on a minimal model of fast linker binding to a slow particle. We show how our framework can be applied to various systems, including those with multiple linkers, stiffening linkers upon binding, or slip bonds with force-dependent unbinding. Importantly, the preservation of detailed balance only sets the ratio of the binding to the unbinding rates, but it does not constrain the detailed expression of binding kinetics. We conclude by discussing how various choices of binding kinetics may affect macroscopic dynamics. 
\end{abstract}

\maketitle

Transient bonds between fast molecules and other slower molecules are found ubiquitously throughout physical systems. Such bonds enable momentum transfer at microscopic scales and are at the root of diverse phenomena, including linkers that tune the material properties of polymers~\cite{schallamach1963theory,leibler1991dynamics,cao2019rheological,lei2020entropy}, shape spatial organization and function of biological systems\edit{\cite{bidone2017morphological,descovich2018cross,korosec2021lawnmower,li2021biomechanics,bihr2012nucleation,bressloff2013stochastic}}, and determine the rate of tethered chemical reactions \cite{goyette2017biophysical}. Stochastic modeling of transient binding of linkers via numerical or theoretical approaches is, therefore, of widespread interest. 

The diversity of phenomena goes hand in hand with a variety of systems and hence of lexical terms: examples include cross-links in polymer meshes~\cite{schallamach1963theory,leibler1991dynamics}, metal-ligand bonds in self-assembled porous materials~\cite{gaillac2017liquid,balestra2022computer}, complementary DNA pairs hybridizing between colloids~\cite{mirkin1996dna,di2013multistep,feng2013specificity,wang2015crystallization,rogers2016using,xia2020linker,cui2022comprehensive,gehrels2022programming}, and fast myosin motors binding to slender actin fibers.\cite{bidone2017morphological,descovich2018cross,maxian2021simulations}
 Henceforth, the word \textit{linker} will refer to any molecule with \textit{(a)} a binding end that can form a bond with another molecule and \textit{(b)} that relaxes \textit{rapidly} to a finite length. In the previous examples, the linker is the polymer linker, ligand, DNA, or myosin molecule. The \textit{bond} will refer to the chemical bond formed between a linker and a slower molecule (or another linker). In the previous examples, the bond is the weak polymer-polymer adhesion, the metal-ligand chemical bond, the hybridized DNA section, or the high-affinity myosin head after ATP hydrolysis. 

 \begin{figure}[b!]
    \centering
    \includegraphics[width = \linewidth]{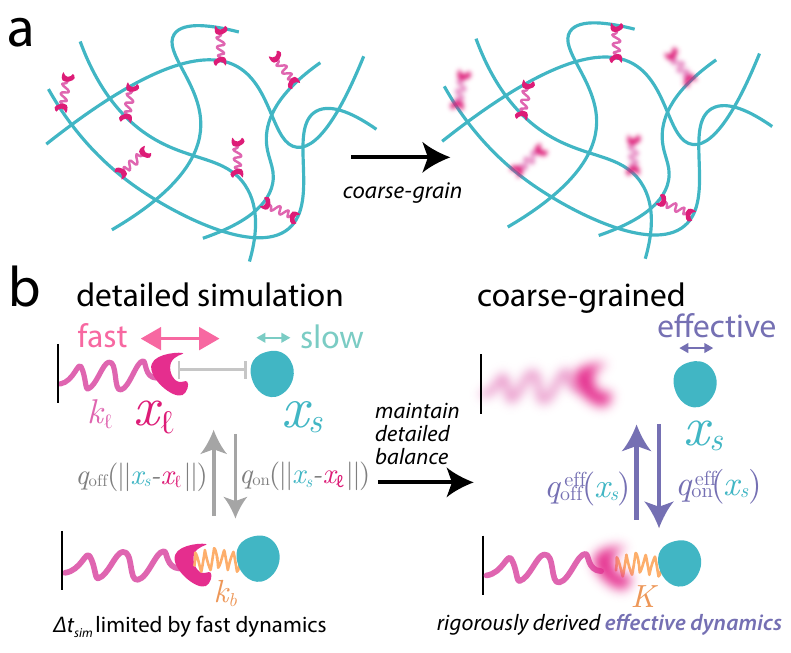}
    \caption{\edit{\textbf{Coarse-graining principle for a fast linker forming a transient bond}. (a) Cartoon of a common coarse-graining procedure where fast linkers (pink) between slow fibers (blue) are modeled only when they are bound. In this work, we study a minimal system where (b) a fast jiggling linker (pink) binds and unbinds rapidly to a binding site on a slow particle with rates $\on$ and $\off$ that depend on the relative distance. The 1D position of the linker is  $\xl$ and that of the slow particle $x_s$. In this paper, we use multiscale \homogenization techniques to coarse-grain the dynamics of the fast linker while preserving detailed balance. In the coarse-grained model, only the slow particle dynamics are specified, with effective kinetic rates $\on^{\rm eff}(x_s)$ and $\off^{\rm eff}(x_s)$ and forces that only depend on the position of the slow particle $x_s$. }}
    \label{fig:cartoon_summary}
\end{figure}

Despite enormous variations in mechanochemical properties of linking molecules, one unifying feature is their diffusion on characteristic timescales much faster than the overall motion of the fibers or objects they link.\edit{\cite{korosec2021substrate,marbach2022nanocaterpillar,PhysRevX.12.031030,bressloff2013stochastic}}. This rapid diffusion of (often many) linking molecules creates disparate time and length scales that must be resolved, often creating a bottleneck for numerical or theoretical investigations. \cite{muller2002coarse,gaillac2017liquid,balestra2022computer} To alleviate this, it is common to rely on coarse-grained descriptions of the linkers, either for their dynamics, binding kinetics, or both. In these coarse-grained scenarios, crosslinkers are replaced by effective laws so that their detailed dynamics need not be considered directly. \cite{fogelson2018enhanced,fogelson2019transport,jana2019translational,noe2018,maxian2021simulations,maxian2022interplay,korosec2018dimensionality,kowalewski2021multivalent,olah2013superdiffusive,merminod2021avidity,mitra2022coarse} For example, if a fast linker binds transiently to a slow particle, a coarse-grained simulation or model would only specify the dynamics of the slow particle; see illustration in Fig.~\ref{fig:cartoon_summary}. While the proposed effective dynamics may be relatively intuitive in some scenarios, there does not seem to be a systematic procedure for justifying and comparing this coarse-graining across different systems. \cite{jana2019translational,maxian2021simulations,maxian2022interplay} This paper provides a mathematical justification for such dynamics and, more importantly, a framework to derive effective dynamics in various settings. 

In establishing the justification for the effective dynamics with fast linkers, several questions arise.  Importantly, when linker dynamics are in thermodynamic equilibrium, with binding and unbinding rates that depend on mutual distance, detailed balance has to be enforced~\cite{marbach2022nanocaterpillar,zhang2022detailed,PhysRevX.12.031030}. An immediate consequence is that the equilibrium distribution of the system is determined by the binding and unbinding rates. As these detailed rates are often unmeasured, a variety of  choices are taken in the literature~\cite{fogelson2018enhanced,fogelson2019transport,jana2019translational,noe2018,maxian2021simulations,maxian2022interplay,korosec2018dimensionality,kowalewski2021multivalent,olah2013superdiffusive}. These choices of specific forms of binding and unbinding spatial dependence introduce ambiguity in interpreting the resulting dynamics. We, therefore, briefly discuss choices of kinetic rates and their consequences in the coarse-graining procedure.

The paper is organized as follows. We first consider a minimal system (Fig.~\ref{fig:cartoon_summary}-b) made of a single fast linker binding to a slow particle (Sec.~\ref{sec:generalSystem}) and catalog the variety of modeling choices at this microscopic level. 
Next, we rigorously coarse grain fast linker dynamics to obtain an effective model for the slow particle (Sec.~\ref{sec:coarsegraining}) that we validate with numerical simulations (Sec.~\ref{sec:numericalValidation}). We then show, with 3 examples, how we can apply our formalism to more complex yet common setups (Sec.~\ref{sec:applications}): we investigate (\textit{i}) 2 fast reactive linkers connecting to each other, (\textit{ii}) a linker that stiffens upon binding, and (\textit{iii}) a slip bond with force-dependent unbinding. 
Finally, we discuss how the choice of microscopic kinetic rates can affect the coarse-grained dynamics both at short and long timescales in nontrivial ways (Sec.~\ref{sec:consequences}). We hope our framework will help to investigate systems with fast transient crosslinkers more systematically.

\section{General system with transient crosslinkers}
\label{sec:generalSystem}

\subsection{Microscopic dynamics}

We consider the motion of a relatively slow particle representing, for example, a slender actin filament, a cell, or a colloid.\edit{\cite{maxian2021simulations,marbach2022nanocaterpillar,PhysRevX.12.031030}} 
The slow particle diffuses, for simplicity, in one spatial dimension (see Fig.~\ref{fig:cartoon_summary}-b, blue particle) and we will later discuss how to extend the results to 3D. The position of the slow particle at time $t$ is $x_s(t)$. The diffusion coefficient of the particle is $D_s = \kT/\gamma_s$, where $k_B$ is Boltzmann's constant, $T$ is temperature, and $\gamma_s$ is the friction coefficient of the particle. The particle evolves in an external potential $\mathcal{U}_s(x_s)$, that could represent connections with other particles. The forces in the unbound state on the particle are thus simply $\Fu = - \partial_s \mathcal{U}_s(x_s)$.

We also track the motion of a relatively fast linker, for example, a myosin head~\cite{maxian2021simulations} or the sticky ends of a single-stranded DNA filament.\cite{marbach2022nanocaterpillar,jana2019translational} The fast linker's position is $\xl$, and the linker diffuses with diffusion coefficient $\Dl = \kT/\gammal$, where $\gammal$ is the friction coefficient of the fast particle (see Fig.~\ref{fig:cartoon_summary}-b, pink linker). 
Here, the linker is attached to another slow object or an immobile surface; for now, we will consider it connected to a fixed point. This assumption ensures that the binding is localized in space. The linker usually resists extension, as it is made of a polymer or a protein that resists uncoiling.\cite{korosec2021substrate,marbach2022nanocaterpillar} It is hence reasonable to assume the linker is submitted to a recoil force, $ - \kl (\xl - \xlO)$ where $k$ is a spring constant,\cite{rubinstein2003polymer} and $\xlO$ is the rest length of the linker. Note that, as long as the force is conservative, extending our approach to other expressions is straightforward.

The unbound dynamics are
\begin{equation}
    \begin{cases}
    \frac{dx_s}{dt}  &= \frac{\Fu(x_s,\xl)}{\gamma_s} + \sqrt{\frac{2 \kT}{\gamma_s}} \eta_s(t)  \\
    &= - \frac{ \partial_s \mathcal{U}_s(x_s)}{\gamma_s} + \sqrt{\frac{2 \kT}{\gamma_s}} \eta_s(t) \\
    \frac{d\xl}{dt}  &= - \frac{\kl}{\gammal} (\xl - \xlO)  + \sqrt{\frac{2 \kT}{\gammal}} \etal(t)
    \end{cases}
        \label{eq:unbounddynamics}
\end{equation}
where the $\eta_i(t)$ are uncorrelated Gaussian white noises, where $\langle \eta_i(t) \rangle = 0$ and $\langle \eta_i(t) \eta_j(t') \rangle = \delta_{ij} \delta(t-t')$ where $\delta_{ij}$ is the Kronecker symbol and $\langle \cdot \rangle$ is an average over realizations of the noise. Without loss of generality, we will shift the domain such that the rest length of the fast variable is at the center of the domain, namely $\xlO = 0$.

The slow particle may transiently bind to the linker  (see Fig.~\ref{fig:cartoon_summary}-b, orange bond). In this entire paper, we will consider that when the bond is formed, it corresponds to a stiff spring with spring constant $\kb$ added between the particle and the linker. Hence, in the bound state, the forces on the slow particle are $\Fb = -\kb(x_s-\xl)  - \partial_s \mathcal{U}_s(x_s)$. 
The bound dynamics are thus
\begin{equation}
    \begin{cases}
     \frac{dx_s}{dt}   &= \frac{\Fb(x_s,\xl)}{\gamma_s} + \sqrt{\frac{2 \kT}{\gamma_s}} \eta_s(t) \\
     &= -\frac{\kb}{\gamma_s}(x_s-\xl) - \frac{ \partial_s \mathcal{U}_s(x_s)}{\gamma_s} + \sqrt{\frac{2 \kT}{\gamma_s}} \eta_s(t) \\
    \frac{d\xl}{dt}  &= \frac{\kb}{\gammal}(x_s-\xl)   - \frac{\kl}{\gammal} (\xl)  + \sqrt{\frac{2 \kT}{\gammal}} \etal(t).
    \end{cases}
    \label{eq:bounddynamics}
\end{equation}

Our model for the bound dynamics is not unique. For example, one could consider that instead of a stiff spring, the bond formed is a rigid rod constraining the dynamics~\cite{marbach2022mass,marbach2022nanocaterpillar,kowalewski2021multivalent,korosec2018dimensionality}. Yet, we choose this spring bond model as it is a simple starting point and because it is physically satisfying. Indeed, with the spring model, the linker and the particle relax towards each other, while in the rigid rod model, they can stay unnaturally far apart.  Ultimately we will consider the dynamics in the limit where the bond is very stiff, corresponding to a so-called \textit{soft} constraint~\cite{morse2003Theory,ciccotti2008projection,holmes2016stochastic}. %

The equilibrium distribution corresponding to these choices of dynamics can be decomposed over the 2 states (bound and unbound) as $\pi = \left( \piu, \pib \right)^T$. The component of the equilibrium distribution corresponding to the unbound  state is
\begin{equation} 
    \piu (x_s,\xl) = \frac{1}{Z_u} 
    e^{-\mathcal{U}_s(x_s)/\kT - \kl (\xl)^2/2\kT} 
\end{equation}
and the bound one is
\begin{equation} 
    \pib (x_s,\xl) = \frac{1}{Z_b} 
    e^{- \kb(x_s - \xl)^2/2\kT -\mathcal{U}_s(x_s)/\kT - \kl (\xl)^2/2\kT},
\end{equation}
where $Z_u$ and $Z_b$ are constant prefactors that are set by a normalization condition on the total equilibrium distribution $\int \dd \xl \dd x_s \left( \piu + \pib \right) = 1$ and by detailed balance, which we turn to now. 

\subsection{Possible kinetic rates and detailed balance}
\label{sec:possibleRates}

We consider that the linker and the particle bind to each other with rate $\on$ and unbind with rate $\off$. To be physically accurate, it is reasonable to assume that both rates may depend on the spatial variables ($x_s,\xl$), \textit{a priori}. While the exact expression of the rates for our coarse-graining approach does not matter, it is crucial to recall how these rates should be specified to satisfy 
detailed balance.

If the system we consider is at equilibrium, the rates \textit{must satisfy detailed balance}~\cite{zhang2022detailed,marbach2022nanocaterpillar}. Here this means the probability flux at equilibrium of going from one state to the other is equal to the inverse flux, namely
\begin{equation}
     \piu (x_s,\xl) \on (x_s,\xl) = \pib (x_s,\xl) \off(x_s,\xl).
     \label{eq:DetailedBalance0}
\end{equation}
Here this relation simplifies to
\begin{equation}
    \frac{\on}{\off} = \frac{Z_u}{Z_b} e^{- \kb(x_s - \xl )^2/2\kT}.
\end{equation}
To make this expression more explicit, we can redefine the constants $Z_u = Z$ and $Z_b = Z \off^0/\on^0$, where $Z$ is a global normalization constant such that $\int \dd \xl \dd x_s \left( \piu + \pib \right) = 1$. Here $\off^0$ and $\on^0$ set the typical range of the kinetic rates and are related via the typical free energy of bond formation $E_0$ such that $\on^0 / \off^0 \equiv e^{-E_0/\kT}$. 
 We obtain 
\begin{equation}
    \frac{\on}{\off} = \frac{\on^0}{\off^0} e^{- \kb(x_s - \xl)^2/2\kT}.
\end{equation}
Since this is the only relation that constrains $\on$ and $\off$, it is clear that \textit{the choice of $\on$ and $\off$ is not unique} and that at least one of the rates has to \textit{depend on space}.

\begin{figure}[h!]
    \centering
    \includegraphics[width=\linewidth]{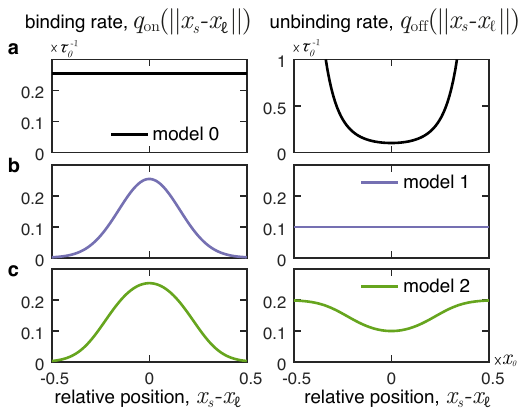}
    \caption{Possible choices of binding and unbinding kinetics agreeing with detailed balance. (a) constant binding rate, (b) constant unbinding rate, and (c) both rates varying; see text for detailed expressions.  Here we chose a bond spring constant $\kb L^2/\kT = 40$ and a fixed macroscopic probability $\Pi_b = 0.5$.  \label{fig:qChoices}}
\end{figure}

\subsubsection{Possible expressions of the rates}
Therefore, there are infinite possibilities in how we can specify rates consistent with detailed balance, that reflects the diversity of choices made in the literature.\edit{\cite{bell1978models,dembo1988reaction,fogelson2018enhanced,fogelson2019transport,jana2019translational,noe2018,maxian2021simulations,maxian2022interplay,korosec2018dimensionality,kowalewski2021multivalent,olah2013superdiffusive,PhysRevX.12.031030,bihr2015multiscale,mitra2022coarse}}. However, we can catalog a few simple commonly chosen examples, see Fig.~\ref{fig:qChoices}, and discuss to what extent they are consistent with physical intuition. 
\begin{itemize} \item (model 0) Unbinding is faster further away from the target, but the binding rate is constant, \begin{equation}
    \begin{cases}
&        \on = \on^0 , \,\, \\
 &        \off(x_s,\xl) = \off^0 e^{ \kb(x_s - \xl + \lb)^2/2\kT}
        \end{cases}
                \label{eq:choiceOff}
    \end{equation}
    \edit{This binding term provides the convenient feature of avoiding the need to resolve detailed dynamics of the fast linker. Further, the unbinding form is motivated by the intuitive observation that bonds break faster when a larger force is exerted. One such example is molecular motor detachment,\cite{berger2019force,klobusicky2020effective,bovyn2021diffusion} although this is often modeled as a slip bond, which we discuss in Sec.~\ref{sec:applications}.} 
    However, we note that this choice suffers from the numerically undesirable feature of the off rate increasing exponentially as a function of distance. 
    \item (model 1) Binding is faster closer to the target, but unbinding is constant, \begin{equation}
    \begin{cases}
        &\on(x_s,\xl) = \on^0 e^{- \kb(x_s - \xl)^2/2\kT}, \,\, \\
        &\off = \off^0,
        \end{cases}
        \label{eq:choiceOn}
    \end{equation}
    which is a model used in Refs.~\onlinecite{fogelson2019transport,maxian2022interplay,mitra2022coarse}. 
    Constant unbinding seems quite unphysical since one would expect the bond to break if the linker and the particle are brought too far apart. 
     We can instead opt for an expression where both rates depend on space, for example:
    \item (model 2) Binding is faster closer to the target, unbinding is faster further away from the target,
    \begin{equation}
    \begin{cases}
        &\on(x_s,\xl) = \displaystyle \frac{\on^0}{1 +  e^{\kb(x_s - \xl)^2/2\kT}}, \,\, \\
        &\off(x_s,\xl) =\displaystyle  \frac{\off^0}{1 +  e^{-\kb(x_s - \xl)^2/2\kT}}
        \end{cases}
                        \label{eq:choiceOnOff}
    \end{equation}
    which solves the issues raised above. 
\end{itemize}
While these are possible choices of the binding rates that agree with detailed balance, these are not the only ones, and one could consider various other kinetics, possibly with a kinetic barrier to overcome to unbind or to bind. 

One might then wonder if the choice of the microscopic kinetic rates affects the long-time dynamics of the system. We raise the question here but we will not attempt to answer it thoroughly. Rather, our goal is, once the microscopic dynamics are properly chosen, to show how to systematically coarse-grain the dynamics of the fast linker.

\subsubsection{Binding models with uniform kinetic rates}

Inspired by a previous model,\cite{goodrich2018enhanced} we also discuss an alternative binding scheme. Specifically, if one wanted the binding and unbinding rates to be uniform, at least over a certain lengthscale, the \textit{only} option is that the dynamics are not specified via Eqs.~\eqref{eq:unbounddynamics} and \eqref{eq:bounddynamics} but have to be changed. We illustrate this point briefly below.

For example, one could define uniform rates, as
\begin{equation}
        \begin{cases}
        &\on = \on^0, \,\, \\
        &\off = \off^0
        \end{cases}
    \end{equation}
where the rates $\on^0$ and $ \off^0$ are non zero. In that case, the bound $\pib$ and unbound $\piu$ parts of the equilibrium distribution must be the same, which constrains the forces to be the same in each state. Otherwise, detailed balance is broken. For example, the bound and unbound dynamics could both satisfy
\begin{equation}
    \begin{cases}
     \frac{dx_s}{dt}  &= -\frac{\kb}{\gamma_s}(x_s-\xl)  - \frac{ \partial_x \mathcal{U}_s(x_s)}{\gamma_s} + \sqrt{\frac{2 \kT}{\gamma_s}} \eta_s(t)  \\
    \frac{d\xl}{dt}  &= \frac{\kb}{\gammal}(x_s-\xl)   - \frac{\kl}{\gammal} (\xl)  + \sqrt{\frac{2 \kT}{\gammal}} \etal(t).
    \end{cases}
    \label{eq:dynamicsGoodrich}
\end{equation}
As a result, the dynamics are not very interesting as they simply amount to a global quadratic confining potential with spring constant $\kb$. 

Tethered dynamics are more in line with physical intuition if the confining potential associated with the bond is restricted to a patch in space. 
Following Ref.~\onlinecite{goodrich2018enhanced,doi1976stochastic}, this can be achieved by defining kinetic rates that are non-zero only when the linker and the particle are close, typically
\begin{equation}
\begin{cases}
     & \on = \on^0 H(|x_s-\xl| < R), \,\, \\
     &\off = \off^0 H(|x_s-\xl| < R)
\end{cases}  
\label{eq:onDoi}
    \end{equation}
    where $R$ is a characteristic distance that sets a reasonable maximum binding distance~\cite{korosec2018dimensionality,kowalewski2021multivalent}. One can define $R = \sqrt{2\kT/\kb}$ where $\kb$ is the spring constant of the bond, and $R$ corresponds, therefore, to thermal vibrations around that bond. When the linker and the particle are nearby $|x_s-\xl| < R$, the dynamics follow Eq.~\eqref{eq:dynamicsGoodrich} while they are given by Eqs.~\eqref{eq:unbounddynamics} and \eqref{eq:bounddynamics} otherwise. This ensures that detailed balance is satisfied in both regions, since when $|x_s-\xl| \geq R$ the kinetic rates are zero.  We refer to models with uniform kinetic rates as in Eq.~\eqref{eq:onDoi}, as Doi models~\cite{agbanusi2014comparison,zhang2019influence,zhang2022detailed} in analogy with reaction models without linkers.

However, such uniform kinetic rates are unphysical in 2 ways: (\textit{i}) bond dissociation is not allowed when the bond is stretched beyond $R$, which seems rather unlikely; and (\textit{ii}) when they are close, the interaction between the linker and particle is the same regardless of the state of the bond -- see Eq.~\eqref{eq:dynamicsGoodrich}), which also seems rather unlikely. 

We will briefly show in Sec.~\ref{sec:consequences} how also at the coarse-grained level, important physical differences arise between Doi models with uniform kinetic rates as in Eq.~\eqref{eq:onDoi}, and models with spatially dependent kinetic rates as say in Eq.~\eqref{eq:choiceOnOff}. %
This means, again, that the choice of kinetic rates affects short-time dynamics at least in non-trivial ways, and hence that is an important assumption of any model.

\section{Coarse-grained dynamics with fast linkers}
\label{sec:coarsegraining}

Our aim is now to rigorously coarse-grain the linker dynamics to obtain the effective long-time dynamics of the particle. This is useful both to gain analytic insight, but also to propose consistent and fast simulation schemes. We use multiscale \homogenization\cite{pavliotis2008multiscale} to coarse-grain the dynamics, a technique which is broadly used in the field to properly average over short length scales and timescales~\cite{marbach2022mass,marbach2022nanocaterpillar,fogelson2018enhanced,fogelson2019transport,klobusicky2020effective,bressloff2013stochastic}. We will show the approach is valid over a broad range of parameters in Sec.~\ref{sec:numericalValidation}. We provide Table~\ref{tab:notations} as a summary of the main notations used throughout the discussion.

\subsection{Set up of the dynamics}

The set of stochastic Eqns.~\eqref{eq:unbounddynamics}-\eqref{eq:bounddynamics} 
defines a Markov process that
is conveniently studied via the Kolmogorov backward  equation~\cite{gardiner1985handbook,pavliotis2008multiscale} on the functions $f(x_s,\xl,t) = \left( f_u(x_s,\xl,t) , f_b(x_s,\xl,t) \right)^T$
\begin{equation}\label{eq:backward}
    \partial_t f = \mathcal{L} f\,, \quad f(x_s,\xl,0) = g(x_s,\xl)\,,
\end{equation}
where $\mathcal{L}$ is the generator of the system and $g$ is any scalar function. Here the functions $f(x_s,\xl,t) = \int p(x_s',\xl',t|x_s,\xl) g(x_s',\xl') \dd \xl' \dd x_s'$ give the expectation of any scalar function $g(x_s(t),\xl(t))$, given an initial condition $x_s(0)=x_s,\xl(0)=\xl$, where $p$ is the probability density that the system evolved from the initial condition to $(x_s',\xl')$ at time $t$. Once we know how such functions $f$ evolve, we may calculate any statistic $g$ of our stochastic process. %
 
The generator $\mathcal{L}$ can be calculated from the forward equation, the Fokker Planck equation, associated with the dynamics Eqns.~\eqref{eq:unbounddynamics}-\eqref{eq:bounddynamics}, see \edit{the Supplementary Information Sec.~1 for further details}. %
The full generator can be written as 
\begin{equation}
    \mathcal{L} = \mathcal{Q} + \mathcal{V}
    \end{equation}
    where
    \begin{align*}
    \mathcal{Q} &= \begin{pmatrix} - \on & \on \\ \off &   -\off \end{pmatrix}, \,\,\, \text{and} \,\, \mathcal{V} = \begin{pmatrix}  \mathcal{V}_u & 0 \\ 0 &    \mathcal{V}_b \end{pmatrix}  \,\, \text{with}
    \\
    \mathcal{V}_u & = - \frac{\kl}{\gammal}\xl \dl + \frac{\kT}{\gammal} \dll - \frac{\partial_x \mathcal{U}(x_s)}{\gamma_s} x_s \partial_s + \frac{\kT}{\gamma_s} \partial_{ss} \\
   \mathcal{V}_b &= 
 - \frac{\kl}{\gammal}\xl \dl + \frac{\kT}{\gammal} \dll - \frac{\partial_x \mathcal{U}(x_s)}{\gamma_s} x_s \partial_s + \frac{\kT}{\gamma_s} \partial_{ss} \\
&\qquad - \frac{\kb}{\gammal}(\xl-x_s)\dl -  \frac{\kb}{\gamma_s}(x_s - \xl)\partial_s. 
\end{align*}

\subsection{Nondimensionalization}

To highlight the ratio between the different temporal and spatial scales at play, we non-dimensionalize our equations. 
Since the particle's motion is much slower than that of the linker, we can identify a small number $\varepsilon = \gammal/\gamma_s$.
We seek the dynamics of the slow particle over a typical \edit{long} timescale $\tau_0$ such that its motion is diffusive, extending over the range $x_0 = \sqrt{D_0 \tau_0}$. 
We now can call $q_{\rm on}' =  q_{\rm on}\tau_0 $ and similarly $\off' = \off \tau_0$, remembering that these are both functions of space. We consider for now that we observe the dynamics over timescales $\tau_0$ where transient binding and unbinding are still relevant such that $\on' = O(1)$. Typical slow dynamics are captured by the timescale $\tau_0$ so again we consider $\kappa = \kl \tau_0/\gamma_s = O(1)$. We also write $\lambda = \kb \tau_0/\gamma_s= O(1)$ for now. Since $\varepsilon = \gammal/\gamma_s$, we have $\kl \tau_0/\gammal = \kl \tau_0/\gamma_s \varepsilon$.

In this scaling, we have the nondimensional generator $ \mathcal{L}' =\frac{1}{\varepsilon}\mathcal{L}'_0 + \mathcal{L}'_1$ such that
\begin{equation*}
    \mathcal{L}'_0  =  \begin{pmatrix} -  \kappa \xl \dl +  \dll  & 0  \\ 0   & -  \kappa \xl \dl   - \lambda(\xl-x_s )\dl + \dll  \end{pmatrix}
\end{equation*}   
and
{\scriptsize\begin{equation*}
     \mathcal{L}'_1= \begin{pmatrix} - \on'  - \frac{ \partial_s \mathcal{U}(x_s)}{k_BT} \partial_s + \partial_{ss} & \on'  \\ \off' &  -\off' -  \lambda (\xl-x_s)\partial_s  - \frac{ \partial_s \mathcal{U}(x_s)}{k_BT} \partial_s + \partial_{ss} \end{pmatrix}.
\end{equation*}} 
In the following, we will drop the $\cdot'$ notations for simplicity. 

\subsection{\Homogenization procedure}

We then look for a solution to $\partial_t f = \mathcal{L}f$ of the form $f = f_0 + \epsilon f_1 + \epsilon^2 f_2$. To first order, we have
\begin{align}
 \mathcal{L}_0 f_0 = \begin{pmatrix} -  \kappa \xl \dl +  \dll  & 0  \\ 0   & -  \kappa \xl \dl   - \lambda(\xl-x_s )\dl + \dll  \end{pmatrix} f_0 = 0
\label{eq:generator1}
\end{align}
Notice that $ -  \kappa \xl - \lambda(\xl-x_s) = - (\kappa+ \lambda) \left( \xl-x_s\frac{\lambda }{\kappa+ \lambda}\right)$
such that the general solution to Eq.~\eqref{eq:generator1} is
\begin{equation}
\begin{split}
    f_0 &=  \begin{pmatrix} g_1(x_s,t) \\  g_2(x_s,t) \end{pmatrix} + \begin{pmatrix} h_1(x_s,t) \\ 0 \end{pmatrix} \int_0^{\xl} e^{\edit{+}\kappa x^2}\, \dd x  \\
    & \qquad + \begin{pmatrix} 0 \\ h_2(x_s,t) \end{pmatrix} \int_0^{\xl} e^{\edit{+}(\kappa+\lambda)(x-\frac{\lambda}{\kappa+ \lambda}x_s)^2}\, \dd x 
    \end{split}
\end{equation}
and for which the associated equilibrium distribution is of the form
\begin{equation}
    \pi_0 = \frac{1}{\sqrt{2\pi}} \begin{pmatrix}
    \alpha e^{-\kappa \xl^2/2} \\ \beta e^{-(\kappa+\lambda) (\xl - \frac{\lambda}{\kappa+ \lambda}x_s)^2/2}
    \end{pmatrix} .
\end{equation}
Here $\alpha$ and $\beta$ are free parameters and are not constrained by detailed balance; they can be any real numbers. In fact, there is no exchange between the bound and the unbound times at these very short timescales. Notice that this means that the \homogenization approach does not know \textit{a priori} that it should preserve detailed balance. 

In any case, we require $\langle f_0 , \pi_0 \rangle_\ell$ to be bound; $\langle f,p  \rangle_\ell$ is the inner product where the integration is only carried over the fast variable $x_\ell$. This imposes $h_1 \equiv  0$ and $h_2 \equiv 0$, and $f_0 = \begin{pmatrix} g_1(x_s,t) & g_2(x_s,t) \end{pmatrix}^T$ is actually independent of the fast variable. 

Seeking the next order $\mathcal{L}_0 f_1 = - \mathcal{L}_1 f_0 + \partial_t f_0$.
A solution exists for $f_1$ if the Fredholm alternative is satisfied~\cite{pavliotis2008multiscale}, namely if $\langle (\partial_t f_0-  \mathcal{L}_1 f_0).\pi_0 \rangle_\ell = 0$ is true for any $\pi_0$ in the nullspace of $\mathcal{L}_0^{\star}$. This corresponds to any real value combinations of $\alpha$ and $\beta$, and hence we may pick the convenient choice of  $(\alpha,\beta) = (1,0)$ and $(\alpha,\beta) = (0,1)$.
We obtain 
\begin{equation}
   \begin{split}
        \partial_t g_1 &= - \frac{\int e^{-\kappa \xl^2/2} q_{\rm on} \dd \xl }{\int e^{-\kappa \xl^2/2} \dd \xl} (g_1- g_2) \edit{- \frac{ \partial_s \mathcal{U}(x_s)}{k_BT} \partial_s g_1} +  \partial_{ss} g_1   \\
        \partial_t g_2 & = \frac{\int e^{-(\kappa+\lambda) (\xl  -\frac{\lambda}{\kappa+ \lambda}x_s)^2/2} q_{\rm off} \dd \xl }{\int e^{-(\kappa+\lambda) (\xl  -\frac{\lambda}{\kappa+ \lambda}x_s)^2/2} \dd \xl }  (g_1  -   g_2) \edit{- \frac{ \partial_s \mathcal{U}(x_s)}{k_BT} \partial_s g_2}   \\&\quad + \partial_{ss} g_2 
        - \frac{\lambda \left( \int (\xl-x_s) e^{-(\kappa+\lambda) (\xl  -\frac{\lambda}{\kappa+ \lambda}x_s)^2/2} d\xl \right) }{\int e^{-(\kappa+\lambda) (\xl  -\frac{\lambda}{\kappa+ \lambda}x_s)^2/2} \dd \xl } \partial_s g_2.
    \end{split}
    \label{eq:coarseGrainedBackward}
\end{equation}

\subsection{Effective kinetic rates and dynamics}
We have just obtained the coarse-grained backward equations for the dynamics.

We can carry over the last line's integral and return to dimensional units, to directly read off the effective on and off rates
\begin{equation}
\begin{cases}
    \on^{\rm eff}(x_s) &=  \frac{\int e^{- \kl \xl^2/2\kT} q_{\rm on} (\xl,x_s) \dd \xl }{\int e^{-\kl \xl^2/2\kT} \dd \xl}, \\
    \off^{\rm eff}(x_s) &=  \frac{\int e^{-(\kl+\kb) (\xl  -\frac{\kb}{\kl+ \kb}x_s)^2/2\kT} q_{\rm off} (\xl,x_s)\dd \xl }{\int e^{-(\kl+\kb) (\xl  -\frac{\kb}{\kl+ \kb}x_s)^2/2\kT} \dd \xl }.
    \label{eq:ratesEff}
    \end{cases}
\end{equation}
We find that, coherently, the effective on and off rates are weighted averages over the distribution of positions of the fast variable in the respective states. \edit{A similar expression for the coarse-grained rates was obtained from a first principles derivation at equilibrium~\cite{PhysRevX.12.031030}.}

Importantly, \edit{beyond the coarse-grained rates}, from Eq.~\eqref{eq:coarseGrainedBackward} we can also read off
the \textit{coarse-grained dynamics} of the slow particle \edit{at $O(1)$ in the small parameter $\varepsilon= \gammal/\gamma_s$,} in the unbound and bound states
\begin{equation}
\begin{cases}
    \frac{dx_s}{dt} &= - \frac{\partial_s \mathcal{U}(x_s)}{\gamma_s}   + \sqrt{\frac{2 \kT}{\gamma_s}} \eta_{\rm u}(t) \,\, \text{(unbound)}\\
    \frac{dx_s}{dt} &= - \frac{\partial_s \mathcal{U}(x_s)}{\gamma_s}  - \frac{K}{\gamma_s} x_s  + \sqrt{\frac{2 \kT}{\gamma_s}} \eta_{\rm b}(t)  \,\, \text{(bound).}
\end{cases}
\label{eq:dynamicsEff}
\end{equation}
where $K = \kl\kb/(\kl + \kb)$ is an effective spring constant.\edit{In Appendix~\ref{app:db_at_coarsegrain} we verify that this coarse graining maintains detailed balance at the macroscopic level.} 

\edit{Overall, the coarse-grained dynamics we obtain are consistent with physical intuition, at this lowest order in $\varepsilon$}. 
The unbound dynamics of the \edit{slow} particle are not changed by the coarse-grained approach. However, in the bound state, an extra recoil force is exerted on the particle, corresponding to a force averaged over the fast-moving linker. Interestingly, the spring constant $K$ of the bond formed at the coarse-grained level corresponds to the effective spring constant corresponding to two springs in series, with spring constants $\kl$ and $\kb$. This is precisely what is expected physically and from the sketch of the setup; see Fig.~\ref{fig:cartoon_summary}-b. One novelty of this calculation is that it that allows one to give meaning to the spring constant $K$ used in effective models such as those used in Refs.~\onlinecite{fogelson2019transport,maxian2022interplay,mitra2022coarse}. \edit{Both in the bound and unbound states the coarse-grained dynamics are damped by the friction coefficient $\gamma_s$. Although this is rather intuitive, notice again that this is only true when $\varepsilon$ is small enough, otherwise the effective friction would be increased by the presence of the linker.\cite{marbach2022nanocaterpillar,fogelson2018enhanced}} 
\editr{All in all, the lowest-order dynamics evolve as if the linker were moving so fast that it loses memory of previous binding and unbinding events.}

\editr{Beyond these $O(1)$ terms in the separation of scales $\varepsilon = \gammal/\gamma_s$, we can derive further terms at $O(\epsilon)$ and beyond, which do account for more and more memory between binding events. This can be done by proceeding with the coarse-graining approach explained above to further terms in the expansion. We report in the Supplementary Information Sec.~2, the coarse-grained dynamics at $O(\varepsilon)$. The effect of memory is 2-fold: it modifies the binding rates which now contain $\on \off$, $\on^2$ and $\off \off$ couplings; and it also modifies the forces. In particular, at $O(\varepsilon)$ there is now a force in the unbound state, which arises from a remnant memory of the recoil force on the bound fast linker as the linker unbinds. Finally, at $O(\varepsilon)$, diffusion in the bound state is now damped by the presence of the linker. Our coarse-graining approach is thus a robust tool to systematically derive coarse-grained dynamics to any order.}

\subsection{General coarse-grained dynamics}

The \homogenization approach can be extended in a straightforward way, following the averaging steps above, to more arbitrary dynamics, and we summarize effective dynamics in full generality in Table~\ref{tab:effectiveQuantities}. %
The initial forces on the slow particle in the bound $\Fb(x_s,\xl)$ and the unbound $\Fu(x_s,\xl)$ states can be arbitrary forces. All forces on the particle and linker should be conservative so as to define an equilibrium distribution. Table~\ref{tab:effectiveQuantities} then reports the general formulas for the \edit{effective dynamics}, effective binding and unbinding rates as well as the effective force on the particle in the unbound and bound states. 
The diffusive part of the slow particle motion is not affected by the coarse-graining procedure, since we assume diffusion coefficients do not depend on space. Finally, the formulas can be extended in a straightforward way to 3D dynamics, and to multiple fast linkers. We will show in Sec.~\ref{sec:applications} how to use these formulas with specific examples.

\begin{table*}
  \centering
\caption{Summary of the effective dynamics considering a coarse-grained fast linker ($\xl$) that can bind to a slow particle ($x_s$). \edit{The bare dynamics of the slow particle are damped by a friction coefficient $\gamma_s$.} For 3D coordinates, the integrals have to be carried over all the spatial dimensions of the fast variable $\xl$ and the force field on the slow particle has to be obtained coordinate by coordinate.}\label{tab:effectiveQuantities}
\rule{\textwidth}{\heavyrulewidth}
\vspace{-\baselineskip}
\begin{flalign}
&\edit{\text{effective dynamics}} &  \edit{\frac{dx_s}{dt} =  \frac{1}{\gamma_s} \mathcal{F}_{\mathrm{u/b}}^{\rm eff}(x_s) + \sqrt{\frac{2\kT}{\gamma_s}} \eta_{\rm u/b}(t)}
&&&\label{eq:dynamics}
\end{flalign}
\vspace{-\baselineskip}
\begin{flalign}
&\text{effective binding rate} &  \displaystyle \on^{\rm eff}(x_s) = \frac{\int \on(x_s,\xl)  \piu (x_s,\xl) d\xl}{\int   \piu (x_s,\xl) d\xl}
&&&\label{eq:TableOn}
\end{flalign}
\vspace{-\baselineskip}
\begin{flalign}
&\text{effective unbinding rate}
&\displaystyle \off^{\rm eff}(x_s) = \frac{\int \off(x_s,\xl)  \pib (x_s,\xl) d\xl}{\int   \pib (x_s,\xl) d\xl}
&&&\label{eq:TableOff}
\end{flalign}
\vspace{-\baselineskip}
\begin{flalign}
&\text{effective force on the slow unbound particle}
&\Fu^{\rm eff}(x_s) = \frac{\int \Fu(\xl,x_s)  \piu (x_s,\xl) d\xl}{\int   \piu (x_s,\xl) d\xl}
&&&\label{eq:TableOffForce}
\end{flalign}
\vspace{-\baselineskip}
\begin{flalign}
&\text{effective force on the slow bound particle}
&\displaystyle\Fb^{\rm eff}(x_s) = \frac{\int \Fb (\xl,x_s)  \pib (x_s,\xl) d\xl}{\int   \pib (x_s,\xl) d\xl}
&&&\label{eq:TableOnForce}
\end{flalign}
\rule{\textwidth}{\heavyrulewidth}
\end{table*}

Notice how each formula from Eq.~(\ref{eq:dynamics}-\ref{eq:TableOnForce}) can be interpreted intuitively. The effective force in a given state or the rate of switching from that state is simply the spatial average weighted by the local probability distribution to be in that state. While this seems consistent \textit{a posteriori}, and consistent with detailed balance at the coarse grained level, the formulas in Eq.~(\ref{eq:TableOn}-\ref{eq:TableOff}) are not the only possible expressions for the effective rates that obey detailed balance \edit{and accurately describe the dynamics at lowest order in $\varepsilon$}. The results of Eq.~(\ref{eq:dynamics}-\ref{eq:TableOnForce}) are therefore not trivial. \edit{Notice that in a related work a similar expression for the coarse-grained rates $\off^{\rm eff}$ and $\on^{\rm eff}$ was obtained using averaging techniques (Eq.~(6) or Eq.~(19) of Ref.~\onlinecite{PhysRevX.12.031030}). However, a crucial addition here is that we provide also the coarse-grained dynamics of the slow particle, through the effective equations of motion Eq.~\eqref{eq:dynamics} and effective forces Eqs.~(\ref{eq:TableOffForce}-\ref{eq:TableOnForce}).}

\edit{Interestingly, the formulas in Table~\ref{tab:effectiveQuantities} are straightforward to compute. One simply needs to know the equilibrium probability distribution of the free and bound linker to obtain the effective dynamics. Specifically, one only needs to know the dependence of the equilibrium distribution on the linker coordinate $\xl$, and not the details of the landscape for the slow particle $x_s$. %
In practice, one could then simulate (or calculate) a free and bound linker, obtain their local probability distributions, and integrate them to obtain the effective dynamics. We note the caveat that the full probability distribution needs to be evaluated if direct interactions between linkers exist.}

\subsection{Limit regimes}

We will now comment on the effective dynamics and kinetic rates obtained for the minimal system of Fig.~\ref{fig:cartoon_summary}-b (Eqs.~\eqref{eq:unbounddynamics} and \eqref{eq:bounddynamics}) in a few limiting regimes of the system parameters \edit{$\kl$ and $\kb$}. In the following, it will be useful to specify the expression of the rates and we will use for example Eq.~\eqref{eq:choiceOn}. 

In all limiting regimes, some variables such as $\on^0/\off^0$ have to be specified with respect to the limits. In practice, the macroscopic probability of being bound, $\Pb$, (or unbound, $\Pu$) can be measured experimentally~\cite{reiter2019force,marbach2022nanocaterpillar}, and is easier to probe than spatially dependent kinetic rates. Hence, we constrain the different functional forms for the kinetic rates to predict the same $\Pb$. 
The macroscopic probability of being bound (respectively unbound) is $\Pb = \int dx_s d\xl \pib (x_s, \xl)$ (resp. $\Pu = \int dx_s d\xl \piu (x_s, \xl)$). It is easy to show that here
\begin{equation}
    \frac{\Pb}{\Pu} =   e^{-E_0/\kT} \sqrt{\frac{K}{\kb}} \frac{\int e^{-\mathcal{U}_s(x_s)/\kT} e^{- k_{\rm eff} x_s^2/2\kT} \dd x_s}{ \int e^{-\mathcal{U}_s(x_s)/\kT}  \dd x_s}
    \label{eq:limitconstraint}
\end{equation} such that the macroscopic bound and unbound probabilities measure in particular the energy of the bond, weighted by the geometry of the system. 
Of course, this does not constrain the local values of the kinetic rates. Rather, it sets the value of some parameters, here of $E_0$.

\subsubsection{Stiff bond.}
\label{sec:stiffbond}
We first investigate the limit regime where the bond is very stiff, \textit{i.e.} $\kb \gg \kl$. This is the so-called ``soft constraint'' limit~\cite{morse2003Theory,ciccotti2008projection,holmes2016stochastic}. In that case, we expect the fast and slow particles are constrained to move synchronously in the bound state.
For this limit to make sense, we need, according to Eq.~\eqref{eq:limitconstraint}, to constrain $\on^0 \sim \sqrt{\kb} \off^0$. 

The effective force in the bound state converges to $\Fu^{\rm eff} = - k_l x_s - \partial_s \mathcal{U}_s$, \textit{i.e.} simply to a recoil force exerted with a spring constant corresponding to that of the fast linker (and the force deriving from the external potential). This makes sense since in the limit $\kb \gg \kl$, since the springs are in series, we expect the weakest spring to take over and $K \sim \kl$. 

The effective kinetic rates, according to Eq.~\eqref{eq:ratesEff} (and choosing the kinetic rates as in Eq.~\eqref{eq:choiceOn}), are   
\begin{equation}
        \on^{\rm eff} = \on^0 \sqrt{\frac{K}{\kb}} e^{-K x_s^2/2\kT}, \,\, \off^{\rm eff}  = \off^0.
    \end{equation}
In the limit regime where $\kb \gg \kl$, we have $\on \sim \off^0  e^{- \kl x_s^2/2\kT}$ and $\off = \off^0$. Hence the on rate remains spatially dependent, and the particle binds more likely when it is closer to the average linker position. It unbinds though at the same rate regardless of the linker position. Similar results may be found with other initial choices of kinetic rates such as with Eq.~\eqref{eq:choiceOff} or \eqref{eq:choiceOnOff}. 

A few models in the literature actually take the spring constant for the binding kinetics to be the linker's spring constant $K=\kl$~\cite{fogelson2019transport,maxian2022interplay,mitra2022coarse}, making the implicit, albeit rather physical, assumption that the bond's spring constant is much stiffer than the linker's. This underlines that the meaning of the parameters in the kinetic rates is underappreciated in the field. 

Notice that here we explored a limit regime after taking the limit of fast linker dynamics. This is not an issue here since the limits commute. Indeed, one could initially consider an infinitely stiff bond, and then take the limit of fast linker dynamics, and get the same result -- see the \edit{Supplementary Information Sec.~1 for further details.} %

\subsubsection{Stiff linker.} 

When the fast linker is stiff, then $\kl \gg \kb$. We then simply have that the force in the bound state is determined by the spring constant of the bond $\Fu^{\rm eff} =  -\kb  x_s - \partial_s\mathcal{U}_s$, as expected since now $K \sim k_b$ in that limit. With the same choice of kinetic rates as in Eq.~\eqref{eq:choiceOn},
we simply have that the effective on rate is faster near the average linker position $\on^{\rm eff} = \on^0 e^{-\frac{ \kb x_s^2}{2\kT} } $ and the effective off rate is constant. All in all, this is a simple consequence of the dynamics of 2 springs in series, when one of the spring constants is stiff compared to the other.

\section{Validation of the coarse-graining approach}
\label{sec:numericalValidation}

We now use numerical simulations to test the derived effective forces and kinetic rates, and determine the range of parameters over which the \homogenization procedure is valid.  %

\subsection{Simulation set up}

We simulate the dynamics specified through Eqs.~\eqref{eq:unbounddynamics}-\eqref{eq:bounddynamics} (with no confining potential on the slow particle $\mathcal{U}(x_s) \equiv 0$). We use the same nondimensional variables $\tau_0$ and $x_0 = \sqrt{D_0 \tau_0} = \sqrt{\kT \tau_0/\gamma_s}$ such that the problem is fully characterized by 5 non-dimensional numbers $\varepsilon, \kappa, \lambda, q_{\rm off}'$ and $\Pb/\Pu$. Here, $\varepsilon = \gammal/\gamma_s$ is not necessarily small and represents the ratio between friction coefficients; $\kappa  = \kl \tau_0/\gamma_s$ is the non-dimensional spring constant of the linker;  $\lambda = \kb \tau_0/\gamma_s$ is that of the bond, $q'_{\rm off}  = \off^0 \tau_0$ the non-dimensional typical off-rate. Finally, the on rate is set by the conservation of macroscopic probability through Eq.~\eqref{eq:limitconstraint}, so by the ratio $\Pb/\Pu$, once a choice of functional forms for the kinetic rates has been made. Here, we will keep Eq.~\eqref{eq:choiceOnOff} as an example. 

We discretize the dynamics of both slow and fast particles with a standard Euler-Maruyama scheme with time step $\Delta t$ until a terminal time of $T=1000\tau_0$ with $M=1000$ simulations for each set of parameters. \edit{Initial configurations are sampled with the equilibrium distribution of the system.} We impose periodic boundary conditions on the slow particle at a non-dimensional distance $L$ so that the slow particle does not escape far away from the domain. The domain size $L=10 x_0$ and $\Delta t = 0.01\tau_0$ are chosen sufficiently large and small respectively such that our results do not depend on the specific value. 

To estimate the effective rates from simulation, we note that this is a doubly-stochastic Poisson process, or a Cox process,\cite{moller2002statistical} because the stochastic dynamics of the particle positions drive the stochastic Poisson events of binding and unbinding. While sophisticated methods for inference on Cox intensities exist \cite{moller2002statistical}, a simple binning approach suffices here. We discretize $x_s$ into bins of width $\Delta x$ and count the number of occurrences of each event in each bin. For bin $i$ with center $x_i$, the estimated flux to the other state is then $\hat{J}(x_i)=N_i/(T M\Delta x)$, where $\hat{\cdot}$ notation means estimate and $N_i$ is the number of events (either binding or unbinding) occurring in bin $i$. Notably, this estimates the macroscopic kinetic rates, for instance, for unbinding $\hat{J}_\mathrm{off}^{\rm eff}(x_s) \approx q^{\rm eff}_{\rm off}(x_s) \pi_{\rm on}^{\rm eff}(x_s)$. The marginal densities of being bound and unbound  $ \pi_{\rm on}^{\rm eff}(x_s)$  and $\pi_{\rm off}^{\rm eff}(x_s)$ are straightforwardly computed by the fraction of time in each state and bin for $x_s$. Then, the microscopic rates are estimated by $\hat{q}^{\rm eff}(x_s) = \hat{J}^{\rm eff}(x_s)/\hat{\pi}^{\rm eff} (x_s)$. Lastly, the forces $\mathcal{F}^{\rm eff}(x_s)$ are taken as the average  $\langle x_s(t+\Delta t)-x_s(t)\rangle$ and accumulated in the corresponding bin for $x_s(t)$.

\begin{figure}[h!]
    \centering
    \includegraphics[width=\linewidth]{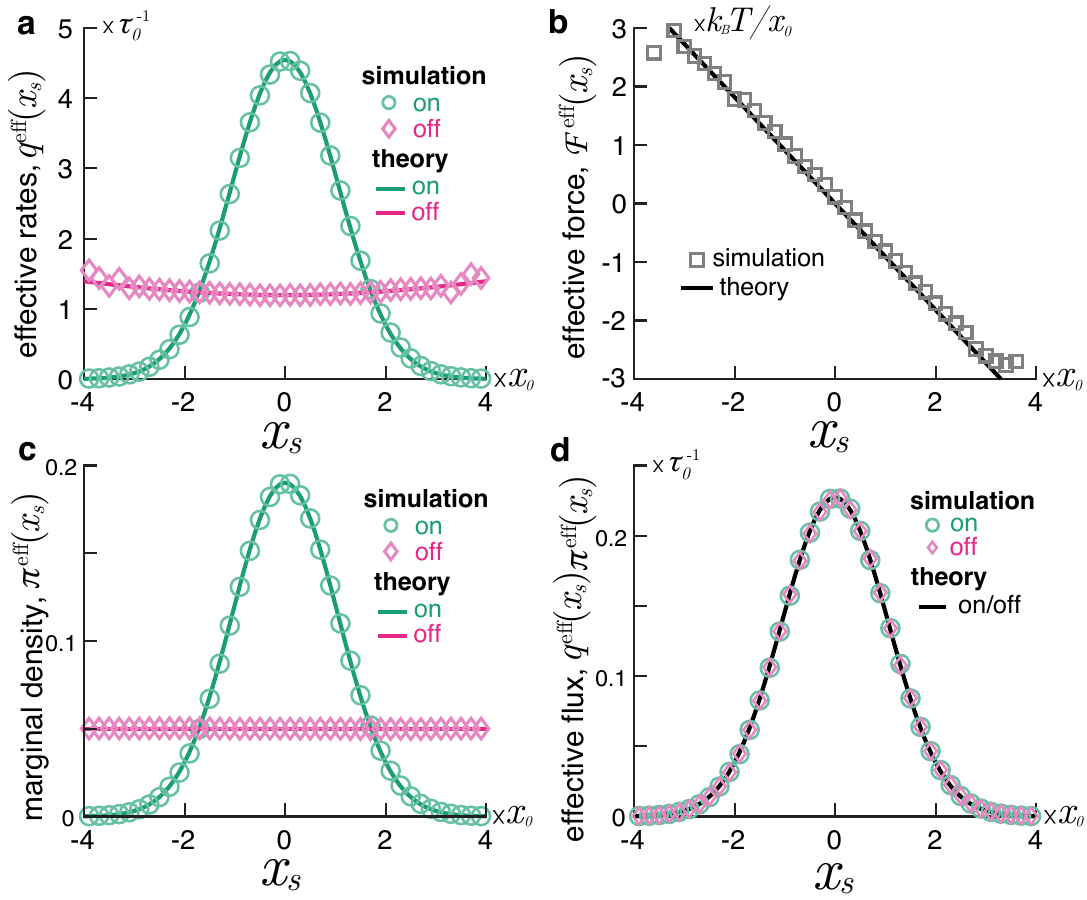}
    \caption{Numerical validation of the coarse-graining procedure, here in the case of a fast linker transiently binding to a slow particle in 1D. Both binding and unbinding are chosen to be spatially dependent, as in model 2, Eq.~\eqref{eq:choiceOnOff}, with numerical parameters $\epsilon = 0.1$, $\kappa = 1$, $\lambda = 10$, $q_{\rm off}^0=1/\tau_0$, and $q_{\rm on}^0$ chosen such that $\Pb/\Pu = 1$ determined by the relation \eqref{eq:limitconstraint}. 1000 Simulations are run until $T=1000\tau_0$.  \textbf{a}: microscropic binding and unbinding effective rates   
\label{fig:explicitcompare} \textbf{b}: effective force. \textbf{c}: effective marginal probabilities of being bound or unbound. \textbf{d}: macroscopic transition rates.  %
Theory curves are obtained with the expressions summarized in Table \ref{tab:effectiveQuantities}.}
\end{figure}

\subsection{Agreement between simulations and \homogenization approach}

\edit{We first compare effective rates and forces at long times.} In Fig.~\ref{fig:explicitcompare}-a and b, the kinetic rates and forces obtained numerically and with \homogenization theory Eq.~\eqref{eq:ratesEff} and \eqref{eq:dynamicsEff} (alternatively with the formulas in Table.~\ref{tab:effectiveQuantities}) are in excellent agreement. 
The effective probability distribution function is also in agreement with the marginal distribution Eq.~\eqref{eq:marginals} (Fig.~\ref{fig:explicitcompare}-c).
Overall, at the coarse-grained level, detailed balance holds numerically as well as analytically (Fig.~\ref{fig:explicitcompare}-d). \edit{While we present in Fig.~\ref{fig:explicitcompare} for a small value of $\varepsilon = 0.1$, similarly good agreement can be obtained for large values of $\varepsilon \simeq 10$ as long as the simulation times are long enough, which is expected from the coarse-graining procedure.}

\begin{figure}[h!]
    \centering
    \includegraphics[width=0.7\linewidth]{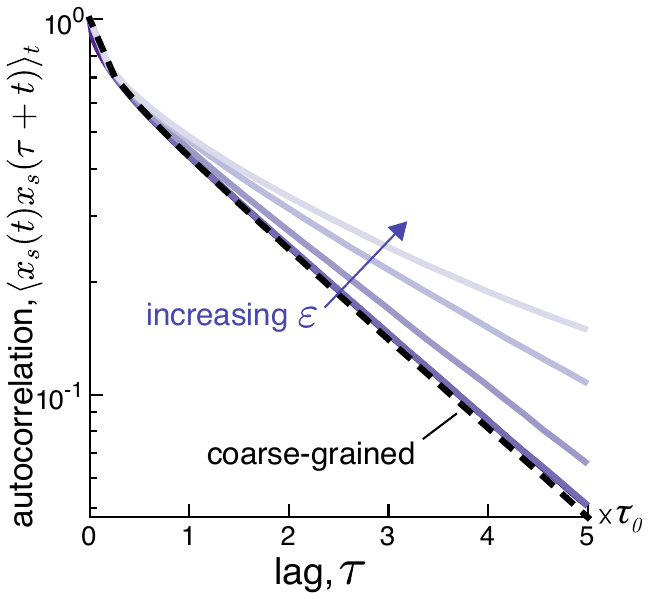}
    \caption{\label{fig:corr_fig} \edit{A comparison between the detailed and coarse-grained model of the autocorrelation of the slow particle position $\langle x_s(t+\tau)x_s(t)\rangle_t$.  In the limit of separation of timescales corresponding to small $\varepsilon$,  the dynamics approach those of the coarse grained-model. Parameter values are the same as in Fig.~\ref{fig:explicitcompare}, with purple curves corresponding to $\epsilon=\{10^{-2},10^{-1},10^0,10^1,10^2\}$. The curves for the lowest two values of $\varepsilon$ are indistinguishable. } }
\end{figure}

The results shown in Fig.~\ref{fig:explicitcompare} compare the effective rates and forces averaged over long times, but do not necessarily validate the short and intermediate time dynamics being correct in the coarse-graining procedure. To validate this, we numerically compute the autocorrelation from trajectories of the explicit microscopic model and coarse-grained as a function of $\varepsilon$. In Fig.~\ref{fig:corr_fig}, the dynamics of the coarse-grained model agree with those of the explicit model only for small $\varepsilon$. At $\varepsilon \gtrsim 1$, the autocorrelation displays significant deviation.  Intuitively, the coarse-grained model loses the memory of recent binding and unbinding events, and therefore has faster decay in correlation. This highlights that the coarse-grained dynamics in Table.~\ref{tab:effectiveQuantities} are only valid when $\varepsilon\lesssim 1$, \textit{i.e.} when the timescales associated with fast linker or slow particle relaxation are disparate enough. \editr{To account for these memory effects, one could add $O(\varepsilon)$ terms to the coarse-grained dynamics, that we derive in Supplementary Information Sec.~2 by continuing the coarse-graining procedure in Sec.~\ref{sec:coarsegraining}.} Altogether, our numerical results thus show the \homogenization approach is robust to infer coarse-grained effective dynamics.

\section{Applications}
\label{sec:applications}

Having shown the validity of the \homogenization approach, we now explore how to use the formulas in Table~\ref{tab:effectiveQuantities} in specific situations. We investigate \textit{(i)} a pair of connecting linkers, \textit{(ii)}  a linker stiffening upon binding, and \textit{(iii)} a slip bond with force-dependent unbinding. %
According to the specific example under scrutiny, we will focus either more on the effective forces or on the effective kinetic rates, and comment on the physical meaning of the results. 

\subsection{Pair of connecting linkers}

\begin{figure}[h!]
    \centering
    \includegraphics[width=\linewidth]{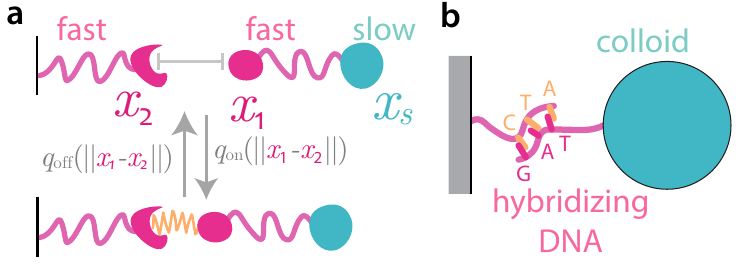}
    \caption{(a) Cartoon of 2 fast linkers that can connect to each other, representing for example (b) 2 complementary DNA strands transiently hybridizing, one of them being attached to a colloid whose motion is of interest. 
    \label{fig:DNA}}
\end{figure}

We first consider 2 fast linkers that can connect to each other (see Fig.~\ref{fig:DNA}). This could represent for example two complementary DNA strands transiently hybridizing, which finds some applications for example in the field of DNA-coated colloids\cite{marbach2022mass,marbach2022nanocaterpillar,jana2019translational,cui2022comprehensive,spinney2022geometrical,lowensohn2022sliding}. We consider that one of the linkers, in position $x_2$, is tethered to a fixed plate while the other, in position $x_1$, is tethered to a mobile slow particle, itself in position $x_s$. Both linkers are described by springs with the same spring constant $\kl$ and fluctuate rapidly.

\vskip2mm

\paragraph*{Unbound dynamics}

The unbound dynamics are
\begin{equation}
    \begin{cases}
    \frac{dx_s}{dt}  &= - \frac{\kl}{\gamma_s} (x_s - x_1)+ \sqrt{\frac{2 \kT}{\gamma_s}} \eta_s(t) \\
    \frac{dx_1}{dt}  &= - \frac{\kl}{\gammal} (x_1 - x_s)   + \sqrt{\frac{2 \kT}{\gammal}} \eta_1(t) \\
    \frac{dx_2}{dt}  &= - \frac{\kl}{\gammal} (x_2) + \sqrt{\frac{2 \kT}{\gammal}} \eta_2(t)
    \end{cases}
        \label{eq:unbounddynamicsEx1}
\end{equation}
where the $\eta_i(t)$ are uncorrelated Gaussian white noises and here we identify the force on the slow particle in the unbound state as
\begin{equation}
    \Fu = - \kl (x_s-x_1).
\end{equation}
The unbound term of the equilibrium distribution corresponding to this choice of dynamics is
\begin{equation} 
    \piu (x_s,x_1,x_2) = \frac{1}{Z_u} 
    e^{- \frac{\kl x_2^2 + \kl (x_s - x_1)^2}{2\kT}}.
\end{equation}

\vskip2mm

\paragraph*{Bound dynamics}

The linkers can transiently form a bond with spring constant $\kb$,
\begin{equation}
     \begin{cases}
    \frac{dx_s}{dt}  &= - \frac{\kl}{\gamma_s} (x_s - x_1)+ \sqrt{\frac{2 \kT}{\gamma_s}} \eta_s(t) \\
    \frac{dx_1}{dt}  &= - \frac{\kl}{\gammal} (x_1 - x_s)  + \sqrt{\frac{2 \kT}{\gammal}} \eta_1(t) + \frac{\kb}{\gammal}(x_2 - x_1)\\
    \frac{dx_2}{dt}  &= - \frac{\kl}{\gammal} (x_2) + \sqrt{\frac{2 \kT}{\gammal}} \eta_2(t) - \frac{\kb}{\gammal}(x_2 - x_1)
    \end{cases}
    \label{eq:bounddynamicsEx1}
\end{equation}
and we identify the force on the slow particle in the bound state as identical to that in the unbound state
\begin{equation}
    \Fb = - \kl (x_s-x_1).
    \label{eq:FboundEx1}
\end{equation}
The bound term of the equilibrium distribution corresponding to this choice of dynamics is
\begin{equation} 
    \pi_{\rm bound} (x_s,x_1,x_2) = \frac{1}{Z_b} 
    e^{- \frac{\kl x_2^2+ \kl (x_s - x_1)^2+ \kb (x_2 - x_1)^2}{2\kT}}.
    \label{eq:piboundEx1}
\end{equation}

\vskip2mm

\paragraph*{Kinetic rates and detailed balance}

Here we will not specify the kinetic rates in detail. However, since we specify the dynamics, the binding and unbinding rates must satisfy detailed balance 
$$
\frac{\on}{\off} =  \frac{\on^0}{\off^0} e^{-k_{\rm\ell}(x_2 - x_1)^2/2\kT}.
$$

\vskip2mm

\paragraph*{Effective force in the unbound state}
We can now use the expressions in Table~\ref{tab:effectiveQuantities} to obtain the effective force. Here, compared to our foundational example with 1 fast linker in Sec.~\ref{sec:generalSystem}, we have 2 fast linkers, hence we need to carry the integral over those 2 fast degrees of freedom.  We find with Eq.~\eqref{eq:TableOffForce}
\begin{equation}
\begin{split}
     &\Fu^{\rm eff}(x_s) \\
     & \quad = \frac{\int \Fu (x_1,x_2,x_s)  \piu (x_1,x_2,x_s) dx_1 dx_2}{\int   \piu (x_1,x_2, x_s) dx_1 dx_2} \\
     & \quad = \frac{\int \left( - \kl (x_s - x_1)  \right)  e^{- \kl x_2^2/2\kT}  e^{- \kl (x_s - x_1)^2/2\kT}  dx_1 dx_2}{\int    e^{- \kl x_2^2/2\kT}  e^{- \kl (x_s - x_1)^2/2\kT}   dx_1 dx_2} \\
     &\quad  = 0
\end{split}
\end{equation}
for symmetry reasons. In the unbound state, quite logically, at the coarse-grained level the particle doesn't feel any effective force from its unbound fast linker. 

\vskip2mm

\paragraph*{Effective force in the bound state}
In the bound state, using Eq.~\eqref{eq:TableOnForce} we find
\begin{equation}
\begin{split}
     \Fb^{\rm eff}(x_s) &= \frac{\int \Fb (x_1,x_2,x_s)  \pi_{\rm bound} (x_1,x_2,x_s) dx_1 dx_2}{\int   \pi_{\rm bound} (x_1,x_2, x_s) dx_1 dx_2} \\
     & = -\frac{\frac{\kl}{2} \kb}{\frac{\kl}{2} + \kb} x_s.
\end{split}
\end{equation}
In the bound state we thus obtain that the effective force on the particle is a spring force. The force is centered around 0 as this is the average position of both springs. The spring constant at the coarse-grained level is $\frac{\frac{\kl}{2} \kb}{\frac{\kl}{2} + \kb}$ which corresponds, logically, to the effective spring constant of $3$ springs in \edit{series}, with spring constants $\kl, \kb, \kl$.

\subsection{Stiffening linker upon binding}
\begin{figure}[h!]
    \centering
    \includegraphics[width=1\linewidth]{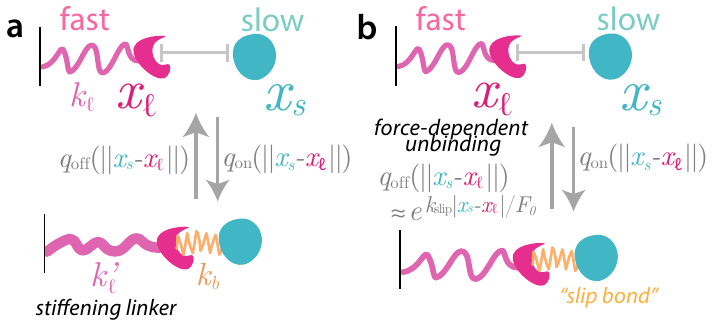}
    \caption{(a) Cartoon of a fast linker stiffening upon binding and (b) a fast linker connecting to the particle via a slip bond with force-dependent unbinding. 
    \label{fig:stiff}}
\end{figure}

We now turn to another example where the linker stiffens when bound -- see Fig.~\ref{fig:stiff}-a. Such stiffening occurs when the linker undergoes conformational changes upon binding~\cite{fogelson2019transport} or else upon single-stranded DNA hybridizing into double-stranded, resulting in a stiffer connection~\cite{feng2013specificity,lee2018modeling,marbach2022nanocaterpillar,xu2011subdiffusion,rogers2016using,gehrels2022programming}. %

\vskip2mm
\paragraph*{Unbound dynamics}

The unbound dynamics are still given by Eq.~\eqref{eq:unbounddynamics}.

\vskip2mm
\paragraph*{Bound dynamics}

The bound dynamics are now changed compared to Eq.~\eqref{eq:bounddynamics}, since the linker now has a stiffer spring constant say $\kl' > \kl$,
\begin{equation}
     \begin{cases}
    \frac{dx_s}{dt}  &= - \frac{\kb}{\gamma_s} (x_s - \xl)+ \sqrt{\frac{2 \kT}{\gamma_s}} \eta_s(t) \\
    \frac{d\xl}{dt}  &=  - \frac{\kl'}{\gammal} \xl - \frac{\kb}{\gammal} (\xl - x_s)   + \sqrt{\frac{2 \kT}{\gammal}} \etal(t).
    \end{cases}
    \label{eq:bounddynamicsEx2}
\end{equation}
The bound term of the equilibrium distribution corresponding to this choice of bound dynamics is
\begin{equation} 
    \pib (x_s,\xl) = \frac{1}{Z_b} 
    e^{- \kl' \xl^2/2\kT}  e^{- \kb (x_s - \xl)^2/2\kT} .
\end{equation}

\vskip2mm
\paragraph*{Kinetic rates}
Again without specifying the kinetic rates in detail, since we set the dynamics, the binding and unbinding rates must satisfy detailed balance and hence are related via 
$$
\frac{\on}{\off} =  \frac{\on^0}{\off^0} e^{-\kb(x_s - \xl)^2/2\kT}  e^{- (\kl'-\kl) \xl^2/2\kT}.
$$
This relation favors binding when the tether is not too extended. Otherwise, the energetic price to pay to ``stiffen'' is too costly. Reciprocally the bond is more likely to break when the tether is quite extended, and the energetic gain to loosen the bond is quite high. One could thus choose, in line with physical intuition,
\begin{equation}
\begin{cases}
    &\on \sim \on^0 e^{-\kb(x_s - \xl)^2/2\kT},\\
    &\off \sim \off^0  e^{(\kl'-\kl) \xl^2/2\kT}.
    \end{cases}
        \label{eq:ratesStiff}
\end{equation}
The detailed prefactors should be specified in agreement with detailed balance.

\vskip2mm
\paragraph*{Effective force in the bound state}

The effective force in the unbound state naturally vanishes. Hence we focus on the effective force in the bound state, again obtained via Eq.~\eqref{eq:TableOnForce}
\begin{equation}
\begin{split}
     \Fb^{\rm eff}(x_s) %
     & = -\frac{\kl' \kb}{\kl' + \kb} x_s.
\end{split}
\end{equation}
The coarse-grained force is that associated with 2 springs in series with spring constants $\kl'$ and $\kb$. However, here it was not obvious \textit{a priori} which of the spring constants for the linker, either $\kl'$ or $\kl$ or a mix of both, should contribute to the force.  

\vskip2mm
\paragraph*{Effective rates}

We now use the rates defined in Eq.~\eqref{eq:ratesStiff} to study coarse-grained kinetic rates as well. Let $K = \kl \kb/(\kl + \kb)$ and $K' = \kl' \kb/(\kl' + \kb)$. Then we obtain 
\begin{equation}
\begin{cases}
    &\on^{\rm eff}(x_s) \sim e^{-K x_s^2 /2\kT} \\
    &\off^{\rm eff}(x_s) \sim e^{(K'-K) x_s^2 /2\kT}. \\
\end{cases}
\end{equation}
The effective binding rate is determined by the typical, unstiff, radius of the interaction,  $\sqrt{\kT/K}$, similarly as in our foundational example in Sec.~\ref{sec:generalSystem}. The unbinding rate is now larger at larger distances, according to how much stiffening occurred. Notice that even if the bond is really stiff, meaning $\kb \gg \kl, \kl'$ then $K' - K \simeq \kl' -\kl$ still converges to a finite value. %

\subsection{Slip bonds with force-dependent unbinding}

Another quite common situation is that of slip bonds with force-dependent unbinding~\cite{bell1978models,fogelson2019transport}  -- see Fig.~\ref{fig:stiff}-b. In these cases the unbinding rate scales as~\cite{bell1978models}
\begin{equation}
\off(x_s,\xl) = \off^0 e^{\ks|x_s - \xl|/F_0}
\label{eq:rates31}
\end{equation}
where $\ks$ is a characteristic bond spring constant and $F_0$ a characteristic force threshold. Above this threshold, the unbinding rate is indeed greatly enhanced. 

\vskip2mm
\paragraph*{Kinetic rates and satisfying detailed balance}
How shall we proceed to model the dynamics with such force-dependent unbinding rate, when the system is still at equilibrium? 
Compared to our foundational example in Sec.~\ref{sec:generalSystem}, here we might \edit{modify} 
the binding rate %
\edit{to satisfy detailed balance.} Notice that this is not the only possibility to satisfy detailed balance and one could also consider alternative bound dynamics. 
\edit{For the sake of the example, here we will consider \begin{equation}
    \on(x_s,\xl) =  \on^0 e^{-k_b(x_s- \xl)^2/2\kT}
\label{eq:rates32}
\end{equation} and an additional force on the bound particle that we specify in Eq.~\eqref{eq:bounddynamicsEx3} below.}
The expressions for the rates are satisfactory since they are coherent with faster binding and slower unbinding when the particle and linker are closer. \edit{Similar choices motivated by their intuitive behavior date back to Bell and Dembo \cite{bell1978models,dembo1988reaction} and are derived in Ref.~\onlinecite{PhysRevX.12.031030} as a coarse-graining of a microscopic model for binding.}  The detailed prefactors should be specified in agreement with detailed balance. 

\vskip2mm
\paragraph*{Unbound dynamics}

The unbound dynamics are still given by Eq.~\eqref{eq:unbounddynamics}. 

\vskip2mm
\paragraph*{Bound dynamics}

The bound dynamics are now changed compared to Eq.~\eqref{eq:bounddynamics}, since the linker and particle are now subjected to additional forces. 
\begin{equation}
     \begin{cases}
    \frac{dx_s}{dt}  &= - \frac{\kb}{\gamma_s} (x_s - \xl) \\
    & \,\, - \frac{\ks}{\gamma_s} \frac{\kT}{F_0} \text{sgn}(x_s - \xl) + \sqrt{\frac{2 \kT}{\gamma_s}} \eta_s(t) \\
    \frac{d\xl}{dt}  &=  - \frac{\kl}{\gammal} \xl - \frac{\kb}{\gammal} (\xl - x_s)  \\
    & \,\,+ \frac{\ks}{\gammal} \frac{\kT}{F_0} \text{sgn}(x_s - \xl)   + \sqrt{\frac{2 \kT}{\gammal}} \etal(t).
    \end{cases}
    \label{eq:bounddynamicsEx3}
\end{equation}
\edit{In higher dimensions, the $\text{sgn}(x) = x/\|x\|$ where $x = x_s - \xl$.} The bound term of the equilibrium distribution corresponding to this choice of bound dynamics is
\begin{equation} 
    \pib (x_s,\xl) = \frac{1}{Z_b} 
    e^{- \kl \xl^2/2\kT}  e^{- \kb (x_s - \xl)^2/2\kT}  e^{- k_s|x_s - \xl|/F_0}.
\end{equation}

\vskip2mm
\paragraph*{Effective force in the bound state}

The effective force in the unbound state naturally vanishes. Hence we focus on the effective force in the bound state, which has a lengthy expression that we do not report here. In the case of a small slip force $\ks \kT / F_0 \ll  \kl |x_s|$ we obtain
\begin{equation}
\begin{split}
     \Fb^{\rm eff}(x_s) 
     & \simeq -\frac{\kl \kb}{\kl + \kb} x_s - \\
     &\frac{\kl }{\kl + \kb} \frac{\ks \kT}{F_0} \text{erf} \left( \sqrt{\frac{\kl}{\kl + \kb} \frac{\kl x_s^2}{2 \kT}} \right).
\end{split}
\end{equation}
The coarse-grained force contains now 2 contributions. The first one is that associated with 2 springs in series with spring constants $\kl$ and $\kb$ that we have seen before. The second one corresponds to the slip bond, which grows stronger when the particle goes further away from the target. Notice that this latter slip bond friction force is screened by a factor $\frac{\kl }{\kl + \kb}$ in the coarse-grained state. In the case of a stiff bond, $\kb \gg \kl$, the slip force is entirely screened, and the particle ``slips''; the linker essentially accommodates changing configurations by rapidly adjusting its length. %

\vskip2mm
\paragraph*{Effective rates}

We now use the rates defined in Eq.~\eqref{eq:rates31} and Eq.~\eqref{eq:rates32} to study coarse-grained kinetic rates as well. Let $K = \kl \kb/(\kl + \kb)$. Then we obtain, in the case of a small slip force $\ks \kT / F_{0}  \ll \kl |x_s|$,
\begin{equation}
\begin{cases}
    &\on^{\rm eff}(x_s) \sim e^{-K x_s^2 /2\kT} \\
    &\off^{\rm eff}(x_s) \sim e^{ \frac{\kl }{\kl + \kb} \frac{\ks |x_s|}{F_{0}}  }. \\
\end{cases}
\end{equation}
The effective binding rate is determined by the typical, unstiff, radius of the interaction,  $\sqrt{\kT/K}$, similarly as in our foundational example in Sec.~\ref{sec:generalSystem}. The unbinding rate is now larger at larger distances, with force-dependent unbinding. %

\section{Macroscopic consequences for the choice of binding kinetics}
\label{sec:consequences}

\subsection{Coarse-graining various microscopic binding models: the Doi model}

How might the effective dynamics change from our foundational example in Sec.~\ref{sec:generalSystem} when we consider the Doi model for binding?

\vskip2mm
\paragraph*{Effective force in the unbound state}
According to Eq.~\eqref{eq:TableOffForce} and \edit{with Eq.~\eqref{eq:dynamicsGoodrich} and Eq.~\eqref{eq:onDoi} describing the Doi model in our context}, the unbound friction force is simply
\begin{equation}
     \Fu^{\rm eff}(x_s) =  - k_b \frac{\int_{x_s - R}^{x_s + R} (x_s-\xl) e^{- \frac{\kb (x_s - \xl)^2 + \kl \xl^2}{2\kT}} \dd \xl }{\int \piu (x_s,\xl)   \dd \xl}
\end{equation}
which has a cumbersome, non-vanishing, expression that we do not report here. 
However we can make a finite perturbation of the obtained expression in the limit of a stiff bond, when $k_b \gg \kl$, and \edit{to simplify further the expression we assume that the radius to bind $R$ is given by the typical spatial scale of the bond} $R = \sqrt{2\kT/k_b}$, such that
\begin{equation}
    \Fu^{\rm eff}(x_s) \simeq - 0.43 \sqrt{\frac{\kl}{k_b}} \kl x_s.
\end{equation}
Since $\xl$ wiggles around 0 and even in the unbound state, close to the linker, the particle feels a recoil force, then it makes sense that the particle can now feel a recoil force everywhere. The magnitude of this force is slightly decreased because the bond can only form when the particle is close enough to the linker. Notice how the scaling of the effective spring constant is entirely nontrivial as $\sqrt{\frac{\kl}{k_b}} \kl$.

\vskip2mm
\paragraph*{Effective force in the bound state}
According to Eq.~\eqref{eq:TableOnForce} the bound friction force is simply
\begin{equation}
    \Fb^{\rm eff}(x_s) =   - \frac{\kl k_b}{\kl + k_b} x_s
\end{equation}
which is similar to the results obtained with our foundational example. 

\vskip2mm
\paragraph*{Effective kinetic rates}
Again, the expression for the effective binding and unbinding rates obtained from Eqs.~\eqref{eq:TableOn} and \eqref{eq:TableOff} is rather cumbersome. When $k_b \gg \kl$, and assuming $R = \sqrt{2\kT/k_b}$, the kinetic rates are smoothly decaying to 0 (instead of a sharp Heaviside function as in the microscopic equations), at a characteristic distance $x_s \simeq \sqrt{\frac{\kT}{\kl}\sqrt{\frac{\kl}{\kb}}}$. Again, one sees how this scaling is entirely nontrivial. Systematic coarse-graining is therefore essential for faster numerical simulations and enhanced theoretical investigations.

\subsection{Macroscopic consequences}

\begin{figure}[h!]
    \centering
    \includegraphics[width=\linewidth]{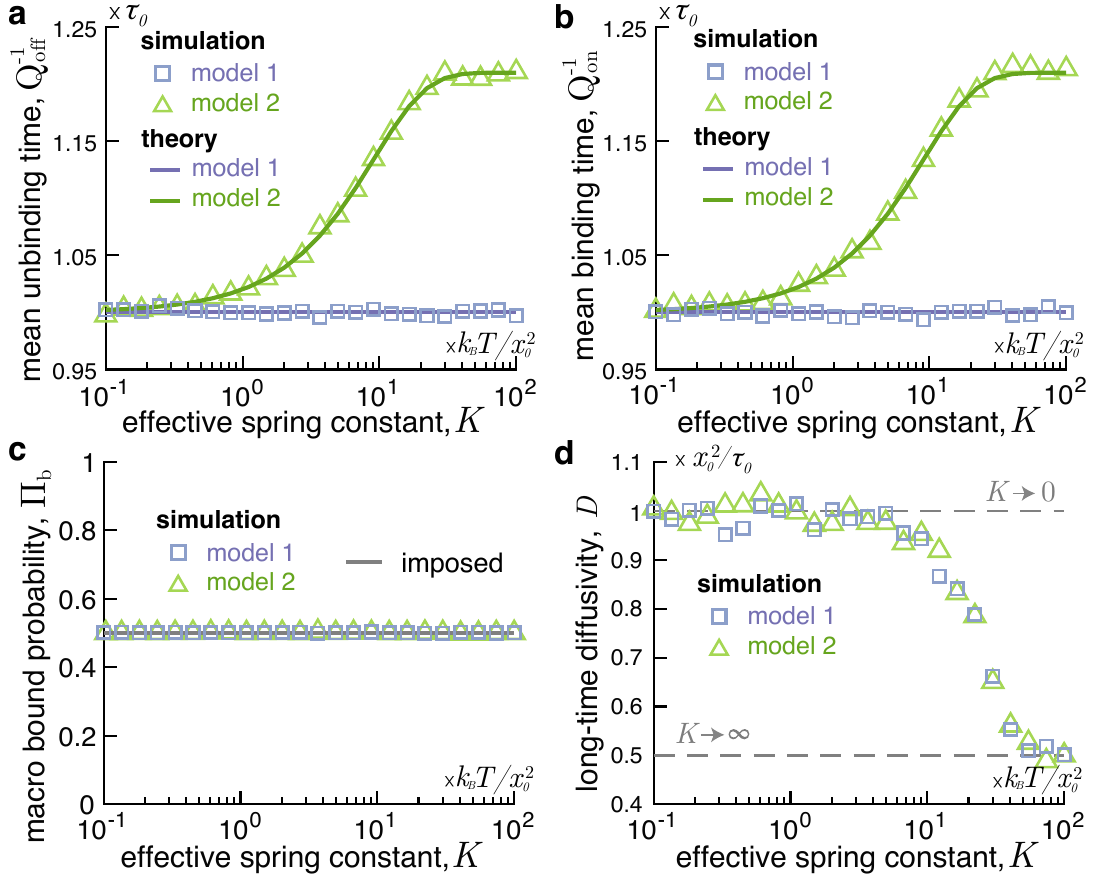}
    \caption{Different choices of $\on(x)$ and $\off(x)$ yield different \edit{macroscopic} dynamics. Model 1 corresponds to Eq.~\eqref{eq:choiceOn} with spatially dependent binding only and model 2 corresponds to Eq.~\eqref{eq:choiceOnOff} with both spatially dependent binding and unbinding. Simulation parameters are %
    $\off^0=0.1  \tau_0^{-1}$ and $\on^0$ chosen such that $\Pi_b/\Pi_u=1$. 1000 simulations are run until $T=1000 \tau_0$.  \textbf{a}: mean unbinding time as a function of the \edit{effective spring constant $K$}, showing divergence between models 1 and 2. Theory curves are obtained with Eq.~\eqref{eq:qoffCoarseGraining2}. \textbf{b}: mean binding time as a function of the \edit{effective spring constant $K$} with similar trend. \textbf{c}: Macroscopic bound probabilities $\Pi_b$ are set to be fixed and are the same for both models. \textbf{d}: Macroscopic diffusion coefficient $D=\lim_{t\to\infty} \langle x_s(t)^2\rangle/(2t)$ for both models. } %
    \label{fig:qonqoff}
\end{figure}

We finish by briefly exploring how the microscopic choices for $\on(x_s,\xl)$ and $\off(x_s,\xl)$ can affect the macroscopic dynamics -- see Sec.~\ref{sec:possibleRates}. To simplify the exploration, here we 
\edit{directly simulate the coarse-grained equations for the slow particle,
using the coarse-grained forces and kinetic rates in Table.~\ref{tab:effectiveQuantities}.
We test the impact of} different initial choices of microscopic binding, specifically with model 1 corresponding to Eq.~\eqref{eq:choiceOn} with spatially dependent binding only and model 2 from Eq.~\eqref{eq:choiceOnOff} with both spatially dependent binding and unbinding. To compare different microscopic models in a sensible fashion, we constrain the macroscopic bound probability $\Pi_b = 0.5$ (see Fig.~\ref{fig:qonqoff}-c) for all microscopic models. We then calculate the macroscopic mean binding and unbinding times for the particle (Fig.~\ref{fig:qonqoff}-a and b), as well as the particle's long-time diffusion coefficient (Fig.~\ref{fig:qonqoff}-d), determined by its mean-squared displacement. Importantly, we vary the effective bond spring constant $K = \kl \kb/(\kb + \kl)$ to probe how model parameters affect macroscopic dynamics. 

We find that both the choice of microscopic binding model and the parameter $K$ significantly affect the macroscopic binding and unbinding times (Figs.~\ref{fig:explicitcompare}-a and b). Since the macroscopic binding probability $\Pi_b$ is fixed, the mean binding and unbinding times are identical, and hence Fig.~\ref{fig:qonqoff}-a and b appear the same. At long times, we expect, as for the coarse-graining over $\xl$, that the kinetic rates do not depend on the position $x_s$ anymore and would verify
\begin{equation}
Q_{\rm off} = \frac{\int \off^{\rm eff}(x_s) \pib^{\rm eff}(x_s) \dd x_s}{ \int  \pib^{\rm eff}(x_s) \dd x_s},
\label{eq:qoffCoarseGraining2}
\end{equation}
and similarly for the binding rate $Q_{\rm on}$, \edit{in agreement with other works.\cite{PhysRevX.12.031030}} Eq.~\eqref{eq:qoffCoarseGraining2} gives the mean unbinding time as $1/Q_{\rm off}$ and reproduces perfectly the numerical results (lines in Fig.~\ref{fig:explicitcompare}-a).

For the different microscopic models, the response to various binding constants $K$ is non-trivial. For model 1, the macroscopic binding and unbinding times remain constant. In model 1, $\off^{\rm eff}(x_s)$ does not depend on space, and, therefore, its macroscopic counterpart $\Off^{\rm eff}$ which verifies Eq.~\eqref{eq:qoffCoarseGraining2}, seems to have no direct dependence on $K$. However, $\on^{\rm eff}(x_s)$ is inherently spatially dependent and hence its macroscopic counterpart may depend on $K$ as well, \textit{via} the associated length scale $\sqrt{\kT/K}$. 
In contrast, with model 2 the macroscopic kinetic rates both increase with the binding constant $K$. When $K$ is larger, the bond is stronger, which decreases the binding time but increases the unbinding time. Hence, the macroscopic dependence where both kinetic rates increase with $K$ is interesting. In fact, to properly coarse-grain to this now macroscopic scale, one should also take into account the probability distribution of being in each state with a given extension. Such entangled behavior requires proper integration beyond simple intuition. Eventually, we find that the macroscopic binding times strongly depend on microscopic choices for the kinetic rates, which means one should use caution when designing such models.

In addition, for both microscopic model choices, the long-time self-diffusion coefficient of the particle depends on $K$ (Fig.~\ref{fig:explicitcompare}-d). In fact, at small $K$ we find $D \rightarrow \kT/\gamma_s$ since the bond is weak enough that it barely affects particle motion. At larger $K$ values, we find $D \rightarrow \kT/2\gamma_s$ since when it is bound, the bond is strong enough that it prevents any motion, and the particle is bound half of the time ($\Pi_b = 0.5$). The dependence on $K$ appears to be similar for both microscopic binding models 1 and 2. In more varied models for binding, there is no reason this should stay true. We leave the general exploration of macroscopic transport properties, such as diffusion coefficients, and their dependence on microscopic binding kinetics, for future work.

\section*{Conclusion}

In this work, we have attempted to unify and justify various coarse-graining approaches for linker dynamics\edit{\cite{bell1978models,dembo1988reaction,fogelson2018enhanced,fogelson2019transport,jana2019translational,noe2018,maxian2021simulations,maxian2022interplay,korosec2018dimensionality,kowalewski2021multivalent,olah2013superdiffusive,bihr2015multiscale,PhysRevX.12.031030,mitra2022coarse}}. In these earlier approaches, we have identified many different choices for how binding and unbinding depend on distance, yet (1) these are often \textit{heuristically} motivated, or \edit{(2) they only provide coarse-grained binding rates and not the full dynamics of the slow particles including forces and friction.}
Here, we have addressed all of these issues by providing a systematic derivation of effective \edit{dynamics, including effective friction,} forces and binding kinetics for linker molecules that obey any detailed microscopic descriptions. 
Our coarse-graining approach is based on \homogenization techniques and preserves detailed balance, in line with our assumption of equilibrium dynamics. We have verified our approach with numerical simulations and found excellent agreement \edit{in both the effective kinetic rates and dynamics after averaging. This averaging analysis hinges on  $\varepsilon=\gammal/\gamma_s$ being small, corresponding to a large separation of timescales, between the fast linker and slow particle relaxation. When $\varepsilon \gtrsim 1 $, the coarse-grained dynamics diverge from the detailed model.} We showed how our general framework may be applied to diverse microscopic scenarios, \edit{including} with 2 linkers binding to each other, a linker stiffening upon binding or a slip bond with force-dependent unbinding. 
Finally, we showed that different microscopic kinetic rates result in fundamentally different dynamics at the macroscopic scale, raising caution in making these choices without care. 

Many choices for effective dynamics in the literature clearly violate detailed balance and hence operate out-of-equilibrium~\edit{\cite{lalitha2021enhanced,fogelson2019transport}}. %
\edit{The coarse-graining in Ref.~\onlinecite{PhysRevX.12.031030} computes the effective rates in a mechanical model with non-equilibrium fluctuations, but relating non-equilibrium rates and dynamics seems to be missing a unified framework. Our systematic approach to coarse-graining may provide a first step toward addressing this.} The authors in Ref.~\onlinecite{walker2022numerical} interestingly note that effective equilibria can arise by switching between non-conservative systems, providing hope that this pursuit is both fruitful and interesting. Thus, extending our method to out-of-equilibrium  systems would be of broad relevance, in particular, to explore transient bonds with activated cleaving that are common in viral linker-mediated motion~\cite{muller2019mobility,ziebert2021influenza} and also artificial nano-motors~\cite{korosec2021lawnmower}. 

Although we have investigated relatively simple setups here, the tools we introduce and the lessons learned are applicable across many more complex systems. The formulas we derived for effective dynamics can be applied with ease via Table~\ref{tab:effectiveQuantities} and provide a baseline for more systematic coarse-grainings found  in simulations and theoretical studies across the literature. One such example is molecular motor binding in intracellular transport. There, coarse-graining ranges from non-spatial effective kinetic rates \cite{miles2018analysis,park2022coarse}, motor linkers obeying a worm-like chain model \cite{spakowitz2005end,mogre2020hitching}, or Doi-like motors that bind when within a specific radius of interaction \cite{bovyn2021diffusion}. Coarse-grained cross-linked cytoskeletal networks are studied extensively \cite{nedelec2007collective,bidone2017morphological,descovich2018cross,lugo2022typical,CohenHendricks2022}, especially in self-organization in the mitotic spindle \cite{peterman2009mitotic,lamson2019theory,hannabuss2019self,gaska2020mitotic} and actomyosin network mechanics \cite{popov2016medyan,freedman2018nonequilibrium}. 
Transient dynamics of (typically coarse-grained) cross-linkers are also fundamental in controlling viral responses in bio-gels \cite{chen2014transient,newby2017blueprint} and building chromosomal territories\cite{walker2019transient}. 
Our framework also readily extends to systems that are not "cross"-linked, such as membrane-filament interactions that drive cell protrusion\cite{mogilner2003force,welf2020actin} and \edit{ adhesions\cite{bihr2012nucleation,bihr2015multiscale,PhysRevX.12.031030}.}

\begin{acknowledgments}
  \noindent  We wish to acknowledge fruitful discussions with Aleksandar Donev and Miranda Holmes-Cerfon.  S.M. received funding from the European Union’s Horizon 2020 research and innovation programme under the Marie Skłodowska-Curie grant agreement 839225, MolecularControl.
\end{acknowledgments}

\edit{
\section*{Supplemental Information}
\noindent See the supplementary information for further mathematical details of the coarse-graining procedure, including calculations of higher order corrections.}

\section*{Data Availability Statement}
\noindent  The data generated by this paper is available upon reasonable request to the authors.

\appendix

\section{Main notations}
\label{app:notations}

We report in the Table~\ref{tab:notations} the main notations used in the paper.%

\begin{table}
  \centering
  \caption{Summary of the main notations used.}
  \rule{\linewidth}{\heavyrulewidth}
  {\scriptsize
    \begin{tabular}{c|l}
      \textbf{Notation} & \textbf{Meaning} \\ \hline
        $x_s$ & coordinate of the slow particle \\
        $\xl$ & coordinate of the fast linker \\
        $x_0$ & characteristic length scale for nondimensionalisation \\
        $\gamma_s$ & friction coefficient on the slow particle \\
        $\gammal$ & friction coefficient on the fast linker \\
        $\varepsilon = \gammal/\gamma_s$ & ratio of friction coefficients \\
        $\kl$ & spring constant describing the linker \\
        $\kb$ & spring constant describing the bond between the particle and the linker \\
        $\kT$ & thermal energy \\
        $\tau_0$ & characteristic timescale for nondimensionalisation \\
        $\pi_{\rm u/b}(x_s,\xl)$ & microscopic unbound (resp. bound) probability distribution \\
        $\pi_{\rm u/b}^{\rm eff}(x_s)$ & coarse-grained unbound (resp. bound)  probability distribution \\
        $\Pi_{\rm u/b}$ & macroscopic unbound (resp. bound) probability  \\
        $\mathcal{F}_{\rm u/b}(x_s,\xl)$ & microscopic force on the slow particle in the unbound (resp. bound) state \\
        $\mathcal{F}^{\rm eff}_{\rm u/b}(x_s)$ & coarse-grained force on the slow particle in the unbound (resp. bound) state  \\
        $q_{\rm on/off}(x_s,\xl)$ & microscopic binding (resp. unbinding) rate \\
        $q^{\rm eff}_{\rm on/off}(x_s)$ & coarse-grained binding (resp. unbinding) rate \\
        $Q_{\rm on/off}$ & macroscopic binding (resp. unbinding) rate \\
      \end{tabular}}
      \label{tab:notations}
   \end{table}

\edit{
\section{Detailed balance at the coarse-grained level}}
\label{app:db_at_coarsegrain}

Finally, for physical consistency, we need to check that the marginal equilibrium distribution,  \textit{i.e.} the equilibrium distribution integrated over the fast degrees of freedom $\pi^{\rm eff}(x_s) = \int  \pi(x_s,\xl) \dd \xl$, is indeed a stationary solution of the effective dynamics obtained. 
The marginal distribution is 
\begin{equation} \label{eq:marginals}
\begin{split}
    \pi^{\rm eff}(x_s) &= \begin{pmatrix}
    \piu^{\rm eff} (x_s) \\
    \pib^{\rm eff} (x_s)
    \end{pmatrix} \\
    &= \frac{\sqrt{2\pi} e^{-\mathcal{U}_s(x_s)/\kT}}{Z} \begin{pmatrix}
    \frac{1}{\sqrt{\kl}}\\
    \frac{\on^0}{\off^0 \sqrt{\kl+\kb}}  e^{- K \frac{x_s^2}{2 \kT}}.  
    \end{pmatrix}
    \end{split}
\end{equation}
It is clear that the marginal distribution of either state is indeed a stationary solution of the dynamics in each state as specified in Eq.~\eqref{eq:dynamicsEff}.

To check that the marginal distribution is a stationary distribution of the dynamics as a whole, we still need to check that detailed balance is satisfied at the coarse-grained level. This is not guaranteed \textit{a priori} since the \homogenization technique does not use at any point that it should preserve detailed balance. 
At this point, it is important to notice that in fact the effective rates are related to the equilibrium probability distribution as
    \begin{equation}
        \on^{\rm eff}(x_s) = \frac{\int \on(x_s,\xl) \piu (x_s,\xl)d\xl}{\int \piu (x_s,\xl)d\xl}
    \end{equation}
    and similarly for $\off^{\rm eff}$. Hence, we simply have that
    \begin{align*}
        \piu^{\rm eff} \on^{\rm eff}(x_s) & =  \piu^{\rm eff} \frac{\int \on(x_s,\xl) \piu(x_s,\xl)d\xl}{\int \piu(x_s,\xl)d\xl} \\
        &= \int \on(x_s,\xl) \piu (x_s,\xl)d\xl \\
        &=\int \off(x_s,\xl) \pib (x_s,\xl)d\xl \\
        &=\int \off(x_s,\xl) \pib (x_s,\xl)d\xl \\
        & \qquad \times \frac{\pib^{\rm eff} }{\int \pib (x_s,\xl)d\xl} \\
        &=  \pib^{\rm eff} \off^{\rm eff}(x_s).
    \end{align*}
    Detailed balance is therefore also true at the coarse-grained level. 
Since the marginal distribution is consistent with the dynamics in each state and with detailed balance, it is indeed the stationary solution to the effective dynamics.

\bibliography{binding}%

%merlin.mbs aipnum4-1.bst 2010-07-25 4.21a (PWD, AO, DPC) hacked
%Control: key (0)
%Control: author (8) initials jnrlst
%Control: editor formatted (1) identically to author
%Control: production of article title (0) allowed
%Control: page (1) range
%Control: year (1) truncated
%Control: production of eprint (0) enabled
\begin{thebibliography}{82}%
\makeatletter
\providecommand \@ifxundefined [1]{%
 \@ifx{#1\undefined}
}%
\providecommand \@ifnum [1]{%
 \ifnum #1\expandafter \@firstoftwo
 \else \expandafter \@secondoftwo
 \fi
}%
\providecommand \@ifx [1]{%
 \ifx #1\expandafter \@firstoftwo
 \else \expandafter \@secondoftwo
 \fi
}%
\providecommand \natexlab [1]{#1}%
\providecommand \enquote  [1]{``#1''}%
\providecommand \bibnamefont  [1]{#1}%
\providecommand \bibfnamefont [1]{#1}%
\providecommand \citenamefont [1]{#1}%
\providecommand \href@noop [0]{\@secondoftwo}%
\providecommand \href [0]{\begingroup \@sanitize@url \@href}%
\providecommand \@href[1]{\@@startlink{#1}\@@href}%
\providecommand \@@href[1]{\endgroup#1\@@endlink}%
\providecommand \@sanitize@url [0]{\catcode `\\12\catcode `\$12\catcode
  `\&12\catcode `\#12\catcode `\^12\catcode `\_12\catcode `\%12\relax}%
\providecommand \@@startlink[1]{}%
\providecommand \@@endlink[0]{}%
\providecommand \url  [0]{\begingroup\@sanitize@url \@url }%
\providecommand \@url [1]{\endgroup\@href {#1}{\urlprefix }}%
\providecommand \urlprefix  [0]{URL }%
\providecommand \Eprint [0]{\href }%
\providecommand \doibase [0]{http://dx.doi.org/}%
\providecommand \selectlanguage [0]{\@gobble}%
\providecommand \bibinfo  [0]{\@secondoftwo}%
\providecommand \bibfield  [0]{\@secondoftwo}%
\providecommand \translation [1]{[#1]}%
\providecommand \BibitemOpen [0]{}%
\providecommand \bibitemStop [0]{}%
\providecommand \bibitemNoStop [0]{.\EOS\space}%
\providecommand \EOS [0]{\spacefactor3000\relax}%
\providecommand \BibitemShut  [1]{\csname bibitem#1\endcsname}%
\let\auto@bib@innerbib\@empty
%</preamble>
\bibitem [{\citenamefont {Schallamach}(1963)}]{schallamach1963theory}%
  \BibitemOpen
  \bibfield  {author} {\bibinfo {author} {\bibfnamefont {A.}~\bibnamefont
  {Schallamach}},\ }\bibfield  {title} {\enquote {\bibinfo {title} {A theory of
  dynamic rubber friction},}\ }\href@noop {} {\bibfield  {journal} {\bibinfo
  {journal} {Wear}\ }\textbf {\bibinfo {volume} {6}},\ \bibinfo {pages}
  {375--382} (\bibinfo {year} {1963})}\BibitemShut {NoStop}%
\bibitem [{\citenamefont {Leibler}, \citenamefont {Rubinstein},\ and\
  \citenamefont {Colby}(1991)}]{leibler1991dynamics}%
  \BibitemOpen
  \bibfield  {author} {\bibinfo {author} {\bibfnamefont {L.}~\bibnamefont
  {Leibler}}, \bibinfo {author} {\bibfnamefont {M.}~\bibnamefont {Rubinstein}},
  \ and\ \bibinfo {author} {\bibfnamefont {R.~H.}\ \bibnamefont {Colby}},\
  }\bibfield  {title} {\enquote {\bibinfo {title} {Dynamics of reversible
  networks},}\ }\href@noop {} {\bibfield  {journal} {\bibinfo  {journal}
  {Macromolecules}\ }\textbf {\bibinfo {volume} {24}},\ \bibinfo {pages}
  {4701--4707} (\bibinfo {year} {1991})}\BibitemShut {NoStop}%
\bibitem [{\citenamefont {Cao}\ and\ \citenamefont
  {Forest}(2019)}]{cao2019rheological}%
  \BibitemOpen
  \bibfield  {author} {\bibinfo {author} {\bibfnamefont {X.-Z.}\ \bibnamefont
  {Cao}}\ and\ \bibinfo {author} {\bibfnamefont {M.~G.}\ \bibnamefont
  {Forest}},\ }\bibfield  {title} {\enquote {\bibinfo {title} {Rheological
  tuning of entangled polymer networks by transient cross-links},}\ }\href@noop
  {} {\bibfield  {journal} {\bibinfo  {journal} {The Journal of Physical
  Chemistry B}\ }\textbf {\bibinfo {volume} {123}},\ \bibinfo {pages}
  {974--982} (\bibinfo {year} {2019})}\BibitemShut {NoStop}%
\bibitem [{\citenamefont {Lei}\ \emph {et~al.}(2020)\citenamefont {Lei},
  \citenamefont {Xia}, \citenamefont {Yang}, \citenamefont {Pica~Ciamarra},\
  and\ \citenamefont {Ni}}]{lei2020entropy}%
  \BibitemOpen
  \bibfield  {author} {\bibinfo {author} {\bibfnamefont {Q.-L.}\ \bibnamefont
  {Lei}}, \bibinfo {author} {\bibfnamefont {X.}~\bibnamefont {Xia}}, \bibinfo
  {author} {\bibfnamefont {J.}~\bibnamefont {Yang}}, \bibinfo {author}
  {\bibfnamefont {M.}~\bibnamefont {Pica~Ciamarra}}, \ and\ \bibinfo {author}
  {\bibfnamefont {R.}~\bibnamefont {Ni}},\ }\bibfield  {title} {\enquote
  {\bibinfo {title} {Entropy-controlled cross-linking in linker-mediated
  vitrimers},}\ }\href@noop {} {\bibfield  {journal} {\bibinfo  {journal}
  {Proceedings of the National Academy of Sciences}\ }\textbf {\bibinfo
  {volume} {117}},\ \bibinfo {pages} {27111--27115} (\bibinfo {year}
  {2020})}\BibitemShut {NoStop}%
\bibitem [{\citenamefont {Bidone}\ \emph {et~al.}(2017)\citenamefont {Bidone},
  \citenamefont {Jung}, \citenamefont {Maruri}, \citenamefont {Borau},
  \citenamefont {Kamm},\ and\ \citenamefont {Kim}}]{bidone2017morphological}%
  \BibitemOpen
  \bibfield  {author} {\bibinfo {author} {\bibfnamefont {T.~C.}\ \bibnamefont
  {Bidone}}, \bibinfo {author} {\bibfnamefont {W.}~\bibnamefont {Jung}},
  \bibinfo {author} {\bibfnamefont {D.}~\bibnamefont {Maruri}}, \bibinfo
  {author} {\bibfnamefont {C.}~\bibnamefont {Borau}}, \bibinfo {author}
  {\bibfnamefont {R.~D.}\ \bibnamefont {Kamm}}, \ and\ \bibinfo {author}
  {\bibfnamefont {T.}~\bibnamefont {Kim}},\ }\bibfield  {title} {\enquote
  {\bibinfo {title} {Morphological transformation and force generation of
  active cytoskeletal networks},}\ }\href@noop {} {\bibfield  {journal}
  {\bibinfo  {journal} {PLoS computational biology}\ }\textbf {\bibinfo
  {volume} {13}},\ \bibinfo {pages} {e1005277} (\bibinfo {year}
  {2017})}\BibitemShut {NoStop}%
\bibitem [{\citenamefont {Descovich}\ \emph {et~al.}(2018)\citenamefont
  {Descovich}, \citenamefont {Cortes}, \citenamefont {Ryan}, \citenamefont
  {Nash}, \citenamefont {Zhang}, \citenamefont {Maddox}, \citenamefont
  {Nedelec},\ and\ \citenamefont {Maddox}}]{descovich2018cross}%
  \BibitemOpen
  \bibfield  {author} {\bibinfo {author} {\bibfnamefont {C.~P.}\ \bibnamefont
  {Descovich}}, \bibinfo {author} {\bibfnamefont {D.~B.}\ \bibnamefont
  {Cortes}}, \bibinfo {author} {\bibfnamefont {S.}~\bibnamefont {Ryan}},
  \bibinfo {author} {\bibfnamefont {J.}~\bibnamefont {Nash}}, \bibinfo {author}
  {\bibfnamefont {L.}~\bibnamefont {Zhang}}, \bibinfo {author} {\bibfnamefont
  {P.~S.}\ \bibnamefont {Maddox}}, \bibinfo {author} {\bibfnamefont
  {F.}~\bibnamefont {Nedelec}}, \ and\ \bibinfo {author} {\bibfnamefont
  {A.~S.}\ \bibnamefont {Maddox}},\ }\bibfield  {title} {\enquote {\bibinfo
  {title} {Cross-linkers both drive and brake cytoskeletal remodeling and
  furrowing in cytokinesis},}\ }\href@noop {} {\bibfield  {journal} {\bibinfo
  {journal} {Molecular Biology of the Cell}\ }\textbf {\bibinfo {volume}
  {29}},\ \bibinfo {pages} {622--631} (\bibinfo {year} {2018})}\BibitemShut
  {NoStop}%
\bibitem [{\citenamefont {Korosec}\ and\ \citenamefont
  {Forde}(2021)}]{korosec2021lawnmower}%
  \BibitemOpen
  \bibfield  {author} {\bibinfo {author} {\bibfnamefont {C.~S.}\ \bibnamefont
  {Korosec}}\ and\ \bibinfo {author} {\bibfnamefont {N.~R.}\ \bibnamefont
  {Forde}},\ }\bibfield  {title} {\enquote {\bibinfo {title} {The lawnmower: an
  artificial protein-based burnt-bridge molecular motor},}\ }\href@noop {}
  {\bibfield  {journal} {\bibinfo  {journal} {arXiv preprint arXiv:2109.10293}\
  } (\bibinfo {year} {2021})}\BibitemShut {NoStop}%
\bibitem [{\citenamefont {Li}\ \emph {et~al.}(2021)\citenamefont {Li},
  \citenamefont {Kamal}, \citenamefont {Stumpf}, \citenamefont {Thibaudau},
  \citenamefont {Sengupta},\ and\ \citenamefont {Smith}}]{li2021biomechanics}%
  \BibitemOpen
  \bibfield  {author} {\bibinfo {author} {\bibfnamefont {L.}~\bibnamefont
  {Li}}, \bibinfo {author} {\bibfnamefont {M.~A.}\ \bibnamefont {Kamal}},
  \bibinfo {author} {\bibfnamefont {B.~H.}\ \bibnamefont {Stumpf}}, \bibinfo
  {author} {\bibfnamefont {F.}~\bibnamefont {Thibaudau}}, \bibinfo {author}
  {\bibfnamefont {K.}~\bibnamefont {Sengupta}}, \ and\ \bibinfo {author}
  {\bibfnamefont {A.-S.}\ \bibnamefont {Smith}},\ }\bibfield  {title} {\enquote
  {\bibinfo {title} {Biomechanics as driver of aggregation of tethers in
  adherent membranes},}\ }\href@noop {} {\bibfield  {journal} {\bibinfo
  {journal} {Soft Matter}\ }\textbf {\bibinfo {volume} {17}},\ \bibinfo {pages}
  {10101--10107} (\bibinfo {year} {2021})}\BibitemShut {NoStop}%
\bibitem [{\citenamefont {Bihr}, \citenamefont {Seifert},\ and\ \citenamefont
  {Smith}(2012)}]{bihr2012nucleation}%
  \BibitemOpen
  \bibfield  {author} {\bibinfo {author} {\bibfnamefont {T.}~\bibnamefont
  {Bihr}}, \bibinfo {author} {\bibfnamefont {U.}~\bibnamefont {Seifert}}, \
  and\ \bibinfo {author} {\bibfnamefont {A.-S.}\ \bibnamefont {Smith}},\
  }\bibfield  {title} {\enquote {\bibinfo {title} {Nucleation of
  ligand-receptor domains in membrane adhesion},}\ }\href@noop {} {\bibfield
  {journal} {\bibinfo  {journal} {Physical review letters}\ }\textbf {\bibinfo
  {volume} {109}},\ \bibinfo {pages} {258101} (\bibinfo {year}
  {2012})}\BibitemShut {NoStop}%
\bibitem [{\citenamefont {Bressloff}\ and\ \citenamefont
  {Newby}(2013)}]{bressloff2013stochastic}%
  \BibitemOpen
  \bibfield  {author} {\bibinfo {author} {\bibfnamefont {P.~C.}\ \bibnamefont
  {Bressloff}}\ and\ \bibinfo {author} {\bibfnamefont {J.~M.}\ \bibnamefont
  {Newby}},\ }\bibfield  {title} {\enquote {\bibinfo {title} {Stochastic models
  of intracellular transport},}\ }\href@noop {} {\bibfield  {journal} {\bibinfo
   {journal} {Reviews of Modern Physics}\ }\textbf {\bibinfo {volume} {85}},\
  \bibinfo {pages} {135} (\bibinfo {year} {2013})}\BibitemShut {NoStop}%
\bibitem [{\citenamefont {Goyette}\ \emph {et~al.}(2017)\citenamefont
  {Goyette}, \citenamefont {Salas}, \citenamefont {Coker-Gordon}, \citenamefont
  {Bridge}, \citenamefont {Isaacson}, \citenamefont {Allard},\ and\
  \citenamefont {Dushek}}]{goyette2017biophysical}%
  \BibitemOpen
  \bibfield  {author} {\bibinfo {author} {\bibfnamefont {J.}~\bibnamefont
  {Goyette}}, \bibinfo {author} {\bibfnamefont {C.~S.}\ \bibnamefont {Salas}},
  \bibinfo {author} {\bibfnamefont {N.}~\bibnamefont {Coker-Gordon}}, \bibinfo
  {author} {\bibfnamefont {M.}~\bibnamefont {Bridge}}, \bibinfo {author}
  {\bibfnamefont {S.~A.}\ \bibnamefont {Isaacson}}, \bibinfo {author}
  {\bibfnamefont {J.}~\bibnamefont {Allard}}, \ and\ \bibinfo {author}
  {\bibfnamefont {O.}~\bibnamefont {Dushek}},\ }\bibfield  {title} {\enquote
  {\bibinfo {title} {Biophysical assay for tethered signaling reactions reveals
  tether-controlled activity for the phosphatase shp-1},}\ }\href@noop {}
  {\bibfield  {journal} {\bibinfo  {journal} {Science Advances}\ }\textbf
  {\bibinfo {volume} {3}},\ \bibinfo {pages} {e1601692} (\bibinfo {year}
  {2017})}\BibitemShut {NoStop}%
\bibitem [{\citenamefont {Gaillac}\ \emph {et~al.}(2017)\citenamefont
  {Gaillac}, \citenamefont {Pullumbi}, \citenamefont {Beyer}, \citenamefont
  {Chapman}, \citenamefont {Keen}, \citenamefont {Bennett},\ and\ \citenamefont
  {Coudert}}]{gaillac2017liquid}%
  \BibitemOpen
  \bibfield  {author} {\bibinfo {author} {\bibfnamefont {R.}~\bibnamefont
  {Gaillac}}, \bibinfo {author} {\bibfnamefont {P.}~\bibnamefont {Pullumbi}},
  \bibinfo {author} {\bibfnamefont {K.~A.}\ \bibnamefont {Beyer}}, \bibinfo
  {author} {\bibfnamefont {K.~W.}\ \bibnamefont {Chapman}}, \bibinfo {author}
  {\bibfnamefont {D.~A.}\ \bibnamefont {Keen}}, \bibinfo {author}
  {\bibfnamefont {T.~D.}\ \bibnamefont {Bennett}}, \ and\ \bibinfo {author}
  {\bibfnamefont {F.-X.}\ \bibnamefont {Coudert}},\ }\bibfield  {title}
  {\enquote {\bibinfo {title} {Liquid metal--organic frameworks},}\ }\href@noop
  {} {\bibfield  {journal} {\bibinfo  {journal} {Nature Materials}\ }\textbf
  {\bibinfo {volume} {16}},\ \bibinfo {pages} {1149--1154} (\bibinfo {year}
  {2017})}\BibitemShut {NoStop}%
\bibitem [{\citenamefont {Balestra}\ and\ \citenamefont
  {Semino}(2022)}]{balestra2022computer}%
  \BibitemOpen
  \bibfield  {author} {\bibinfo {author} {\bibfnamefont {S.~R.}\ \bibnamefont
  {Balestra}}\ and\ \bibinfo {author} {\bibfnamefont {R.}~\bibnamefont
  {Semino}},\ }\bibfield  {title} {\enquote {\bibinfo {title} {Computer
  simulation of the early stages of self-assembly and thermal decomposition of
  zif-8},}\ }\href@noop {} {\bibfield  {journal} {\bibinfo  {journal} {The
  Journal of Chemical Physics}\ }\textbf {\bibinfo {volume} {157}},\ \bibinfo
  {pages} {184502} (\bibinfo {year} {2022})}\BibitemShut {NoStop}%
\bibitem [{\citenamefont {Mirkin}\ \emph {et~al.}(1996)\citenamefont {Mirkin},
  \citenamefont {Letsinger}, \citenamefont {Mucic},\ and\ \citenamefont
  {Storhoff}}]{mirkin1996dna}%
  \BibitemOpen
  \bibfield  {author} {\bibinfo {author} {\bibfnamefont {C.~A.}\ \bibnamefont
  {Mirkin}}, \bibinfo {author} {\bibfnamefont {R.~L.}\ \bibnamefont
  {Letsinger}}, \bibinfo {author} {\bibfnamefont {R.~C.}\ \bibnamefont
  {Mucic}}, \ and\ \bibinfo {author} {\bibfnamefont {J.~J.}\ \bibnamefont
  {Storhoff}},\ }\bibfield  {title} {\enquote {\bibinfo {title} {A dna-based
  method for rationally assembling nanoparticles into macroscopic materials},}\
  }\href@noop {} {\bibfield  {journal} {\bibinfo  {journal} {Nature}\ }\textbf
  {\bibinfo {volume} {382}},\ \bibinfo {pages} {607--609} (\bibinfo {year}
  {1996})}\BibitemShut {NoStop}%
\bibitem [{\citenamefont {Di~Michele}\ \emph {et~al.}(2013)\citenamefont
  {Di~Michele}, \citenamefont {Varrato}, \citenamefont {Kotar}, \citenamefont
  {Nathan}, \citenamefont {Foffi},\ and\ \citenamefont
  {Eiser}}]{di2013multistep}%
  \BibitemOpen
  \bibfield  {author} {\bibinfo {author} {\bibfnamefont {L.}~\bibnamefont
  {Di~Michele}}, \bibinfo {author} {\bibfnamefont {F.}~\bibnamefont {Varrato}},
  \bibinfo {author} {\bibfnamefont {J.}~\bibnamefont {Kotar}}, \bibinfo
  {author} {\bibfnamefont {S.~H.}\ \bibnamefont {Nathan}}, \bibinfo {author}
  {\bibfnamefont {G.}~\bibnamefont {Foffi}}, \ and\ \bibinfo {author}
  {\bibfnamefont {E.}~\bibnamefont {Eiser}},\ }\bibfield  {title} {\enquote
  {\bibinfo {title} {Multistep kinetic self-assembly of dna-coated colloids},}\
  }\href@noop {} {\bibfield  {journal} {\bibinfo  {journal} {Nature
  communications}\ }\textbf {\bibinfo {volume} {4}},\ \bibinfo {pages} {2007}
  (\bibinfo {year} {2013})}\BibitemShut {NoStop}%
\bibitem [{\citenamefont {Feng}\ \emph {et~al.}(2013)\citenamefont {Feng},
  \citenamefont {Pontani}, \citenamefont {Dreyfus}, \citenamefont {Chaikin},\
  and\ \citenamefont {Brujic}}]{feng2013specificity}%
  \BibitemOpen
  \bibfield  {author} {\bibinfo {author} {\bibfnamefont {L.}~\bibnamefont
  {Feng}}, \bibinfo {author} {\bibfnamefont {L.-L.}\ \bibnamefont {Pontani}},
  \bibinfo {author} {\bibfnamefont {R.}~\bibnamefont {Dreyfus}}, \bibinfo
  {author} {\bibfnamefont {P.}~\bibnamefont {Chaikin}}, \ and\ \bibinfo
  {author} {\bibfnamefont {J.}~\bibnamefont {Brujic}},\ }\bibfield  {title}
  {\enquote {\bibinfo {title} {Specificity, flexibility and valence of dna
  bonds guide emulsion architecture},}\ }\href@noop {} {\bibfield  {journal}
  {\bibinfo  {journal} {Soft Matter}\ }\textbf {\bibinfo {volume} {9}},\
  \bibinfo {pages} {9816--9823} (\bibinfo {year} {2013})}\BibitemShut {NoStop}%
\bibitem [{\citenamefont {Wang}\ \emph {et~al.}(2015)\citenamefont {Wang},
  \citenamefont {Wang}, \citenamefont {Zheng}, \citenamefont {Ducrot},
  \citenamefont {Yodh}, \citenamefont {Weck},\ and\ \citenamefont
  {Pine}}]{wang2015crystallization}%
  \BibitemOpen
  \bibfield  {author} {\bibinfo {author} {\bibfnamefont {Y.}~\bibnamefont
  {Wang}}, \bibinfo {author} {\bibfnamefont {Y.}~\bibnamefont {Wang}}, \bibinfo
  {author} {\bibfnamefont {X.}~\bibnamefont {Zheng}}, \bibinfo {author}
  {\bibfnamefont {{\'E}.}~\bibnamefont {Ducrot}}, \bibinfo {author}
  {\bibfnamefont {J.~S.}\ \bibnamefont {Yodh}}, \bibinfo {author}
  {\bibfnamefont {M.}~\bibnamefont {Weck}}, \ and\ \bibinfo {author}
  {\bibfnamefont {D.~J.}\ \bibnamefont {Pine}},\ }\bibfield  {title} {\enquote
  {\bibinfo {title} {Crystallization of dna-coated colloids},}\ }\href@noop {}
  {\bibfield  {journal} {\bibinfo  {journal} {Nature Communications}\ }\textbf
  {\bibinfo {volume} {6}},\ \bibinfo {pages} {1--8} (\bibinfo {year}
  {2015})}\BibitemShut {NoStop}%
\bibitem [{\citenamefont {Rogers}, \citenamefont {Shih},\ and\ \citenamefont
  {Manoharan}(2016)}]{rogers2016using}%
  \BibitemOpen
  \bibfield  {author} {\bibinfo {author} {\bibfnamefont {W.~B.}\ \bibnamefont
  {Rogers}}, \bibinfo {author} {\bibfnamefont {W.~M.}\ \bibnamefont {Shih}}, \
  and\ \bibinfo {author} {\bibfnamefont {V.~N.}\ \bibnamefont {Manoharan}},\
  }\bibfield  {title} {\enquote {\bibinfo {title} {Using dna to program the
  self-assembly of colloidal nanoparticles and microparticles},}\ }\href@noop
  {} {\bibfield  {journal} {\bibinfo  {journal} {Nature Reviews Materials}\
  }\textbf {\bibinfo {volume} {1}},\ \bibinfo {pages} {1--14} (\bibinfo {year}
  {2016})}\BibitemShut {NoStop}%
\bibitem [{\citenamefont {Xia}\ \emph {et~al.}(2020)\citenamefont {Xia},
  \citenamefont {Hu}, \citenamefont {Ciamarra},\ and\ \citenamefont
  {Ni}}]{xia2020linker}%
  \BibitemOpen
  \bibfield  {author} {\bibinfo {author} {\bibfnamefont {X.}~\bibnamefont
  {Xia}}, \bibinfo {author} {\bibfnamefont {H.}~\bibnamefont {Hu}}, \bibinfo
  {author} {\bibfnamefont {M.~P.}\ \bibnamefont {Ciamarra}}, \ and\ \bibinfo
  {author} {\bibfnamefont {R.}~\bibnamefont {Ni}},\ }\bibfield  {title}
  {\enquote {\bibinfo {title} {Linker-mediated self-assembly of mobile
  dna-coated colloids},}\ }\href@noop {} {\bibfield  {journal} {\bibinfo
  {journal} {Science Advances}\ }\textbf {\bibinfo {volume} {6}},\ \bibinfo
  {pages} {eaaz6921} (\bibinfo {year} {2020})}\BibitemShut {NoStop}%
\bibitem [{\citenamefont {Cui}\ \emph {et~al.}(2022)\citenamefont {Cui},
  \citenamefont {Marbach}, \citenamefont {Zheng}, \citenamefont
  {Holmes-Cerfon},\ and\ \citenamefont {Pine}}]{cui2022comprehensive}%
  \BibitemOpen
  \bibfield  {author} {\bibinfo {author} {\bibfnamefont {F.}~\bibnamefont
  {Cui}}, \bibinfo {author} {\bibfnamefont {S.}~\bibnamefont {Marbach}},
  \bibinfo {author} {\bibfnamefont {J.~A.}\ \bibnamefont {Zheng}}, \bibinfo
  {author} {\bibfnamefont {M.}~\bibnamefont {Holmes-Cerfon}}, \ and\ \bibinfo
  {author} {\bibfnamefont {D.~J.}\ \bibnamefont {Pine}},\ }\bibfield  {title}
  {\enquote {\bibinfo {title} {Comprehensive view of microscopic interactions
  between dna-coated colloids},}\ }\href@noop {} {\bibfield  {journal}
  {\bibinfo  {journal} {Nature communications}\ }\textbf {\bibinfo {volume}
  {13}},\ \bibinfo {pages} {1--10} (\bibinfo {year} {2022})}\BibitemShut
  {NoStop}%
\bibitem [{\citenamefont {Gehrels}\ \emph {et~al.}(2022)\citenamefont
  {Gehrels}, \citenamefont {Rogers}, \citenamefont {Zeravcic},\ and\
  \citenamefont {Manoharan}}]{gehrels2022programming}%
  \BibitemOpen
  \bibfield  {author} {\bibinfo {author} {\bibfnamefont {E.~W.}\ \bibnamefont
  {Gehrels}}, \bibinfo {author} {\bibfnamefont {W.~B.}\ \bibnamefont {Rogers}},
  \bibinfo {author} {\bibfnamefont {Z.}~\bibnamefont {Zeravcic}}, \ and\
  \bibinfo {author} {\bibfnamefont {V.~N.}\ \bibnamefont {Manoharan}},\
  }\bibfield  {title} {\enquote {\bibinfo {title} {Programming directed motion
  with {DNA}-grafted particles},}\ }\href@noop {} {\bibfield  {journal}
  {\bibinfo  {journal} {ACS Nano}\ } (\bibinfo {year} {2022})}\BibitemShut
  {NoStop}%
\bibitem [{\citenamefont {Maxian}\ \emph {et~al.}(2021)\citenamefont {Maxian},
  \citenamefont {Pel{\'a}ez}, \citenamefont {Mogilner},\ and\ \citenamefont
  {Donev}}]{maxian2021simulations}%
  \BibitemOpen
  \bibfield  {author} {\bibinfo {author} {\bibfnamefont {O.}~\bibnamefont
  {Maxian}}, \bibinfo {author} {\bibfnamefont {R.~P.}\ \bibnamefont
  {Pel{\'a}ez}}, \bibinfo {author} {\bibfnamefont {A.}~\bibnamefont
  {Mogilner}}, \ and\ \bibinfo {author} {\bibfnamefont {A.}~\bibnamefont
  {Donev}},\ }\bibfield  {title} {\enquote {\bibinfo {title} {Simulations of
  dynamically cross-linked actin networks: morphology, rheology, and
  hydrodynamic interactions},}\ }\href@noop {} {\bibfield  {journal} {\bibinfo
  {journal} {PLoS Computational Biology}\ }\textbf {\bibinfo {volume} {17}},\
  \bibinfo {pages} {e1009240} (\bibinfo {year} {2021})}\BibitemShut {NoStop}%
\bibitem [{\citenamefont {Korosec}\ \emph {et~al.}(2021)\citenamefont
  {Korosec}, \citenamefont {Jindal}, \citenamefont {Schneider}, \citenamefont
  {de~la Barca}, \citenamefont {Zuckermann}, \citenamefont {Forde},\ and\
  \citenamefont {Emberly}}]{korosec2021substrate}%
  \BibitemOpen
  \bibfield  {author} {\bibinfo {author} {\bibfnamefont {C.~S.}\ \bibnamefont
  {Korosec}}, \bibinfo {author} {\bibfnamefont {L.}~\bibnamefont {Jindal}},
  \bibinfo {author} {\bibfnamefont {M.}~\bibnamefont {Schneider}}, \bibinfo
  {author} {\bibfnamefont {I.~C.}\ \bibnamefont {de~la Barca}}, \bibinfo
  {author} {\bibfnamefont {M.~J.}\ \bibnamefont {Zuckermann}}, \bibinfo
  {author} {\bibfnamefont {N.~R.}\ \bibnamefont {Forde}}, \ and\ \bibinfo
  {author} {\bibfnamefont {E.}~\bibnamefont {Emberly}},\ }\bibfield  {title}
  {\enquote {\bibinfo {title} {Substrate stiffness tunes the dynamics of
  polyvalent rolling motors},}\ }\href@noop {} {\bibfield  {journal} {\bibinfo
  {journal} {Soft Matter}\ }\textbf {\bibinfo {volume} {17}},\ \bibinfo {pages}
  {1468--1479} (\bibinfo {year} {2021})}\BibitemShut {NoStop}%
\bibitem [{\citenamefont {Marbach}, \citenamefont {Zheng},\ and\ \citenamefont
  {Holmes-Cerfon}(2022)}]{marbach2022nanocaterpillar}%
  \BibitemOpen
  \bibfield  {author} {\bibinfo {author} {\bibfnamefont {S.}~\bibnamefont
  {Marbach}}, \bibinfo {author} {\bibfnamefont {J.~A.}\ \bibnamefont {Zheng}},
  \ and\ \bibinfo {author} {\bibfnamefont {M.}~\bibnamefont {Holmes-Cerfon}},\
  }\bibfield  {title} {\enquote {\bibinfo {title} {The nanocaterpillar's random
  walk: diffusion with ligand--receptor contacts},}\ }\href@noop {} {\bibfield
  {journal} {\bibinfo  {journal} {Soft Matter}\ }\textbf {\bibinfo {volume}
  {18}},\ \bibinfo {pages} {3130--3146} (\bibinfo {year} {2022})}\BibitemShut
  {NoStop}%
\bibitem [{\citenamefont {Jane\v{s}}\ \emph {et~al.}(2022)\citenamefont
  {Jane\v{s}}, \citenamefont {Monzel}, \citenamefont {Schmidt}, \citenamefont
  {Merkel}, \citenamefont {Seifert}, \citenamefont {Sengupta},\ and\
  \citenamefont {Smith}}]{PhysRevX.12.031030}%
  \BibitemOpen
  \bibfield  {author} {\bibinfo {author} {\bibfnamefont {J.~A.}\ \bibnamefont
  {Jane\v{s}}}, \bibinfo {author} {\bibfnamefont {C.}~\bibnamefont {Monzel}},
  \bibinfo {author} {\bibfnamefont {D.}~\bibnamefont {Schmidt}}, \bibinfo
  {author} {\bibfnamefont {R.}~\bibnamefont {Merkel}}, \bibinfo {author}
  {\bibfnamefont {U.}~\bibnamefont {Seifert}}, \bibinfo {author} {\bibfnamefont
  {K.}~\bibnamefont {Sengupta}}, \ and\ \bibinfo {author} {\bibfnamefont
  {A.-S.}\ \bibnamefont {Smith}},\ }\bibfield  {title} {\enquote {\bibinfo
  {title} {First-principle coarse-graining framework for scale-free bell-like
  association and dissociation rates in thermal and active systems},}\ }\href
  {\doibase 10.1103/PhysRevX.12.031030} {\bibfield  {journal} {\bibinfo
  {journal} {Physical Review X}\ }\textbf {\bibinfo {volume} {12}},\ \bibinfo
  {pages} {031030} (\bibinfo {year} {2022})}\BibitemShut {NoStop}%
\bibitem [{\citenamefont {M{\"u}ller-Plathe}(2002)}]{muller2002coarse}%
  \BibitemOpen
  \bibfield  {author} {\bibinfo {author} {\bibfnamefont {F.}~\bibnamefont
  {M{\"u}ller-Plathe}},\ }\bibfield  {title} {\enquote {\bibinfo {title}
  {Coarse-graining in polymer simulation: from the atomistic to the mesoscopic
  scale and back},}\ }\href@noop {} {\bibfield  {journal} {\bibinfo  {journal}
  {ChemPhysChem}\ }\textbf {\bibinfo {volume} {3}},\ \bibinfo {pages}
  {754--769} (\bibinfo {year} {2002})}\BibitemShut {NoStop}%
\bibitem [{\citenamefont {Fogelson}\ and\ \citenamefont
  {Keener}(2018)}]{fogelson2018enhanced}%
  \BibitemOpen
  \bibfield  {author} {\bibinfo {author} {\bibfnamefont {B.}~\bibnamefont
  {Fogelson}}\ and\ \bibinfo {author} {\bibfnamefont {J.~P.}\ \bibnamefont
  {Keener}},\ }\bibfield  {title} {\enquote {\bibinfo {title} {Enhanced
  nucleocytoplasmic transport due to competition for elastic binding sites},}\
  }\href@noop {} {\bibfield  {journal} {\bibinfo  {journal} {Biophysical
  Journal}\ }\textbf {\bibinfo {volume} {115}},\ \bibinfo {pages} {108--116}
  (\bibinfo {year} {2018})}\BibitemShut {NoStop}%
\bibitem [{\citenamefont {Fogelson}\ and\ \citenamefont
  {Keener}(2019)}]{fogelson2019transport}%
  \BibitemOpen
  \bibfield  {author} {\bibinfo {author} {\bibfnamefont {B.}~\bibnamefont
  {Fogelson}}\ and\ \bibinfo {author} {\bibfnamefont {J.~P.}\ \bibnamefont
  {Keener}},\ }\bibfield  {title} {\enquote {\bibinfo {title} {Transport
  facilitated by rapid binding to elastic tethers},}\ }\href@noop {} {\bibfield
   {journal} {\bibinfo  {journal} {SIAM Journal on Applied Mathematics}\
  }\textbf {\bibinfo {volume} {79}},\ \bibinfo {pages} {1405--1422} (\bibinfo
  {year} {2019})}\BibitemShut {NoStop}%
\bibitem [{\citenamefont {Jana}\ and\ \citenamefont
  {Mognetti}(2019)}]{jana2019translational}%
  \BibitemOpen
  \bibfield  {author} {\bibinfo {author} {\bibfnamefont {P.~K.}\ \bibnamefont
  {Jana}}\ and\ \bibinfo {author} {\bibfnamefont {B.~M.}\ \bibnamefont
  {Mognetti}},\ }\bibfield  {title} {\enquote {\bibinfo {title} {Translational
  and rotational dynamics of colloidal particles interacting through reacting
  linkers},}\ }\href@noop {} {\bibfield  {journal} {\bibinfo  {journal}
  {Physical Review E}\ }\textbf {\bibinfo {volume} {100}},\ \bibinfo {pages}
  {060601} (\bibinfo {year} {2019})}\BibitemShut {NoStop}%
\bibitem [{\citenamefont {Fr{\"o}hner}\ and\ \citenamefont
  {No{\'e}}(2018)}]{noe2018}%
  \BibitemOpen
  \bibfield  {author} {\bibinfo {author} {\bibfnamefont {C.}~\bibnamefont
  {Fr{\"o}hner}}\ and\ \bibinfo {author} {\bibfnamefont {F.}~\bibnamefont
  {No{\'e}}},\ }\bibfield  {title} {\enquote {\bibinfo {title} {Reversible
  interacting-particle reaction dynamics},}\ }\href@noop {} {\bibfield
  {journal} {\bibinfo  {journal} {The Journal of Physical Chemistry B}\
  }\textbf {\bibinfo {volume} {122}},\ \bibinfo {pages} {11240--11250}
  (\bibinfo {year} {2018})}\BibitemShut {NoStop}%
\bibitem [{\citenamefont {Maxian}, \citenamefont {Donev},\ and\ \citenamefont
  {Mogilner}(2022)}]{maxian2022interplay}%
  \BibitemOpen
  \bibfield  {author} {\bibinfo {author} {\bibfnamefont {O.}~\bibnamefont
  {Maxian}}, \bibinfo {author} {\bibfnamefont {A.}~\bibnamefont {Donev}}, \
  and\ \bibinfo {author} {\bibfnamefont {A.}~\bibnamefont {Mogilner}},\
  }\bibfield  {title} {\enquote {\bibinfo {title} {Interplay between brownian
  motion and cross-linking controls bundling dynamics in actin networks},}\
  }\href@noop {} {\bibfield  {journal} {\bibinfo  {journal} {Biophysical
  journal}\ }\textbf {\bibinfo {volume} {121}},\ \bibinfo {pages} {1230--1245}
  (\bibinfo {year} {2022})}\BibitemShut {NoStop}%
\bibitem [{\citenamefont {Korosec}, \citenamefont {Zuckermann},\ and\
  \citenamefont {Forde}(2018)}]{korosec2018dimensionality}%
  \BibitemOpen
  \bibfield  {author} {\bibinfo {author} {\bibfnamefont {C.~S.}\ \bibnamefont
  {Korosec}}, \bibinfo {author} {\bibfnamefont {M.~J.}\ \bibnamefont
  {Zuckermann}}, \ and\ \bibinfo {author} {\bibfnamefont {N.~R.}\ \bibnamefont
  {Forde}},\ }\bibfield  {title} {\enquote {\bibinfo {title}
  {Dimensionality-dependent crossover in motility of polyvalent burnt-bridges
  ratchets},}\ }\href@noop {} {\bibfield  {journal} {\bibinfo  {journal}
  {Physical Review E}\ }\textbf {\bibinfo {volume} {98}},\ \bibinfo {pages}
  {032114} (\bibinfo {year} {2018})}\BibitemShut {NoStop}%
\bibitem [{\citenamefont {Kowalewski}, \citenamefont {Forde},\ and\
  \citenamefont {Korosec}(2021)}]{kowalewski2021multivalent}%
  \BibitemOpen
  \bibfield  {author} {\bibinfo {author} {\bibfnamefont {A.}~\bibnamefont
  {Kowalewski}}, \bibinfo {author} {\bibfnamefont {N.~R.}\ \bibnamefont
  {Forde}}, \ and\ \bibinfo {author} {\bibfnamefont {C.~S.}\ \bibnamefont
  {Korosec}},\ }\bibfield  {title} {\enquote {\bibinfo {title} {Multivalent
  diffusive transport},}\ }\href@noop {} {\bibfield  {journal} {\bibinfo
  {journal} {The Journal of Physical Chemistry B}\ }\textbf {\bibinfo {volume}
  {125}},\ \bibinfo {pages} {6857--6863} (\bibinfo {year} {2021})}\BibitemShut
  {NoStop}%
\bibitem [{\citenamefont {Olah}\ and\ \citenamefont
  {Stefanovic}(2013)}]{olah2013superdiffusive}%
  \BibitemOpen
  \bibfield  {author} {\bibinfo {author} {\bibfnamefont {M.~J.}\ \bibnamefont
  {Olah}}\ and\ \bibinfo {author} {\bibfnamefont {D.}~\bibnamefont
  {Stefanovic}},\ }\bibfield  {title} {\enquote {\bibinfo {title}
  {Superdiffusive transport by multivalent molecular walkers moving under
  load},}\ }\href@noop {} {\bibfield  {journal} {\bibinfo  {journal} {Physical
  Review E}\ }\textbf {\bibinfo {volume} {87}},\ \bibinfo {pages} {062713}
  (\bibinfo {year} {2013})}\BibitemShut {NoStop}%
\bibitem [{\citenamefont {Merminod}\ \emph {et~al.}(2021)\citenamefont
  {Merminod}, \citenamefont {Edison}, \citenamefont {Fang}, \citenamefont
  {Hagan},\ and\ \citenamefont {Rogers}}]{merminod2021avidity}%
  \BibitemOpen
  \bibfield  {author} {\bibinfo {author} {\bibfnamefont {S.}~\bibnamefont
  {Merminod}}, \bibinfo {author} {\bibfnamefont {J.~R.}\ \bibnamefont
  {Edison}}, \bibinfo {author} {\bibfnamefont {H.}~\bibnamefont {Fang}},
  \bibinfo {author} {\bibfnamefont {M.~F.}\ \bibnamefont {Hagan}}, \ and\
  \bibinfo {author} {\bibfnamefont {W.~B.}\ \bibnamefont {Rogers}},\ }\bibfield
   {title} {\enquote {\bibinfo {title} {Avidity and surface mobility in
  multivalent ligand-receptor binding},}\ }\href@noop {} {\bibfield  {journal}
  {\bibinfo  {journal} {Nanoscale}\ } (\bibinfo {year} {2021})}\BibitemShut
  {NoStop}%
\bibitem [{\citenamefont {Mitra}\ \emph {et~al.}(2022)\citenamefont {Mitra},
  \citenamefont {Chang}, \citenamefont {McMullen}, \citenamefont {Puchall},
  \citenamefont {Brujic},\ and\ \citenamefont {Hocky}}]{mitra2022coarse}%
  \BibitemOpen
  \bibfield  {author} {\bibinfo {author} {\bibfnamefont {G.}~\bibnamefont
  {Mitra}}, \bibinfo {author} {\bibfnamefont {C.}~\bibnamefont {Chang}},
  \bibinfo {author} {\bibfnamefont {A.}~\bibnamefont {McMullen}}, \bibinfo
  {author} {\bibfnamefont {D.}~\bibnamefont {Puchall}}, \bibinfo {author}
  {\bibfnamefont {J.}~\bibnamefont {Brujic}}, \ and\ \bibinfo {author}
  {\bibfnamefont {G.~M.}\ \bibnamefont {Hocky}},\ }\bibfield  {title} {\enquote
  {\bibinfo {title} {A coarse-grained simulation model for self-assembly of
  liquid droplets featuring explicit mobile binders},}\ }\href@noop {}
  {\bibfield  {journal} {\bibinfo  {journal} {arXiv preprint arXiv:2212.11946}\
  } (\bibinfo {year} {2022})}\BibitemShut {NoStop}%
\bibitem [{\citenamefont {Zhang}\ and\ \citenamefont
  {Isaacson}(2022)}]{zhang2022detailed}%
  \BibitemOpen
  \bibfield  {author} {\bibinfo {author} {\bibfnamefont {Y.}~\bibnamefont
  {Zhang}}\ and\ \bibinfo {author} {\bibfnamefont {S.~A.}\ \bibnamefont
  {Isaacson}},\ }\bibfield  {title} {\enquote {\bibinfo {title} {Detailed
  balance for particle models of reversible reactions in bounded domains},}\
  }\href@noop {} {\bibfield  {journal} {\bibinfo  {journal} {The Journal of
  Chemical Physics}\ }\textbf {\bibinfo {volume} {156}},\ \bibinfo {pages}
  {204105} (\bibinfo {year} {2022})}\BibitemShut {NoStop}%
\bibitem [{\citenamefont {Rubinstein}, \citenamefont {Colby}\ \emph
  {et~al.}(2003)\citenamefont {Rubinstein}, \citenamefont {Colby} \emph
  {et~al.}}]{rubinstein2003polymer}%
  \BibitemOpen
  \bibfield  {author} {\bibinfo {author} {\bibfnamefont {M.}~\bibnamefont
  {Rubinstein}}, \bibinfo {author} {\bibfnamefont {R.~H.}\ \bibnamefont
  {Colby}},  \emph {et~al.},\ }\href@noop {} {\emph {\bibinfo {title} {Polymer
  physics}}},\ Vol.~\bibinfo {volume} {23}\ (\bibinfo  {publisher} {Oxford
  university press New York},\ \bibinfo {year} {2003})\BibitemShut {NoStop}%
\bibitem [{\citenamefont {Marbach}\ and\ \citenamefont
  {Holmes-Cerfon}(2022)}]{marbach2022mass}%
  \BibitemOpen
  \bibfield  {author} {\bibinfo {author} {\bibfnamefont {S.}~\bibnamefont
  {Marbach}}\ and\ \bibinfo {author} {\bibfnamefont {M.}~\bibnamefont
  {Holmes-Cerfon}},\ }\bibfield  {title} {\enquote {\bibinfo {title} {Mass
  changes the diffusion coefficient of particles with ligand-receptor contacts
  in the overdamped limit},}\ }\href@noop {} {\bibfield  {journal} {\bibinfo
  {journal} {Physical Review Letters}\ }\textbf {\bibinfo {volume} {129}},\
  \bibinfo {pages} {048003} (\bibinfo {year} {2022})}\BibitemShut {NoStop}%
\bibitem [{\citenamefont {Morse}(2003)}]{morse2003Theory}%
  \BibitemOpen
  \bibfield  {author} {\bibinfo {author} {\bibfnamefont {D.~C.}\ \bibnamefont
  {Morse}},\ }\bibfield  {title} {\enquote {\bibinfo {title} {Theory of
  constrained brownian motion},}\ }in\ \href {\doibase 10.1002/0471484237.ch2}
  {\emph {\bibinfo {booktitle} {Advances in Chemical Physics}}}\ (\bibinfo
  {publisher} {John Wiley \& Sons, Ltd},\ \bibinfo {year} {2003})\
  Chap.~\bibinfo {chapter} {2}, pp.\ \bibinfo {pages} {65--189}\BibitemShut
  {NoStop}%
\bibitem [{\citenamefont {Ciccotti}, \citenamefont {Lelievre},\ and\
  \citenamefont {Vanden-Eijnden}(2008)}]{ciccotti2008projection}%
  \BibitemOpen
  \bibfield  {author} {\bibinfo {author} {\bibfnamefont {G.}~\bibnamefont
  {Ciccotti}}, \bibinfo {author} {\bibfnamefont {T.}~\bibnamefont {Lelievre}},
  \ and\ \bibinfo {author} {\bibfnamefont {E.}~\bibnamefont {Vanden-Eijnden}},\
  }\bibfield  {title} {\enquote {\bibinfo {title} {Projection of diffusions on
  submanifolds: Application to mean force computation},}\ }\href@noop {}
  {\bibfield  {journal} {\bibinfo  {journal} {Communications on Pure and
  Applied Mathematics: A Journal Issued by the Courant Institute of
  Mathematical Sciences}\ }\textbf {\bibinfo {volume} {61}},\ \bibinfo {pages}
  {371--408} (\bibinfo {year} {2008})}\BibitemShut {NoStop}%
\bibitem [{\citenamefont {Holmes-Cerfon}(2016)}]{holmes2016stochastic}%
  \BibitemOpen
  \bibfield  {author} {\bibinfo {author} {\bibfnamefont {M.}~\bibnamefont
  {Holmes-Cerfon}},\ }\bibfield  {title} {\enquote {\bibinfo {title}
  {Stochastic disks that roll},}\ }\href@noop {} {\bibfield  {journal}
  {\bibinfo  {journal} {Physical Review E}\ }\textbf {\bibinfo {volume} {94}},\
  \bibinfo {pages} {052112} (\bibinfo {year} {2016})}\BibitemShut {NoStop}%
\bibitem [{\citenamefont {Bell}(1978)}]{bell1978models}%
  \BibitemOpen
  \bibfield  {author} {\bibinfo {author} {\bibfnamefont {G.~I.}\ \bibnamefont
  {Bell}},\ }\bibfield  {title} {\enquote {\bibinfo {title} {Models for the
  specific adhesion of cells to cells: a theoretical framework for adhesion
  mediated by reversible bonds between cell surface molecules.}}\ }\href@noop
  {} {\bibfield  {journal} {\bibinfo  {journal} {Science}\ }\textbf {\bibinfo
  {volume} {200}},\ \bibinfo {pages} {618--627} (\bibinfo {year}
  {1978})}\BibitemShut {NoStop}%
\bibitem [{\citenamefont {Dembo}\ \emph {et~al.}(1988)\citenamefont {Dembo},
  \citenamefont {Torney}, \citenamefont {Saxman},\ and\ \citenamefont
  {Hammer}}]{dembo1988reaction}%
  \BibitemOpen
  \bibfield  {author} {\bibinfo {author} {\bibfnamefont {M.}~\bibnamefont
  {Dembo}}, \bibinfo {author} {\bibfnamefont {D.}~\bibnamefont {Torney}},
  \bibinfo {author} {\bibfnamefont {K.}~\bibnamefont {Saxman}}, \ and\ \bibinfo
  {author} {\bibfnamefont {D.}~\bibnamefont {Hammer}},\ }\bibfield  {title}
  {\enquote {\bibinfo {title} {The reaction-limited kinetics of
  membrane-to-surface adhesion and detachment},}\ }\href@noop {} {\bibfield
  {journal} {\bibinfo  {journal} {Proceedings of the Royal Society of London.
  Series B. Biological Sciences}\ }\textbf {\bibinfo {volume} {234}},\ \bibinfo
  {pages} {55--83} (\bibinfo {year} {1988})}\BibitemShut {NoStop}%
\bibitem [{\citenamefont {Bihr}, \citenamefont {Seifert},\ and\ \citenamefont
  {Smith}(2015)}]{bihr2015multiscale}%
  \BibitemOpen
  \bibfield  {author} {\bibinfo {author} {\bibfnamefont {T.}~\bibnamefont
  {Bihr}}, \bibinfo {author} {\bibfnamefont {U.}~\bibnamefont {Seifert}}, \
  and\ \bibinfo {author} {\bibfnamefont {A.-S.}\ \bibnamefont {Smith}},\
  }\bibfield  {title} {\enquote {\bibinfo {title} {Multiscale approaches to
  protein-mediated interactions between membranes—relating microscopic and
  macroscopic dynamics in radially growing adhesions},}\ }\href@noop {}
  {\bibfield  {journal} {\bibinfo  {journal} {New Journal of Physics}\ }\textbf
  {\bibinfo {volume} {17}},\ \bibinfo {pages} {083016} (\bibinfo {year}
  {2015})}\BibitemShut {NoStop}%
\bibitem [{\citenamefont {Berger}, \citenamefont {Klumpp},\ and\ \citenamefont
  {Lipowsky}(2019)}]{berger2019force}%
  \BibitemOpen
  \bibfield  {author} {\bibinfo {author} {\bibfnamefont {F.}~\bibnamefont
  {Berger}}, \bibinfo {author} {\bibfnamefont {S.}~\bibnamefont {Klumpp}}, \
  and\ \bibinfo {author} {\bibfnamefont {R.}~\bibnamefont {Lipowsky}},\
  }\bibfield  {title} {\enquote {\bibinfo {title} {Force-dependent unbinding
  rate of molecular motors from stationary optical trap data},}\ }\href@noop {}
  {\bibfield  {journal} {\bibinfo  {journal} {Nano letters}\ }\textbf {\bibinfo
  {volume} {19}},\ \bibinfo {pages} {2598--2602} (\bibinfo {year}
  {2019})}\BibitemShut {NoStop}%
\bibitem [{\citenamefont {Klobusicky}, \citenamefont {Fricks},\ and\
  \citenamefont {Kramer}(2020)}]{klobusicky2020effective}%
  \BibitemOpen
  \bibfield  {author} {\bibinfo {author} {\bibfnamefont {J.~J.}\ \bibnamefont
  {Klobusicky}}, \bibinfo {author} {\bibfnamefont {J.}~\bibnamefont {Fricks}},
  \ and\ \bibinfo {author} {\bibfnamefont {P.~R.}\ \bibnamefont {Kramer}},\
  }\bibfield  {title} {\enquote {\bibinfo {title} {Effective behavior of
  cooperative and nonidentical molecular motors},}\ }\href@noop {} {\bibfield
  {journal} {\bibinfo  {journal} {Research in the mathematical sciences}\
  }\textbf {\bibinfo {volume} {7}},\ \bibinfo {pages} {1--49} (\bibinfo {year}
  {2020})}\BibitemShut {NoStop}%
\bibitem [{\citenamefont {Bovyn}\ \emph {et~al.}(2021)\citenamefont {Bovyn},
  \citenamefont {Janakaloti~Narayanareddy}, \citenamefont {Gross},\ and\
  \citenamefont {Allard}}]{bovyn2021diffusion}%
  \BibitemOpen
  \bibfield  {author} {\bibinfo {author} {\bibfnamefont {M.}~\bibnamefont
  {Bovyn}}, \bibinfo {author} {\bibfnamefont {B.~R.}\ \bibnamefont
  {Janakaloti~Narayanareddy}}, \bibinfo {author} {\bibfnamefont
  {S.}~\bibnamefont {Gross}}, \ and\ \bibinfo {author} {\bibfnamefont
  {J.}~\bibnamefont {Allard}},\ }\bibfield  {title} {\enquote {\bibinfo {title}
  {Diffusion of kinesin motors on cargo can enhance binding and run lengths
  during intracellular transport},}\ }\href@noop {} {\bibfield  {journal}
  {\bibinfo  {journal} {Molecular Biology of the Cell}\ }\textbf {\bibinfo
  {volume} {32}},\ \bibinfo {pages} {984--994} (\bibinfo {year}
  {2021})}\BibitemShut {NoStop}%
\bibitem [{\citenamefont {Goodrich}, \citenamefont {Brenner},\ and\
  \citenamefont {Ribbeck}(2018)}]{goodrich2018enhanced}%
  \BibitemOpen
  \bibfield  {author} {\bibinfo {author} {\bibfnamefont {C.~P.}\ \bibnamefont
  {Goodrich}}, \bibinfo {author} {\bibfnamefont {M.~P.}\ \bibnamefont
  {Brenner}}, \ and\ \bibinfo {author} {\bibfnamefont {K.}~\bibnamefont
  {Ribbeck}},\ }\bibfield  {title} {\enquote {\bibinfo {title} {Enhanced
  diffusion by binding to the crosslinks of a polymer gel},}\ }\href@noop {}
  {\bibfield  {journal} {\bibinfo  {journal} {Nature Communications}\ }\textbf
  {\bibinfo {volume} {9}},\ \bibinfo {pages} {1--8} (\bibinfo {year}
  {2018})}\BibitemShut {NoStop}%
\bibitem [{\citenamefont {Doi}(1976)}]{doi1976stochastic}%
  \BibitemOpen
  \bibfield  {author} {\bibinfo {author} {\bibfnamefont {M.}~\bibnamefont
  {Doi}},\ }\bibfield  {title} {\enquote {\bibinfo {title} {Stochastic theory
  of diffusion-controlled reaction},}\ }\href@noop {} {\bibfield  {journal}
  {\bibinfo  {journal} {Journal of Physics A: Mathematical and General}\
  }\textbf {\bibinfo {volume} {9}},\ \bibinfo {pages} {1479} (\bibinfo {year}
  {1976})}\BibitemShut {NoStop}%
\bibitem [{\citenamefont {Agbanusi}\ and\ \citenamefont
  {Isaacson}(2014)}]{agbanusi2014comparison}%
  \BibitemOpen
  \bibfield  {author} {\bibinfo {author} {\bibfnamefont {I.~C.}\ \bibnamefont
  {Agbanusi}}\ and\ \bibinfo {author} {\bibfnamefont {S.~A.}\ \bibnamefont
  {Isaacson}},\ }\bibfield  {title} {\enquote {\bibinfo {title} {A comparison
  of bimolecular reaction models for stochastic reaction--diffusion systems},}\
  }\href@noop {} {\bibfield  {journal} {\bibinfo  {journal} {Bulletin of
  Mathematical Biology}\ }\textbf {\bibinfo {volume} {76}},\ \bibinfo {pages}
  {922--946} (\bibinfo {year} {2014})}\BibitemShut {NoStop}%
\bibitem [{\citenamefont {Zhang}\ \emph {et~al.}(2019)\citenamefont {Zhang},
  \citenamefont {Clemens}, \citenamefont {Goyette}, \citenamefont {Allard},
  \citenamefont {Dushek},\ and\ \citenamefont {Isaacson}}]{zhang2019influence}%
  \BibitemOpen
  \bibfield  {author} {\bibinfo {author} {\bibfnamefont {Y.}~\bibnamefont
  {Zhang}}, \bibinfo {author} {\bibfnamefont {L.}~\bibnamefont {Clemens}},
  \bibinfo {author} {\bibfnamefont {J.}~\bibnamefont {Goyette}}, \bibinfo
  {author} {\bibfnamefont {J.}~\bibnamefont {Allard}}, \bibinfo {author}
  {\bibfnamefont {O.}~\bibnamefont {Dushek}}, \ and\ \bibinfo {author}
  {\bibfnamefont {S.~A.}\ \bibnamefont {Isaacson}},\ }\bibfield  {title}
  {\enquote {\bibinfo {title} {The influence of molecular reach and diffusivity
  on the efficacy of membrane-confined reactions},}\ }\href@noop {} {\bibfield
  {journal} {\bibinfo  {journal} {Biophysical Journal}\ }\textbf {\bibinfo
  {volume} {117}},\ \bibinfo {pages} {1189--1201} (\bibinfo {year}
  {2019})}\BibitemShut {NoStop}%
\bibitem [{\citenamefont {Pavliotis}\ and\ \citenamefont
  {Stuart}(2008)}]{pavliotis2008multiscale}%
  \BibitemOpen
  \bibfield  {author} {\bibinfo {author} {\bibfnamefont {G.}~\bibnamefont
  {Pavliotis}}\ and\ \bibinfo {author} {\bibfnamefont {A.}~\bibnamefont
  {Stuart}},\ }\href@noop {} {\emph {\bibinfo {title} {Multiscale methods:
  averaging and homogenization}}}\ (\bibinfo  {publisher} {Springer Science \&
  Business Media},\ \bibinfo {year} {2008})\BibitemShut {NoStop}%
\bibitem [{\citenamefont {Gardiner}\ \emph {et~al.}(1985)\citenamefont
  {Gardiner} \emph {et~al.}}]{gardiner1985handbook}%
  \BibitemOpen
  \bibfield  {author} {\bibinfo {author} {\bibfnamefont {C.~W.}\ \bibnamefont
  {Gardiner}} \emph {et~al.},\ }\href@noop {} {\emph {\bibinfo {title}
  {Handbook of stochastic methods}}},\ Vol.~\bibinfo {volume} {3}\ (\bibinfo
  {publisher} {springer Berlin},\ \bibinfo {year} {1985})\BibitemShut {NoStop}%
\bibitem [{\citenamefont {Reiter-Scherer}\ \emph {et~al.}(2019)\citenamefont
  {Reiter-Scherer}, \citenamefont {Cuellar-Camacho}, \citenamefont {Bhatia},
  \citenamefont {Haag}, \citenamefont {Herrmann}, \citenamefont {Lauster},\
  and\ \citenamefont {Rabe}}]{reiter2019force}%
  \BibitemOpen
  \bibfield  {author} {\bibinfo {author} {\bibfnamefont {V.}~\bibnamefont
  {Reiter-Scherer}}, \bibinfo {author} {\bibfnamefont {J.~L.}\ \bibnamefont
  {Cuellar-Camacho}}, \bibinfo {author} {\bibfnamefont {S.}~\bibnamefont
  {Bhatia}}, \bibinfo {author} {\bibfnamefont {R.}~\bibnamefont {Haag}},
  \bibinfo {author} {\bibfnamefont {A.}~\bibnamefont {Herrmann}}, \bibinfo
  {author} {\bibfnamefont {D.}~\bibnamefont {Lauster}}, \ and\ \bibinfo
  {author} {\bibfnamefont {J.~P.}\ \bibnamefont {Rabe}},\ }\bibfield  {title}
  {\enquote {\bibinfo {title} {Force spectroscopy shows dynamic binding of
  influenza hemagglutinin and neuraminidase to sialic acid},}\ }\href@noop {}
  {\bibfield  {journal} {\bibinfo  {journal} {Biophysical journal}\ }\textbf
  {\bibinfo {volume} {116}},\ \bibinfo {pages} {1037--1048} (\bibinfo {year}
  {2019})}\BibitemShut {NoStop}%
\bibitem [{\citenamefont {M{\o}ller}\ and\ \citenamefont
  {Waagepetersen}(2002)}]{moller2002statistical}%
  \BibitemOpen
  \bibfield  {author} {\bibinfo {author} {\bibfnamefont {J.}~\bibnamefont
  {M{\o}ller}}\ and\ \bibinfo {author} {\bibfnamefont {R.~P.}\ \bibnamefont
  {Waagepetersen}},\ }\bibfield  {title} {\enquote {\bibinfo {title}
  {Statistical inference for cox processes},}\ }in\ \href@noop {} {\emph
  {\bibinfo {booktitle} {Spatial Cluster Modelling}}}\ (\bibinfo  {publisher}
  {Chapman \& Hall},\ \bibinfo {year} {2002})\BibitemShut {NoStop}%
\bibitem [{\citenamefont {Spinney}, \citenamefont {Lee},\ and\ \citenamefont
  {Morris}(2022)}]{spinney2022geometrical}%
  \BibitemOpen
  \bibfield  {author} {\bibinfo {author} {\bibfnamefont {R.~E.}\ \bibnamefont
  {Spinney}}, \bibinfo {author} {\bibfnamefont {L.}~\bibnamefont {Lee}}, \ and\
  \bibinfo {author} {\bibfnamefont {R.~G.}\ \bibnamefont {Morris}},\ }\bibfield
   {title} {\enquote {\bibinfo {title} {Geometrical patterning of receptor
  sites controls kinetics via many-body effects in bivalent systems},}\
  }\href@noop {} {\bibfield  {journal} {\bibinfo  {journal} {Physical Review
  Research}\ }\textbf {\bibinfo {volume} {4}},\ \bibinfo {pages} {L042028}
  (\bibinfo {year} {2022})}\BibitemShut {NoStop}%
\bibitem [{\citenamefont {Lowensohn}\ \emph {et~al.}(2022)\citenamefont
  {Lowensohn}, \citenamefont {Stevens}, \citenamefont {Goldstein},\ and\
  \citenamefont {Mognetti}}]{lowensohn2022sliding}%
  \BibitemOpen
  \bibfield  {author} {\bibinfo {author} {\bibfnamefont {J.}~\bibnamefont
  {Lowensohn}}, \bibinfo {author} {\bibfnamefont {L.}~\bibnamefont {Stevens}},
  \bibinfo {author} {\bibfnamefont {D.}~\bibnamefont {Goldstein}}, \ and\
  \bibinfo {author} {\bibfnamefont {B.~M.}\ \bibnamefont {Mognetti}},\
  }\bibfield  {title} {\enquote {\bibinfo {title} {Sliding across a surface:
  particles with fixed and mobile ligands},}\ }\href@noop {} {\bibfield
  {journal} {\bibinfo  {journal} {The Journal of Chemical Physics}\ }\textbf
  {\bibinfo {volume} {156}},\ \bibinfo {pages} {164902} (\bibinfo {year}
  {2022})}\BibitemShut {NoStop}%
\bibitem [{\citenamefont {Lee-Thorp}\ and\ \citenamefont
  {Holmes-Cerfon}(2018)}]{lee2018modeling}%
  \BibitemOpen
  \bibfield  {author} {\bibinfo {author} {\bibfnamefont {J.~P.}\ \bibnamefont
  {Lee-Thorp}}\ and\ \bibinfo {author} {\bibfnamefont {M.}~\bibnamefont
  {Holmes-Cerfon}},\ }\bibfield  {title} {\enquote {\bibinfo {title} {Modeling
  the relative dynamics of dna-coated colloids},}\ }\href@noop {} {\bibfield
  {journal} {\bibinfo  {journal} {Soft Matter}\ }\textbf {\bibinfo {volume}
  {14}},\ \bibinfo {pages} {8147--8159} (\bibinfo {year} {2018})}\BibitemShut
  {NoStop}%
\bibitem [{\citenamefont {Xu}\ \emph {et~al.}(2011)\citenamefont {Xu},
  \citenamefont {Feng}, \citenamefont {Sha}, \citenamefont {Seeman},\ and\
  \citenamefont {Chaikin}}]{xu2011subdiffusion}%
  \BibitemOpen
  \bibfield  {author} {\bibinfo {author} {\bibfnamefont {Q.}~\bibnamefont
  {Xu}}, \bibinfo {author} {\bibfnamefont {L.}~\bibnamefont {Feng}}, \bibinfo
  {author} {\bibfnamefont {R.}~\bibnamefont {Sha}}, \bibinfo {author}
  {\bibfnamefont {N.}~\bibnamefont {Seeman}}, \ and\ \bibinfo {author}
  {\bibfnamefont {P.}~\bibnamefont {Chaikin}},\ }\bibfield  {title} {\enquote
  {\bibinfo {title} {Subdiffusion of a sticky particle on a surface},}\
  }\href@noop {} {\bibfield  {journal} {\bibinfo  {journal} {Physical Review
  Letters}\ }\textbf {\bibinfo {volume} {106}},\ \bibinfo {pages} {228102}
  (\bibinfo {year} {2011})}\BibitemShut {NoStop}%
\bibitem [{\citenamefont {Lalitha~Sridhar}\ \emph {et~al.}(2021)\citenamefont
  {Lalitha~Sridhar}, \citenamefont {Dunagin}, \citenamefont {Koo},
  \citenamefont {Hough},\ and\ \citenamefont {Vernerey}}]{lalitha2021enhanced}%
  \BibitemOpen
  \bibfield  {author} {\bibinfo {author} {\bibfnamefont {S.}~\bibnamefont
  {Lalitha~Sridhar}}, \bibinfo {author} {\bibfnamefont {J.}~\bibnamefont
  {Dunagin}}, \bibinfo {author} {\bibfnamefont {K.}~\bibnamefont {Koo}},
  \bibinfo {author} {\bibfnamefont {L.}~\bibnamefont {Hough}}, \ and\ \bibinfo
  {author} {\bibfnamefont {F.}~\bibnamefont {Vernerey}},\ }\bibfield  {title}
  {\enquote {\bibinfo {title} {Enhanced diffusion by reversible binding to
  active polymers},}\ }\href@noop {} {\bibfield  {journal} {\bibinfo  {journal}
  {Macromolecules}\ }\textbf {\bibinfo {volume} {54}},\ \bibinfo {pages}
  {1850--1858} (\bibinfo {year} {2021})}\BibitemShut {NoStop}%
\bibitem [{\citenamefont {Walker}\ and\ \citenamefont
  {Newhall}(2022)}]{walker2022numerical}%
  \BibitemOpen
  \bibfield  {author} {\bibinfo {author} {\bibfnamefont {B.~L.}\ \bibnamefont
  {Walker}}\ and\ \bibinfo {author} {\bibfnamefont {K.~A.}\ \bibnamefont
  {Newhall}},\ }\bibfield  {title} {\enquote {\bibinfo {title} {Numerical
  computation of effective thermal equilibria in stochastically switching
  langevin systems},}\ }\href@noop {} {\bibfield  {journal} {\bibinfo
  {journal} {Physical Review E}\ }\textbf {\bibinfo {volume} {105}},\ \bibinfo
  {pages} {064113} (\bibinfo {year} {2022})}\BibitemShut {NoStop}%
\bibitem [{\citenamefont {M\"{u}ller}\ \emph {et~al.}(2019)\citenamefont
  {M\"{u}ller}, \citenamefont {Lauster}, \citenamefont {Wildenauer},
  \citenamefont {Herrmann},\ and\ \citenamefont {Block}}]{muller2019mobility}%
  \BibitemOpen
  \bibfield  {author} {\bibinfo {author} {\bibfnamefont {M.}~\bibnamefont
  {M\"{u}ller}}, \bibinfo {author} {\bibfnamefont {D.}~\bibnamefont {Lauster}},
  \bibinfo {author} {\bibfnamefont {H.~H.}\ \bibnamefont {Wildenauer}},
  \bibinfo {author} {\bibfnamefont {A.}~\bibnamefont {Herrmann}}, \ and\
  \bibinfo {author} {\bibfnamefont {S.}~\bibnamefont {Block}},\ }\bibfield
  {title} {\enquote {\bibinfo {title} {Mobility-based quantification of
  multivalent virus-receptor interactions: New insights into influenza a virus
  binding mode},}\ }\href@noop {} {\bibfield  {journal} {\bibinfo  {journal}
  {Nano Letters}\ }\textbf {\bibinfo {volume} {19}},\ \bibinfo {pages}
  {1875--1882} (\bibinfo {year} {2019})}\BibitemShut {NoStop}%
\bibitem [{\citenamefont {Ziebert}\ and\ \citenamefont
  {Kuli{\'c}}(2021)}]{ziebert2021influenza}%
  \BibitemOpen
  \bibfield  {author} {\bibinfo {author} {\bibfnamefont {F.}~\bibnamefont
  {Ziebert}}\ and\ \bibinfo {author} {\bibfnamefont {I.~M.}\ \bibnamefont
  {Kuli{\'c}}},\ }\bibfield  {title} {\enquote {\bibinfo {title} {How
  influenza’s spike motor works},}\ }\href@noop {} {\bibfield  {journal}
  {\bibinfo  {journal} {Physical Review Letters}\ }\textbf {\bibinfo {volume}
  {126}},\ \bibinfo {pages} {218101} (\bibinfo {year} {2021})}\BibitemShut
  {NoStop}%
\bibitem [{\citenamefont {Miles}, \citenamefont {Lawley},\ and\ \citenamefont
  {Keener}(2018)}]{miles2018analysis}%
  \BibitemOpen
  \bibfield  {author} {\bibinfo {author} {\bibfnamefont {C.~E.}\ \bibnamefont
  {Miles}}, \bibinfo {author} {\bibfnamefont {S.~D.}\ \bibnamefont {Lawley}}, \
  and\ \bibinfo {author} {\bibfnamefont {J.~P.}\ \bibnamefont {Keener}},\
  }\bibfield  {title} {\enquote {\bibinfo {title} {Analysis of nonprocessive
  molecular motor transport using renewal reward theory},}\ }\href@noop {}
  {\bibfield  {journal} {\bibinfo  {journal} {SIAM Journal on Applied
  Mathematics}\ }\textbf {\bibinfo {volume} {78}},\ \bibinfo {pages}
  {2511--2532} (\bibinfo {year} {2018})}\BibitemShut {NoStop}%
\bibitem [{\citenamefont {Park}, \citenamefont {Singh},\ and\ \citenamefont
  {Fai}(2022)}]{park2022coarse}%
  \BibitemOpen
  \bibfield  {author} {\bibinfo {author} {\bibfnamefont {Y.}~\bibnamefont
  {Park}}, \bibinfo {author} {\bibfnamefont {P.}~\bibnamefont {Singh}}, \ and\
  \bibinfo {author} {\bibfnamefont {T.~G.}\ \bibnamefont {Fai}},\ }\bibfield
  {title} {\enquote {\bibinfo {title} {Coarse-grained stochastic model of
  myosin-driven vesicles into dendritic spines},}\ }\href@noop {} {\bibfield
  {journal} {\bibinfo  {journal} {SIAM Journal on Applied Mathematics}\
  }\textbf {\bibinfo {volume} {82}},\ \bibinfo {pages} {793--820} (\bibinfo
  {year} {2022})}\BibitemShut {NoStop}%
\bibitem [{\citenamefont {Spakowitz}\ and\ \citenamefont
  {Wang}(2005)}]{spakowitz2005end}%
  \BibitemOpen
  \bibfield  {author} {\bibinfo {author} {\bibfnamefont {A.~J.}\ \bibnamefont
  {Spakowitz}}\ and\ \bibinfo {author} {\bibfnamefont {Z.-G.}\ \bibnamefont
  {Wang}},\ }\bibfield  {title} {\enquote {\bibinfo {title} {End-to-end
  distance vector distribution with fixed end orientations for the wormlike
  chain model},}\ }\href@noop {} {\bibfield  {journal} {\bibinfo  {journal}
  {Physical Review E}\ }\textbf {\bibinfo {volume} {72}},\ \bibinfo {pages}
  {041802} (\bibinfo {year} {2005})}\BibitemShut {NoStop}%
\bibitem [{\citenamefont {Mogre}\ \emph {et~al.}(2020)\citenamefont {Mogre},
  \citenamefont {Christensen}, \citenamefont {Niman}, \citenamefont
  {Reck-Peterson},\ and\ \citenamefont {Koslover}}]{mogre2020hitching}%
  \BibitemOpen
  \bibfield  {author} {\bibinfo {author} {\bibfnamefont {S.~S.}\ \bibnamefont
  {Mogre}}, \bibinfo {author} {\bibfnamefont {J.~R.}\ \bibnamefont
  {Christensen}}, \bibinfo {author} {\bibfnamefont {C.~S.}\ \bibnamefont
  {Niman}}, \bibinfo {author} {\bibfnamefont {S.~L.}\ \bibnamefont
  {Reck-Peterson}}, \ and\ \bibinfo {author} {\bibfnamefont {E.~F.}\
  \bibnamefont {Koslover}},\ }\bibfield  {title} {\enquote {\bibinfo {title}
  {Hitching a ride: mechanics of transport initiation through linker-mediated
  hitchhiking},}\ }\href@noop {} {\bibfield  {journal} {\bibinfo  {journal}
  {Biophysical Journal}\ }\textbf {\bibinfo {volume} {118}},\ \bibinfo {pages}
  {1357--1369} (\bibinfo {year} {2020})}\BibitemShut {NoStop}%
\bibitem [{\citenamefont {Nedelec}\ and\ \citenamefont
  {Foethke}(2007)}]{nedelec2007collective}%
  \BibitemOpen
  \bibfield  {author} {\bibinfo {author} {\bibfnamefont {F.}~\bibnamefont
  {Nedelec}}\ and\ \bibinfo {author} {\bibfnamefont {D.}~\bibnamefont
  {Foethke}},\ }\bibfield  {title} {\enquote {\bibinfo {title} {Collective
  langevin dynamics of flexible cytoskeletal fibers},}\ }\href@noop {}
  {\bibfield  {journal} {\bibinfo  {journal} {New Journal of Physics}\ }\textbf
  {\bibinfo {volume} {9}},\ \bibinfo {pages} {427} (\bibinfo {year}
  {2007})}\BibitemShut {NoStop}%
\bibitem [{\citenamefont {Lugo}, \citenamefont {Saikia},\ and\ \citenamefont
  {Nedelec}(2022)}]{lugo2022typical}%
  \BibitemOpen
  \bibfield  {author} {\bibinfo {author} {\bibfnamefont {C.~A.}\ \bibnamefont
  {Lugo}}, \bibinfo {author} {\bibfnamefont {E.}~\bibnamefont {Saikia}}, \ and\
  \bibinfo {author} {\bibfnamefont {F.}~\bibnamefont {Nedelec}},\ }\bibfield
  {title} {\enquote {\bibinfo {title} {A typical workflow to simulate
  cytoskeletal systems with cytosim},}\ }\href@noop {} {\bibfield  {journal}
  {\bibinfo  {journal} {arXiv preprint arXiv:2205.13852}\ } (\bibinfo {year}
  {2022})}\BibitemShut {NoStop}%
\bibitem [{\citenamefont {Cohen}\ and\ \citenamefont
  {Hendricks}(2022)}]{CohenHendricks2022}%
  \BibitemOpen
  \bibfield  {author} {\bibinfo {author} {\bibfnamefont {O.~T.}\ \bibnamefont
  {Cohen}}\ and\ \bibinfo {author} {\bibfnamefont {A.~G.}\ \bibnamefont
  {Hendricks}},\ }\bibfield  {title} {\enquote {\bibinfo {title} {Two dominant
  timescales of cytoskeletal crosslinking in the viscoelastic response of the
  cytoplasm},}\ }\href {\doibase 10.1103/PhysRevResearch.4.043167} {\bibfield
  {journal} {\bibinfo  {journal} {Physical Review Research}\ }\textbf {\bibinfo
  {volume} {4}},\ \bibinfo {pages} {043167} (\bibinfo {year}
  {2022})}\BibitemShut {NoStop}%
\bibitem [{\citenamefont {Peterman}\ and\ \citenamefont
  {Scholey}(2009)}]{peterman2009mitotic}%
  \BibitemOpen
  \bibfield  {author} {\bibinfo {author} {\bibfnamefont {E.~J.}\ \bibnamefont
  {Peterman}}\ and\ \bibinfo {author} {\bibfnamefont {J.~M.}\ \bibnamefont
  {Scholey}},\ }\bibfield  {title} {\enquote {\bibinfo {title} {Mitotic
  microtubule crosslinkers: insights from mechanistic studies},}\ }\href@noop
  {} {\bibfield  {journal} {\bibinfo  {journal} {Current Biology}\ }\textbf
  {\bibinfo {volume} {19}},\ \bibinfo {pages} {R1089--R1094} (\bibinfo {year}
  {2009})}\BibitemShut {NoStop}%
\bibitem [{\citenamefont {Lamson}\ \emph {et~al.}(2019)\citenamefont {Lamson},
  \citenamefont {Edelmaier}, \citenamefont {Glaser},\ and\ \citenamefont
  {Betterton}}]{lamson2019theory}%
  \BibitemOpen
  \bibfield  {author} {\bibinfo {author} {\bibfnamefont {A.~R.}\ \bibnamefont
  {Lamson}}, \bibinfo {author} {\bibfnamefont {C.~J.}\ \bibnamefont
  {Edelmaier}}, \bibinfo {author} {\bibfnamefont {M.~A.}\ \bibnamefont
  {Glaser}}, \ and\ \bibinfo {author} {\bibfnamefont {M.~D.}\ \bibnamefont
  {Betterton}},\ }\bibfield  {title} {\enquote {\bibinfo {title} {Theory of
  cytoskeletal reorganization during cross-linker-mediated mitotic spindle
  assembly},}\ }\href@noop {} {\bibfield  {journal} {\bibinfo  {journal}
  {Biophysical Journal}\ }\textbf {\bibinfo {volume} {116}},\ \bibinfo {pages}
  {1719--1731} (\bibinfo {year} {2019})}\BibitemShut {NoStop}%
\bibitem [{\citenamefont {Hannabuss}\ \emph {et~al.}(2019)\citenamefont
  {Hannabuss}, \citenamefont {Lera-Ramirez}, \citenamefont {Cade},
  \citenamefont {Fourniol}, \citenamefont {N{\'e}d{\'e}lec},\ and\
  \citenamefont {Surrey}}]{hannabuss2019self}%
  \BibitemOpen
  \bibfield  {author} {\bibinfo {author} {\bibfnamefont {J.}~\bibnamefont
  {Hannabuss}}, \bibinfo {author} {\bibfnamefont {M.}~\bibnamefont
  {Lera-Ramirez}}, \bibinfo {author} {\bibfnamefont {N.~I.}\ \bibnamefont
  {Cade}}, \bibinfo {author} {\bibfnamefont {F.~J.}\ \bibnamefont {Fourniol}},
  \bibinfo {author} {\bibfnamefont {F.}~\bibnamefont {N{\'e}d{\'e}lec}}, \ and\
  \bibinfo {author} {\bibfnamefont {T.}~\bibnamefont {Surrey}},\ }\bibfield
  {title} {\enquote {\bibinfo {title} {Self-organization of minimal anaphase
  spindle midzone bundles},}\ }\href@noop {} {\bibfield  {journal} {\bibinfo
  {journal} {Current Biology}\ }\textbf {\bibinfo {volume} {29}},\ \bibinfo
  {pages} {2120--2130} (\bibinfo {year} {2019})}\BibitemShut {NoStop}%
\bibitem [{\citenamefont {Gaska}\ \emph {et~al.}(2020)\citenamefont {Gaska},
  \citenamefont {Armstrong}, \citenamefont {Alfieri},\ and\ \citenamefont
  {Forth}}]{gaska2020mitotic}%
  \BibitemOpen
  \bibfield  {author} {\bibinfo {author} {\bibfnamefont {I.}~\bibnamefont
  {Gaska}}, \bibinfo {author} {\bibfnamefont {M.~E.}\ \bibnamefont
  {Armstrong}}, \bibinfo {author} {\bibfnamefont {A.}~\bibnamefont {Alfieri}},
  \ and\ \bibinfo {author} {\bibfnamefont {S.}~\bibnamefont {Forth}},\
  }\bibfield  {title} {\enquote {\bibinfo {title} {The mitotic crosslinking
  protein prc1 acts like a mechanical dashpot to resist microtubule sliding},}\
  }\href@noop {} {\bibfield  {journal} {\bibinfo  {journal} {Developmental
  Cell}\ }\textbf {\bibinfo {volume} {54}},\ \bibinfo {pages} {367--378}
  (\bibinfo {year} {2020})}\BibitemShut {NoStop}%
\bibitem [{\citenamefont {Popov}, \citenamefont {Komianos},\ and\ \citenamefont
  {Papoian}(2016)}]{popov2016medyan}%
  \BibitemOpen
  \bibfield  {author} {\bibinfo {author} {\bibfnamefont {K.}~\bibnamefont
  {Popov}}, \bibinfo {author} {\bibfnamefont {J.}~\bibnamefont {Komianos}}, \
  and\ \bibinfo {author} {\bibfnamefont {G.~A.}\ \bibnamefont {Papoian}},\
  }\bibfield  {title} {\enquote {\bibinfo {title} {Medyan: Mechanochemical
  simulations of contraction and polarity alignment in actomyosin networks},}\
  }\href@noop {} {\bibfield  {journal} {\bibinfo  {journal} {PLoS Computational
  Biology}\ }\textbf {\bibinfo {volume} {12}},\ \bibinfo {pages} {e1004877}
  (\bibinfo {year} {2016})}\BibitemShut {NoStop}%
\bibitem [{\citenamefont {Freedman}\ \emph {et~al.}(2018)\citenamefont
  {Freedman}, \citenamefont {Hocky}, \citenamefont {Banerjee},\ and\
  \citenamefont {Dinner}}]{freedman2018nonequilibrium}%
  \BibitemOpen
  \bibfield  {author} {\bibinfo {author} {\bibfnamefont {S.~L.}\ \bibnamefont
  {Freedman}}, \bibinfo {author} {\bibfnamefont {G.~M.}\ \bibnamefont {Hocky}},
  \bibinfo {author} {\bibfnamefont {S.}~\bibnamefont {Banerjee}}, \ and\
  \bibinfo {author} {\bibfnamefont {A.~R.}\ \bibnamefont {Dinner}},\ }\bibfield
   {title} {\enquote {\bibinfo {title} {Nonequilibrium phase diagrams for
  actomyosin networks},}\ }\href@noop {} {\bibfield  {journal} {\bibinfo
  {journal} {Soft Matter}\ }\textbf {\bibinfo {volume} {14}},\ \bibinfo {pages}
  {7740--7747} (\bibinfo {year} {2018})}\BibitemShut {NoStop}%
\bibitem [{\citenamefont {Chen}\ \emph {et~al.}(2014)\citenamefont {Chen},
  \citenamefont {McKinley}, \citenamefont {Wang}, \citenamefont {Shi},
  \citenamefont {Mucha}, \citenamefont {Forest},\ and\ \citenamefont
  {Lai}}]{chen2014transient}%
  \BibitemOpen
  \bibfield  {author} {\bibinfo {author} {\bibfnamefont {A.}~\bibnamefont
  {Chen}}, \bibinfo {author} {\bibfnamefont {S.~A.}\ \bibnamefont {McKinley}},
  \bibinfo {author} {\bibfnamefont {S.}~\bibnamefont {Wang}}, \bibinfo {author}
  {\bibfnamefont {F.}~\bibnamefont {Shi}}, \bibinfo {author} {\bibfnamefont
  {P.~J.}\ \bibnamefont {Mucha}}, \bibinfo {author} {\bibfnamefont {M.~G.}\
  \bibnamefont {Forest}}, \ and\ \bibinfo {author} {\bibfnamefont {S.~K.}\
  \bibnamefont {Lai}},\ }\bibfield  {title} {\enquote {\bibinfo {title}
  {Transient antibody-mucin interactions produce a dynamic molecular shield
  against viral invasion},}\ }\href@noop {} {\bibfield  {journal} {\bibinfo
  {journal} {Biophysical Journal}\ }\textbf {\bibinfo {volume} {106}},\
  \bibinfo {pages} {2028--2036} (\bibinfo {year} {2014})}\BibitemShut {NoStop}%
\bibitem [{\citenamefont {Newby}\ \emph {et~al.}(2017)\citenamefont {Newby},
  \citenamefont {Schiller}, \citenamefont {Wessler}, \citenamefont {Edelstein},
  \citenamefont {Forest},\ and\ \citenamefont {Lai}}]{newby2017blueprint}%
  \BibitemOpen
  \bibfield  {author} {\bibinfo {author} {\bibfnamefont {J.}~\bibnamefont
  {Newby}}, \bibinfo {author} {\bibfnamefont {J.~L.}\ \bibnamefont {Schiller}},
  \bibinfo {author} {\bibfnamefont {T.}~\bibnamefont {Wessler}}, \bibinfo
  {author} {\bibfnamefont {J.}~\bibnamefont {Edelstein}}, \bibinfo {author}
  {\bibfnamefont {M.~G.}\ \bibnamefont {Forest}}, \ and\ \bibinfo {author}
  {\bibfnamefont {S.~K.}\ \bibnamefont {Lai}},\ }\bibfield  {title} {\enquote
  {\bibinfo {title} {A blueprint for robust crosslinking of mobile species in
  biogels with weakly adhesive molecular anchors},}\ }\href@noop {} {\bibfield
  {journal} {\bibinfo  {journal} {Nature Communications}\ }\textbf {\bibinfo
  {volume} {8}},\ \bibinfo {pages} {1--10} (\bibinfo {year}
  {2017})}\BibitemShut {NoStop}%
\bibitem [{\citenamefont {Walker}\ \emph {et~al.}(2019)\citenamefont {Walker},
  \citenamefont {Taylor}, \citenamefont {Lawrimore}, \citenamefont {Hult},
  \citenamefont {Adalsteinsson}, \citenamefont {Bloom},\ and\ \citenamefont
  {Forest}}]{walker2019transient}%
  \BibitemOpen
  \bibfield  {author} {\bibinfo {author} {\bibfnamefont {B.}~\bibnamefont
  {Walker}}, \bibinfo {author} {\bibfnamefont {D.}~\bibnamefont {Taylor}},
  \bibinfo {author} {\bibfnamefont {J.}~\bibnamefont {Lawrimore}}, \bibinfo
  {author} {\bibfnamefont {C.}~\bibnamefont {Hult}}, \bibinfo {author}
  {\bibfnamefont {D.}~\bibnamefont {Adalsteinsson}}, \bibinfo {author}
  {\bibfnamefont {K.}~\bibnamefont {Bloom}}, \ and\ \bibinfo {author}
  {\bibfnamefont {M.~G.}\ \bibnamefont {Forest}},\ }\bibfield  {title}
  {\enquote {\bibinfo {title} {Transient crosslinking kinetics optimize gene
  cluster interactions},}\ }\href@noop {} {\bibfield  {journal} {\bibinfo
  {journal} {PLoS Computational Biology}\ }\textbf {\bibinfo {volume} {15}},\
  \bibinfo {pages} {e1007124} (\bibinfo {year} {2019})}\BibitemShut {NoStop}%
\bibitem [{\citenamefont {Mogilner}\ and\ \citenamefont
  {Oster}(2003)}]{mogilner2003force}%
  \BibitemOpen
  \bibfield  {author} {\bibinfo {author} {\bibfnamefont {A.}~\bibnamefont
  {Mogilner}}\ and\ \bibinfo {author} {\bibfnamefont {G.}~\bibnamefont
  {Oster}},\ }\bibfield  {title} {\enquote {\bibinfo {title} {Force generation
  by actin polymerization ii: the elastic ratchet and tethered filaments},}\
  }\href@noop {} {\bibfield  {journal} {\bibinfo  {journal} {Biophysical
  Journal}\ }\textbf {\bibinfo {volume} {84}},\ \bibinfo {pages} {1591--1605}
  (\bibinfo {year} {2003})}\BibitemShut {NoStop}%
\bibitem [{\citenamefont {Welf}\ \emph {et~al.}(2020)\citenamefont {Welf},
  \citenamefont {Miles}, \citenamefont {Huh}, \citenamefont {Sapoznik},
  \citenamefont {Chi}, \citenamefont {Driscoll}, \citenamefont {Isogai},
  \citenamefont {Noh}, \citenamefont {Weems}, \citenamefont {Pohlkamp} \emph
  {et~al.}}]{welf2020actin}%
  \BibitemOpen
  \bibfield  {author} {\bibinfo {author} {\bibfnamefont {E.~S.}\ \bibnamefont
  {Welf}}, \bibinfo {author} {\bibfnamefont {C.~E.}\ \bibnamefont {Miles}},
  \bibinfo {author} {\bibfnamefont {J.}~\bibnamefont {Huh}}, \bibinfo {author}
  {\bibfnamefont {E.}~\bibnamefont {Sapoznik}}, \bibinfo {author}
  {\bibfnamefont {J.}~\bibnamefont {Chi}}, \bibinfo {author} {\bibfnamefont
  {M.~K.}\ \bibnamefont {Driscoll}}, \bibinfo {author} {\bibfnamefont
  {T.}~\bibnamefont {Isogai}}, \bibinfo {author} {\bibfnamefont
  {J.}~\bibnamefont {Noh}}, \bibinfo {author} {\bibfnamefont {A.~D.}\
  \bibnamefont {Weems}}, \bibinfo {author} {\bibfnamefont {T.}~\bibnamefont
  {Pohlkamp}},  \emph {et~al.},\ }\bibfield  {title} {\enquote {\bibinfo
  {title} {Actin-membrane release initiates cell protrusions},}\ }\href@noop {}
  {\bibfield  {journal} {\bibinfo  {journal} {Developmental Cell}\ }\textbf
  {\bibinfo {volume} {55}},\ \bibinfo {pages} {723--736} (\bibinfo {year}
  {2020})}\BibitemShut {NoStop}%
\end{thebibliography}%


%merlin.mbs apsrev4-1.bst 2010-07-25 4.21a (PWD, AO, DPC) hacked
%Control: key (0)
%Control: author (8) initials jnrlst
%Control: editor formatted (1) identically to author
%Control: production of article title (-1) disabled
%Control: page (0) single
%Control: year (1) truncated
%Control: production of eprint (0) enabled
\begin{thebibliography}{5}%
\makeatletter
\providecommand \@ifxundefined [1]{%
 \@ifx{#1\undefined}
}%
\providecommand \@ifnum [1]{%
 \ifnum #1\expandafter \@firstoftwo
 \else \expandafter \@secondoftwo
 \fi
}%
\providecommand \@ifx [1]{%
 \ifx #1\expandafter \@firstoftwo
 \else \expandafter \@secondoftwo
 \fi
}%
\providecommand \natexlab [1]{#1}%
\providecommand \enquote  [1]{``#1''}%
\providecommand \bibnamefont  [1]{#1}%
\providecommand \bibfnamefont [1]{#1}%
\providecommand \citenamefont [1]{#1}%
\providecommand \href@noop [0]{\@secondoftwo}%
\providecommand \href [0]{\begingroup \@sanitize@url \@href}%
\providecommand \@href[1]{\@@startlink{#1}\@@href}%
\providecommand \@@href[1]{\endgroup#1\@@endlink}%
\providecommand \@sanitize@url [0]{\catcode `\\12\catcode `\$12\catcode
  `\&12\catcode `\#12\catcode `\^12\catcode `\_12\catcode `\%12\relax}%
\providecommand \@@startlink[1]{}%
\providecommand \@@endlink[0]{}%
\providecommand \url  [0]{\begingroup\@sanitize@url \@url }%
\providecommand \@url [1]{\endgroup\@href {#1}{\urlprefix }}%
\providecommand \urlprefix  [0]{URL }%
\providecommand \Eprint [0]{\href }%
\providecommand \doibase [0]{http://dx.doi.org/}%
\providecommand \selectlanguage [0]{\@gobble}%
\providecommand \bibinfo  [0]{\@secondoftwo}%
\providecommand \bibfield  [0]{\@secondoftwo}%
\providecommand \translation [1]{[#1]}%
\providecommand \BibitemOpen [0]{}%
\providecommand \bibitemStop [0]{}%
\providecommand \bibitemNoStop [0]{.\EOS\space}%
\providecommand \EOS [0]{\spacefactor3000\relax}%
\providecommand \BibitemShut  [1]{\csname bibitem#1\endcsname}%
\let\auto@bib@innerbib\@empty
%</preamble>
\bibitem [{\citenamefont {Gardiner}\ \emph {et~al.}(1985)\citenamefont
  {Gardiner} \emph {et~al.}}]{gardiner1985handbook}%
  \BibitemOpen
  \bibfield  {author} {\bibinfo {author} {\bibfnamefont {C.~W.}\ \bibnamefont
  {Gardiner}} \emph {et~al.},\ }\href@noop {} {\emph {\bibinfo {title}
  {Handbook of stochastic methods}}},\ Vol.~\bibinfo {volume} {3}\ (\bibinfo
  {publisher} {springer Berlin},\ \bibinfo {year} {1985})\BibitemShut {NoStop}%
\bibitem [{\citenamefont {Pavliotis}\ and\ \citenamefont
  {Stuart}(2008)}]{pavliotis2008multiscale}%
  \BibitemOpen
  \bibfield  {author} {\bibinfo {author} {\bibfnamefont {G.}~\bibnamefont
  {Pavliotis}}\ and\ \bibinfo {author} {\bibfnamefont {A.}~\bibnamefont
  {Stuart}},\ }\href@noop {} {\emph {\bibinfo {title} {Multiscale methods:
  averaging and homogenization}}}\ (\bibinfo  {publisher} {Springer Science \&
  Business Media},\ \bibinfo {year} {2008})\BibitemShut {NoStop}%
\bibitem [{\citenamefont {Morse}(2003)}]{morse2003Theory}%
  \BibitemOpen
  \bibfield  {author} {\bibinfo {author} {\bibfnamefont {D.~C.}\ \bibnamefont
  {Morse}},\ }in\ \href {\doibase 10.1002/0471484237.ch2} {\emph {\bibinfo
  {booktitle} {Advances in Chemical Physics}}}\ (\bibinfo  {publisher} {John
  Wiley \& Sons, Ltd},\ \bibinfo {year} {2003})\ Chap.~\bibinfo {chapter} {2},
  pp.\ \bibinfo {pages} {65--189}\BibitemShut {NoStop}%
\bibitem [{\citenamefont {Holmes-Cerfon}(2016)}]{holmes2016stochastic}%
  \BibitemOpen
  \bibfield  {author} {\bibinfo {author} {\bibfnamefont {M.}~\bibnamefont
  {Holmes-Cerfon}},\ }\href@noop {} {\bibfield  {journal} {\bibinfo  {journal}
  {Physical Review E}\ }\textbf {\bibinfo {volume} {94}},\ \bibinfo {pages}
  {052112} (\bibinfo {year} {2016})}\BibitemShut {NoStop}%
\bibitem [{\citenamefont {Ciccotti}\ \emph {et~al.}(2008)\citenamefont
  {Ciccotti}, \citenamefont {Lelievre},\ and\ \citenamefont
  {Vanden-Eijnden}}]{ciccotti2008projection}%
  \BibitemOpen
  \bibfield  {author} {\bibinfo {author} {\bibfnamefont {G.}~\bibnamefont
  {Ciccotti}}, \bibinfo {author} {\bibfnamefont {T.}~\bibnamefont {Lelievre}},
  \ and\ \bibinfo {author} {\bibfnamefont {E.}~\bibnamefont {Vanden-Eijnden}},\
  }\href@noop {} {\bibfield  {journal} {\bibinfo  {journal} {Communications on
  Pure and Applied Mathematics: A Journal Issued by the Courant Institute of
  Mathematical Sciences}\ }\textbf {\bibinfo {volume} {61}},\ \bibinfo {pages}
  {371} (\bibinfo {year} {2008})}\BibitemShut {NoStop}%
\end{thebibliography}%

\end{document}

\title[Supplementary Information]{\normalsize{\textit{Supplementary Information}} \\ Coarse-grained dynamics of transiently-bound fast linkers}

\author{S. Marbach and C. E. Miles}

\maketitle

%\eqref{eq:unbounddynamics}
%\eqref{eq:bounddynamics}
%\eqref{eq:backward}
%\ref{sec:stiffbond}
%\eqref{eq:coarseGrainedBackward}
%(1) (2) (14) II F 1 (19)

\tableofcontents

\section{Extended coarse-graining calculations}

\subsection{Obtaining the generator from the dynamics}
\label{app:generator}

The set of stochastic Eqns.~(1)-(2) in the main text 
defines a Markov process that
is conveniently studied via the Fokker-Planck equation and its adjoint, the Kolmogorov backward  equation~\cite{gardiner1985handbook,pavliotis2008multiscale}. 
Let $p(x_s,\xl,t) = \left( p_u(x_s,\xl,t), p_b(x_s,\xl,t) \right)^T$ be the probability density function of finding the system at time $t$ and positions $x_s,\xl$ in the unbound or bound states. We obtain from Eqns.(1)-(2) the Fokker-Plank equation 
\begin{equation}\label{eq:FPE}
    \partial_t p = \mathcal{L}^{\star} p\,,
\end{equation}
with $\mathcal{L}^{\star} = \mathcal{Q}^{\star} + \mathcal{V}^{\star}$ where
\begin{align*}  
    \mathcal{Q}^{\star} &= \begin{pmatrix}  - q_{\rm on} & q_{\rm off} \\  q_{\rm on} & - q_{\rm off} \end{pmatrix}, 
    \mathcal{V}^{\star} = \begin{pmatrix}   \mathcal{V}_u^{\star} & 0 \\  0 &  \mathcal{V}_b^{\star} \end{pmatrix}, \\
    \mathcal{V}_u^{\star} &= \dl \left( \frac{\kl}{\gammal} \xl  + \frac{\kT}{\gammal} \dl \right) + \partial_s (\partial_s \frac{\mathcal{U}_s (x_s)}{\gamma_s})  + \frac{\kT}{\gamma_s} \partial_{ss} \\
    \mathcal{V}_b^{\star} &=
    \left( \frac{\kl}{\gammal} \xl - \frac{\kb}{\gammal} (\xl - x_s)  + \frac{\kT}{\gammal} \dl \right) + \\
    & \,\,\, \partial_s (\partial_s \frac{\mathcal{U}_s (x_s)}{\gamma_s} + \frac{\kb}{\gamma_s} (\xl - x_s))  + \frac{\kT}{\gamma_s} \partial_{ss}
\end{align*}
with an appropriate initial condition. 
% \begin{eqnarray}  
%     & \partial_t p = \mathcal{L}^{\star} p \,\, \mathrm{with} \,\,\,  \mathcal{L}^{\star} = \mathcal{U}^{\star} + \mathcal{Q}^{\star},   \nonumber \\
%     \mathcal{U}^{\star} &= \mathrm{diag} \begin{pmatrix} \partial_l \left( \frac{k}{\gamma} (l-l_0) + \frac{k_B T}{\gamma} \partial_l\right) + \frac{k_B T}{\Gamma} \partial_{xx} \\
%     (\partial_l -  \partial_x) \left( \frac{k}{\Gamma + \gamma} (l-l_0) + \frac{k_B T}{\Gamma + \gamma} (\partial_l -  \partial_x)  \right)
%     \end{pmatrix}, \nonumber \\
%     & \mathrm{and} \,\, \mathcal{Q}^{\star} = \begin{pmatrix}  - q_{\rm on} & q_{\rm off} \\  q_{\rm on} & - q_{\rm off} \end{pmatrix}.
%   \label{eq:FPE}
% \end{eqnarray}
Additionally, we require the flux in either state to vanish at infinity, to conserve total probability. The stationary solution of Eq.~\eqref{eq:FPE} is $\pi = \left( \piu , \pib \right)^T $, the equilibrium probability density of the system. As we have seen, it satisfies detailed balance. 
 
While probability densities have an intuitive physical meaning, when \homogenization it is often easier -- and mathematically better posed -- to consider the adjoint of the Fokker-Planck equation and the corresponding dual functions. These are functions $f(x_s,\xl,t) =$$\int p(x_s',\xl',t|x_s,\xl) g(x_s',\xl') \dd \xl' \dd x_s'$ that give the expectation of any scalar function $g(x_s(t),\xl(t))$, given an initial condition $x_s(0)=x_s,\xl(0)=\xl$. 
Writing $f(x_s,\xl,t) = \left( f_u(x_s,\xl,t) , f_b(x_s,\xl,t) \right)^T$, we have that $f$ satisfies the Kolmogorov backward equation~\cite{gardiner1985handbook} Eq.~(14) in the main text. 
Here, $\mathcal{L}$ is the adjoint operator of $\mathcal{L}^{\star}$, defined by the operator that satisfies $\langle f , \mathcal{L}^{\star} p \rangle = \langle \mathcal{L} f , p \rangle$ for any probability density $p$ and statistic $f$, where $\langle f , p \rangle = \iint (f_u p_u + f_b p_b) \dd \xl \dd x_s$ is the inner product. 

\subsection{Commuting the limits of effective dynamics and stiff bond.}
\label{app:limits}

We now show the limits of a stiff spring constant for the bond and of coarse-graining fast linker dynamics commute. Since in the text we started with coarse-graining, \textit{i.e.} the limit of a fast linker, we here first take the limit of stiff spring constant for the bond $\kb \gg \kl$. This limit corresponds to a so-called soft constraint, where the linker and the slow particle move synchronously when they are bound. The dynamics in the bound state with this case can be readily obtained, by projecting the unbound dynamics~\cite{morse2003Theory,holmes2016stochastic,ciccotti2008projection}
\begin{equation}
    \frac{dx_s}{dt} = \frac{d\xl}{dt} = - \frac{\kl\xl + \partial_x\mathcal{U}(x_s)}{\gamma_s + \gammal} + \sqrt{\frac{2\kT}{\gamma_s + \gammal}} \eta_s(t).
\end{equation}
We obtain the generator
\begin{equation}
    \mathcal{L} = \mathcal{Q} + \mathcal{V}
\end{equation}
where
\begin{align}
\mathcal{V}_u &= - \frac{\kl}{\gammal}\xl \dl+ \frac{\kT}{\gammal} \dll - \frac{\partial_x \mathcal{U}(x_s)}{\gamma_s} x_s \partial_s + \frac{\kT}{\gamma_s} \partial_{ss} \\
\mathcal{V}_b &=  - \frac{\kl\xl + \partial_x \mathcal{U}(x_s)}{\gammal + \gamma_s} (\dl + \partial_s) + \frac{\kT}{\gammal+ \gamma_s} (\dl + \partial_s)^2
\end{align}
and we have to be careful that the bound state is only defined on the line $\xl = x_s$. As a result, the equilibrium probability distribution in the unbound state contains a contribution $\delta(\xl - x_s)$ and hence, the force in the bound state is simply 
\begin{equation}
        \Fb^{\rm eff}(x_s) =   - \kl x_s.
\end{equation}
This corresponds exactly with the effective force in the bound state when the limits are performed the other way around, see Sec.~II.F.1 in the main text on the model with a stiff bond.

\edit{\section{Coarse-graining at the subsequent order}
\label{app:nextorder}
In this section, we extend the averaging procedure to determine the dynamics in the next order of $\varepsilon$. }

\edit{\paragraph{Framework to obtain $O(\varepsilon)$ coarse-grained dynamics}
Recall from Eq.~(19) in the main text, that the dynamics at $O(1)$ in $\varepsilon$ satisfy, for  $f_0 = \begin{pmatrix}
g_1(x_s,t) & g_2(x_s,t)    
\end{pmatrix}^T$ any test function, 
\begin{equation}
   \begin{split}
        \partial_t g_1 &= - \oneff (g_1- g_2) - \frac{ \partial_s \mathcal{U}(x_s)}{k_BT} \partial_s g_1 +  \partial_{ss} g_1   \\
        \partial_t g_2 & = \offeff  (g_1  -   g_2) - \frac{ \partial_s \mathcal{U}(x_s)}{k_BT} \partial_s g_2 +  \partial_{ss} g_2 - \lambda \frac{\kappa}{\kappa + \lambda} x_s \partial_s g_2.
    \end{split}
    \label{eq:coarseGrainedBackward2}
\end{equation}
 To obtain dynamics at the following order, namely $O(\varepsilon)$, we need to find $f_1$, such that $\mathcal{L}_0 f_1 = - \mathcal{L}_1 f_0 + \partial_t f_0$. We know that such a solution $f_1$ exists since $f_0$ satisfies the closure equation, or Fredholm alternative, $\langle \left( \mathcal{L}_1 f_0- \partial_t f_0 \right)\cdot \pi_0 \rangle_{\ell} = 0$, which is equivalent to Eq.~\eqref{eq:coarseGrainedBackward2} above. Then, we will seek a solution for $f_2$ as $\mathcal{L}_0 f_2 =  \partial_t f_1 -  \mathcal{L}_1 f_1$ which exists only if $f_1$ satisfies Fredholm's alternative $\langle ( \partial_t f_1 -\mathcal{L}_1 f_1).\pi_0 \rangle_\ell = 0$. This final closure equation together with the previous order will yield the dynamics at $O(\varepsilon)$ as  $\langle (\partial_t f_0 + \varepsilon \partial_t f_1 -  \mathcal{L}_1 f_0 -  \varepsilon\mathcal{L}_1 f_1).\pi_0 \rangle_\ell = 0$}

\edit{\paragraph{Derivation of $f_1$.} We look for $f_1$ as $f_1 = \begin{pmatrix}
m_1(x_s,\xl,t) & m_2(x_s,\xl,t)    
\end{pmatrix}^T$ where $m_1$ and $m_2$ are $2$ functions to be determined. We have $\mathcal{L}_0 f_1 = \partial_t f_0 - \mathcal{L}_1 f_0$. Using Eq.~\eqref{eq:coarseGrainedBackward2} to expand $\partial_t f_0$, we obtain $2$ separate equations for $m_1$ and $m_2$ as
\begin{align*}
  \begin{cases} 
  - \kappa \xl \dl m_1 + \dll m_1 &=
        - (\on(x_s,\xl) - \oneff(x_s)) (g_1 - g_2), \\
       - \kappa \xl \dl m_2 - \lambda (\xl - x_s) \dl m_2 + \dll m_2 &=  (\off(x_s,\xl) - \offeff(x_s)) (g_1 - g_2) - \lambda\left( \xl - \frac{\lambda}{\kappa + \lambda}x_s  \right) \partial_s g_2 .
    \end{cases}
\end{align*}
To find $m_1$ and $m_2$ we essentially need to solve second order equations of the form
\begin{equation}
    - k (\xl - \xl^0) \dl m + \dll m = f(\xl)
    \label{eq:generic}
\end{equation}
where $f(\xl)$ is a smooth function of $\xl$ and $m(x_s,\xl,t)$. For $m_1$, notice that $k = \kappa$ and $\xl^0 = 0$ while for $m_2$ we have $k = \kappa + \lambda$ and $\xl^0 = x_s \frac{\lambda}{\kappa + \lambda}$.
A special solution of Eq.~\eqref{eq:generic} is 
\begin{align*}
    m(\xl) &=  \int_0^{\xl} \left(  \int_0^{x} f(x') e^{-k(x' - \xl^0)^2/2} \dd x' \right) e^{k(x - \xl^0)^2/2}  \dd x + A
\end{align*}
where $A$ is a constant. To avoid terms in $f_1$, that are already contained at $O(1)$ in $f_0$, we should verify that $\langle f_1 \cdot \pi_0 \rangle_{\ell} = 0$. Continuing on with the generic Eq.~\eqref{eq:generic}, we can check the condition for $m$ as $\langle m \cdot \pi_0 \rangle_{\ell} = 0$ and we have $\pi_0 \sim  e^{-k(x - \xl^0)^2/2}$. We obtain
\begin{align*}
    m(\xl) =  &\int_0^{\xl} \left(  \int_{-\infty}^{x} f(x') e^{-k(x' - \xl^0)^2/2} \dd x' \right) e^{k(x - \xl^0)^2/2}  \dd x  \bigg]  \\
    &- \sqrt{\frac{k}{2\pi}} \int_{-\infty}^{\infty} \bigg[ \int_0^{x''} \left(  \int_{-\infty}^{x} f(x') e^{-k(x' - \xl^0)^2/2} \dd x' \right) e^{k(x - \xl^0)^2/2}  \dd x  \bigg]  e^{-k(x'' - \xl^0)^2/2}  \dd x'' ,
\end{align*}
which can alternatively be written in a more compact form as
\begin{align*}
    m(\xl) =  \int_{-\infty}^{\infty} \bigg[ \int_0^{x''} \left(  \int_{-\infty}^{x} f(x') e^{-k(x' - \xl^0)^2/2} \dd x' \right) e^{k(x - \xl^0)^2/2}  \dd x  \bigg] \times  \\
    \left( \delta(x'' - \xl)  -  \sqrt{\frac{k}{2\pi}} e^{-k(x'' - \xl^0)^2/2} \right) \dd x'' .
    \end{align*}}
%\edit{which can alternatively be written

\edit{We obtain a full expression for $f_1$ by inserting the right variables for $k$ and $\xl$ in the above expression. We find
\begin{equation}
\begin{split}
    m_1(x_s,\xl,t) =& 
         -\delta Q_{\rm on}(x_s,\xl) (g_1(x_s,t) - g_2(x_s,t)) 
         \end{split}
\end{equation}
where 
\begin{equation}
\begin{split}
    \delta Q_{\rm on}(x_s,\xl) &= \bigg(\int_{-\infty}^{\infty} \bigg[ \int_0^{x''} \left(  \int_{-\infty}^{x} \delta \on(x',x_s) e^{-\kappa (x')^2/2} \dd x' \right) e^{\kappa(x)^2/2}  \dd x  \bigg] \times \\
         & \left( \delta(x'' - \xl)  - \sqrt{\frac{\kappa}{2\pi}} e^{-\kappa(x'')^2/2} \right) \dd x''  \bigg)    
\end{split}
\end{equation}
using $\delta \on = \on(x_s,\xl) - \oneff(x_s)$ and similarly for $\delta \off$. For $m_2$ we find
\begin{equation}
\begin{split}
    m_2 = \delta Q_{\rm off}(x_s,\xl)  (g_1(x_s,t) - g_2(x_s,t)) + \frac{\lambda}{\lambda+ \kappa} \left( \xl -  x_s \frac{\lambda}{\lambda + \kappa} \right) \partial_s g_2 (x_s,t).
         \end{split}
\end{equation}
where 
\begin{equation}
    \begin{split}
        \delta Q_{\rm off}(x_s,\xl) &= \bigg(\int_{-\infty}^{\infty} \bigg[ \int_0^{x''} \left(  \int_{-\infty}^{x} \delta \off(x',x_s) e^{-(\kappa + \lambda)\left(x' - x_s \frac{\lambda}{\lambda + \kappa}\right)^2/2} \dd x' \right) e^{(\kappa + \lambda)\left(x - x_s \frac{\lambda}{\lambda + \kappa}\right)^2/2}  \dd x  \bigg] \times \\
         & \left( \delta(x'' - \xl)  - \sqrt{\frac{\kappa+\lambda}{2\pi}} e^{-(\kappa + \lambda)\left(x'' - x_s \frac{\lambda}{\lambda + \kappa}\right)^2/2} \right) \dd x''  \bigg).
    \end{split}
\end{equation}
}

\edit{\paragraph{Fredholm alternative to obtain the $O(\varepsilon)$ dynamics.} We now calculate $\langle \left(\partial_t f_1 - \mathcal{L}_1 f_1 \right) \cdot \pi_0 \rangle_{\ell}$. It is rather straightforward to show
\begin{equation}
    \langle \partial_t f_1 \cdot \pi_0 \rangle_{\ell} = 0
\end{equation}
since we imposed $\langle f_1 \cdot \pi_0 \rangle_{\ell} = 0$. Similarly any term as $\langle \partial_s m_1 \cdot \pi_0 \rangle = 0$ or $\langle \partial_{ss} m_1 \cdot \pi_0 \rangle = 0$, and $\langle \partial_s m_2 \cdot \pi_0 \rangle = 0$ or $\langle \partial_{ss} m_2 \cdot \pi_0 \rangle = 0$. 
The remaining terms yield the $O(\varepsilon)$ dynamics as $\langle (\partial_t f_0 + \varepsilon\partial_t f_1 - \mathcal{L}_1 f_0 - \varepsilon \mathcal{L}_1 f_1) \cdot \pi_0 \rangle_{\ell} = 0$.

The modified dynamics for $g_1$, \textit{i.e.} the unbound dynamics, are
\begin{equation}
    \begin{split}
        \partial_t g_1& = O(1)  \\
        &+ \frac{\varepsilon}{\int \piu(x_s,\xl)  \dd \xl} \int \piu(x_s,\xl) \left( - \on(x_s,\xl) \left[\delta Q_{\rm on} (x_s,\xl) - \delta Q_{\rm off} (x_s,\xl) \right] \right) \dd \xl \,\, \times (g_1 - g_2) \\
        & + \frac{\varepsilon}{\int \piu(x_s,\xl)  \dd \xl} \int \piu(x_s,\xl)  \on(x_s,\xl) \left[\frac{\lambda}{\lambda + \kappa} \left(\xl - x_s \frac{\lambda}{\lambda+\kappa} \right)\right] \dd \xl \,\, \times \partial_s g_2 
    \end{split}
\end{equation}
where we abbreviated by $O(1)$ the $O(1)$ dynamics for $g_1$ reported on the right hand side of Eq.~\eqref{eq:coarseGrainedBackward2}. We analyze the additional terms in $O(\varepsilon)$ briefly. }

The first term at $O(\varepsilon)$ is a contribution to the binding rate that is modified compared to a simple integral of $\on(x_s,\xl)$ by $\delta Q_{\rm on} (x_s,\xl) - \delta Q_{\rm off} (x_s,\xl)$. This latter factor accounts roughly for states that are close to the average state and that are not bound yet, and therefore quantifies how memory affects binding. 

Interestingly, in the \textit{unbound} state, there is now the appearance of a source term, akin to a force (notice that it goes as $\partial_s g_2$ and not $\partial_s g_1$), corresponding to the second term in $O(\varepsilon)$. This ``force'' is is a remnant of the recoil force on the bound fast linker because again there is a slight memory in the system. This additional force in the bound state
\begin{equation}
    \mathcal{F}^{\rm eff, 1}_u = \frac{\varepsilon}{\int \pi_u(x_s,\xl) \dd \xl}  \int \dd \xl \, \pi_u(x_s,\xl) \on (x_s,\xl) \bigg( \frac{\lambda}{\lambda+ \kappa} \left( \xl -  x_s \frac{\lambda}{\lambda + \kappa} \right)  \bigg)
\end{equation}
averages, when $\on$ is uniform, simply to $\sim - k x_s$ where $k$ is some effective spring constant whose lengthy expression we do not report here.

The modified dynamics for $g_2$, \textit{i.e.} the bound dynamics, are
\begin{equation}
    \begin{split}
        \partial_t g_2& = O(1)  \\
        &+ \frac{\varepsilon}{\int \pib(x_s,\xl)  \dd \xl} \int \pib(x_s,\xl) \left( - \off(x_s,\xl) \left[\delta Q_{\rm off} (x_s,\xl) - \delta Q_{\rm on} (x_s,\xl) \right] \right) \dd \xl \,\, \times (g_1 - g_2) \\
        & - \frac{\varepsilon}{\int \pib(x_s,\xl)  \dd \xl} \int \pib(x_s,\xl)  \off(x_s,\xl) \left[\frac{\lambda}{\lambda + \kappa} \left(\xl - x_s \frac{\lambda}{\lambda+\kappa} \right)\right] \dd \xl \,\, \times \partial_s g_2 \\
        & - \frac{\varepsilon}{\int \pib(x_s,\xl)  \dd \xl} \int \pib(x_s,\xl) \delta Q_{\rm off} (x_s,\xl) \lambda  \xl  \dd \xl \,\, \times (\partial_s g_2  - \partial_s g_1) \\
        & - \frac{\varepsilon}{\int \pib(x_s,\xl)  \dd \xl} \int \pib(x_s,\xl) \partial_s [\delta Q_{\rm off} (x_s,\xl) ]\lambda  \xl  \dd \xl \,\, \times (g_2  -  g_1) \\
        & + \frac{\varepsilon}{\int \pib(x_s,\xl)  \dd \xl} \int \pib(x_s,\xl)  \lambda \xl  \left( \frac{\lambda}{\lambda+\kappa} \right)^2   \dd \xl \,\,  \times\partial_{s} g_2 \\
         & - \frac{\varepsilon}{\int \pib(x_s,\xl)  \dd \xl} \int \pib(x_s,\xl)  \lambda \xl \left[\frac{\lambda}{\lambda + \kappa} \left(\xl - x_s \frac{\lambda}{\lambda+\kappa} \right)\right]    \dd \xl \,\, \times\partial_{ss} g_2 
    \end{split}
    \label{eq:dtg2}
\end{equation}
where we abbreviated by $O(1)$ the $O(1)$ dynamics for $g_2$ reported on the right hand side of Eq.~\eqref{eq:coarseGrainedBackward2}. 

We analyze these 6 additional terms briefly. The 1st term and the 4th correspond to modified unbinding rates due to memory effects. The 2nd, 3rd and 5th terms (the latter of which simplifies to $+ \varepsilon \lambda (\lambda/(\lambda+\kappa))^3 x_s \partial_s g_2$) correspond to forces on the bound particle, which modify the $O(1)$ force term by accounting for some degree of memory of unbound relaxation of the spring. Notice again that one of these force terms contains a $\partial_s g_1$ coupling which acts as a source for $g_2$ and translates memory effect of the force on the unbound linker feeding into the bound dynamics.

The last term can be integrated to 
\begin{equation}
- \varepsilon \left(\frac{\lambda}{\lambda+ \kappa}\right)^2 \partial_{ss} g_2
\end{equation}
which means the diffusion in the bound state is slowed down. Taking the usual limit of a very stiff bond $\lambda \gg \kappa$ we obtain the effective diffusion in the bound state, back in dimensional units, $(k_BT/\gamma_s) \times (1 - \gammal/\gamma_s)$. Notice that this updated diffusion coefficient corresponds to the order 1 Taylor series of $k_BT/(\gamma_s + \gammal)$, which, through the projection method detailed in Sec.~\ref{app:limits}, is indeed the diffusion coefficient in the bound state, before the assumption $\gammal \ll \gamma_s$ is made. This additional friction on the particle is also visible in the forces, namely through the 5th term of Eq.~\eqref{eq:dtg2}. In contrast, when the bond is weak $\kappa \gg \lambda$, there is no additional friction on the bound particle and the particle diffuses as fast in the bound and unbound states. In summary, due to the springs-in-series structure, $\lambda$ can be seen as a filter for friction on the particle: if stiff ($\lambda \gg \kappa$) increased friction is felt on the particle, but otherwise ($\kappa \gg \lambda$) the additional friction is absorbed into the bond and not felt.

Overall, our coarse-graining method can be used systematically to derive dynamics at any order in $\varepsilon$. To investigate the relevance of the terms to capture the dynamics, one could conduct simulations with and without these additional terms, and with the full dynamics and compare their relative relaxation. These simulations may shed more light also on the relevance of these cross-force terms. We leave more detailed investigation for future work. 

\bibliography{binding}